# Revealing evolving affinity between Coulombic reversibility and hysteretic Li-Si phase transformations

**K. Ogata[1,2*§], SH. Joen[1*§], DS. Ko[1*], IS. Jung[1], JH. Kim[1], K. Ito[3], Y. Kubo[3], K. Takei[1], S. Saito[2], YH. Cho[1], HS. Park[1], JH. Jang[1], HG. Kim[1], JH. Kim[1], YS. Kim[1], M. Koh[1], K.Uosaki[3], SG. Doo[1], YI. Hwang[1], SS. Han[1§]**

1. Samsung Advanced Institute of Technology (SAIT), Samsung Electronics, Samsung-ro 130, Suwon, Gyeonggi-do, 16678, Korea

2. Samsung Research Institute of Japan (SRJ), Samsung Electronics, 2-1-11, Senba-nishi, Mino-shi, Osaka-fu, 562-0036, Japan

3. C4GR-GREEN, National Institute for Materials Science (NIMS), 1-1 Namiki, Tsukuba, Ibaraki, 305-0044, Japan

*These authors equally contributed to the work

§Corresponding authors

## Abstract

Nano-structured silicon anodes are attractive alternatives to graphite in Li-ion batteries. Despite recent remarkable progresses in numerous Si/C composites, the commercialisation with significance is still limited. One of the most critical issues remained to understand is fundamentals on Li–Si Coulombic efficiency (CE). Particularly, it is key to quantitatively and qualitatively resolve CE alterations and evolutions by the various Li–Si structural changes over longer cycling. However, such work is surprisingly scarce. Here, we provide new findings that iterating the hysteretic *amorphous–crystalline* Li-Si phase transformations accumulatively governs CE evolutions, the manner of which is numerically distinguished from incremental *amorphous* Li-Si volume changes. The iterations, usually featured as capacity degradation factors, can form the most efficient CE profiles over hundreds of cycles, i.e. minimising accumulative irreversible Li consumption, among the given Li–Si reaction sequences. Combined with atomistic probing methodologies, we show that the iteration drastically alters electrochemical and structural characteristics, which is synchronised with the CE behaviours.

(158 words)

## Introduction

Si is an extremely attractive candidate for replacing commonly used graphite (Gr) as a Li-ion battery (LIB) negative electrode owing to its significantly higher specific capacity (~3579 mAh/g at room temperature, assuming $Li_{3.75}Si$). [1] However, the high capacity is associated with huge volume changes (~270–300%),[1] which results in capacity loss and prolonged irreversible reactions. Toward major use of Si, a number of elaborately engineered Si composites have recently been examined, [2-15] which can reasonably accommodate volume changes and retain the capacity over hundreds or thousands cycles.[2,3,11-13,15,16] Further, recent analytic developments via *in situ* and *ex situ* methodologies have also deepened the mechanistic understandings. [17-30] These studies show that *crystalline*-Si (*c*-Si) is converted into *amorphous*-$Li_xSi$ (*a*-$Li_xSi$) phases during the first lithiation, which involves large asymmetric volume changes [22] owning to different Li reaction rate constants at different *c*-Si facets.[31] Upon lithiation, *a*-$Li_xSi$ transforms inhomogeneously into metastable *crystalline*-$Li_{3.75}Si$ (*c*-$Li_{3.75}Si$) at low voltages (<70 mV versus Li), [21,23-27,30,32] and over-lithiated phases such as *c*-$Li_{3.75+\delta}Si$, [23,25] at room temperature, which are associated with a large overpotential on delithiation (430–450 mV),[26,30] different interfacial formation,[33] and extra capacity loss.[7,26,30] Thus, Li-Si processes on (de)lithiation can be either asymmetric or symmetric with *c*-$Li_{3.75(+\delta)}Si$ presence or absence, respectively, involving complex Li-Si energetics and Si clusters interplays. Despite all these insights, the anodes with higher Si concentrations still struggle for emerging on the market.

One of most critical bottlenecks remained to be explored toward emergence of the Si-dominated anode is capacity loss via prolonged Li irreversible consumption in the Li-Si processes, often quantified by Coulombic efficiency (CE, delithiation/lithiation capacity ratio).[34-36] This is because Li atom supply in pragmatic full cells is limited by the cathode loading, unlike the unlimited supply in half cells. Also because the volume changes and the capacity decay are somehow manageable for longer cycling by state-of-the-art commercial Si/C composites, but CE is not. This is serious particularly when Si volumetric density in the composites gets higher. CE is strongly associated with by-product formation (solid electrolyte interphase; SEI) [33,37-39] at the Si–electrolyte interface and/or Li trap in Si owing to the unique volume changes on (de)lithiation. Hence, one intuitive strategy to achieve higher CE is to limit excessive electrolyte invasion into Si interface by forming protective shells/coating around Si. [2,16,40] Nevertheless, electrolyte can still invade due to Li-ions' transport nature (coupled with organic components) and/or due to gradual composites' deformation upon iterative volume changes even with internal-pore-engineered structures. Hence, scenarios that Si can be anyway exposed to electrolytes need to be ultimately considered for understanding CE fundamentals.

One of the most basic is to understand evolving CE alterations by different Li–Si reaction paths over longer cycles, being fully exposed to representative electrolytes. More specifically, the key is to quantitatively separate CE alterations by incremental *amorphous* Li–Si volume changes and that by *amorphous-crystalline* (*a-c*)-Li–Si phase transformations. However, such CE insights remain surprisingly scarce. Achieving this necessitates a few prerequisites, which are hardly considered in the previous works. Firstly, CE accuracy due to instrumental precision and electrode reproducibility needs to be well-defined and constrained small enough to argue potentially subtle CE alterations. Secondly, it is necessary to reference empirical Si reversible capacity in the first cycle to the theoretical one such that Li–Si lithiation depth in the following cycles is numerically controlled by capacity-controlled depth of discharge (DOD). Thus, the abrupt *a-c* transformation can occur near DOD 100%. Only after achieving this, the different Li–Si reaction paths, i.e. *amorphous* Li–Si volume changes and the *a-c* phase transformations, can be quantitatively and qualitatively separated. A misunderstanding often arises that exploring the *a-c* transformation is not the matter because practical state of charge (SOC) for the anode in full cell systems is usually less than 100 %, i.e. $x < 3.75$ in $Li_xSi$ on *average*. However, this is not really the case with realistic current density because the Li–Si processes involve large Li-concentration inhomogeneity across the electrodes and overpotentials under kinetic cycling conditions (see Methods). Hence, revealing such CE alterations by a deeper range of DOD is of significant importance.

Here, we quantitatively demonstrate how CE is altered by incremental *a*-Li–Si structural changes and by the *a-c* transformations by precise controls over DOD % in a range of Li-rich Li–Si phases. The electrodes are designed to satisfy the prerequisites; CE error is constrained within ±0.1% and that the empirical and theoretical reversible capacities agree by ~99 % accuracy (see Methods). Also, presence/absence of the *a-c* transformation can be abruptly controlled between DOD95 and DOD100%. The electrochemical probing on CE alterations is further combined with atomistic probing methodologies such as *ex situ* X-ray absorption fine structure (XAFS), *ex situ* magic-angle-spinning solid-state nuclear magnetic resonance (MAS *ss*-NMR), density functional theory (DFT), *ex situ* X-ray diffraction (XRD), *ex situ* transmission electron microscopy (TEM/STEM), and *ex situ* X-ray photoelectron spectroscopy (XPS). Complimentarily combined all these together, alterations of the structural/interfacial characteristics and the consequent Li–Si reaction mechanism are clearly associated with CE behaviours.

## Experimental

### ***Reference electrode fabrication and cycling conditions***

Two different active materials, in a form of spherical secondary particles, are fabricated via spray-dry method with different Si concentrations (named type-A and -B, respectively) as shown in Figure 1a–d, which consist of commercially available poly-crystalline Si nano-particles, multi-wall carbon nanotubes (MWCNT), with/without flake-type graphite (Gr). The theoretically calculated reversible capacities agree with empirical ones by ~99 % accuracy (see Methods) and other physical parameters in active materials are also well controlled (Supplementary Table S 1). Li–Si processes dominate capacity portions in type-A and -B electrodes, being larger than 94.7% and 99.5%, respectively. As shown in Figure 1b–d, the active materials are designed with porous open structures for easy wetting with the electrolyte. Accordingly these electrodes undergo relatively abrupt *a-c* phase transformations between DOD95% and 100 % even at higher current-rates (Figure 1e,f and Supplementary Fig.S 1,2). CE errors are overall constrained within ±0.1 % as shown in Supplementary Fig.S 3 (see Methods). Electrodes are cycled in 2032-type coin half cells under constant current constant voltage (CCCV) cycling on lithiation (held at 0.01 V until current reaches 0.01 C), and constant current (CC) during delithiation (1.5 V cutoff). For the first two cycles, all the electrodes are cycled at 0.1C and 0.2C to fully amorphise *c*-Si. This is followed by DOD control regime (DOD70–100%) cycled at 1C, in which DOD X% is referenced by the lithiation capacity profile of DOD100% in staircase manners as shown in Figure 2a,b and Supplementary Fig.S 4a,b. X in DOD X% is incremented from 70–100 % with a constant pitch so that the effect of the *a-c* transformation and the mere volume change on CE can be separated. For all the structural probing, a recovery cycling condition (at 0.02-0.1C under DOD100% protocol) is inserted every $20^{th}$ cycle regardless of ever-cycled DOD pathways. Our electrochemical and structural probing schemes are summarised in Supplementary Table 2.

### ***Li-Si electrochemical processes for different DOD pathways***

To identify the Li–Si processes from the *dQ/dV* profiles of the electrodes, we use the same notation as in our previous study. [25] Si#d-X and Si#c-X are Li–Si processes, and Gr#d-X and Gr#c-X are Li–Gr processes; #d-X and #c-X denote the $X^{th}$ discharge (lithiation) and charge (delithiation) processes in the half cells, respectively, which is summarised in Supplementary Table S **2**. Electrochemical probing scheme and new findings on Li di processes are summarised in Supplementary Table S **3** and 4. Note that Liy Table S g scin type-A electrode can be separated from Li–Si processes (Supplementary Fig.S 5). Li-Si processes for the first two cycles in *dQ/dV* (Figure 2c–f and Supplementary Fig.S 4c–f) show asymmetric reaction sequence, which is in a good agreement with the previous studies [25,27,41] (see Methods). For the first 44 cycles under DOD100% cycling protocol, the asymmetric sequence gradually changes to symmetric one. On lithiation, Si#d2 ($a$-Si → $a$-$Li_{2.0}Si$) and Si#d3 ($a$-$Li_{2.0}Si$→ $a$-$Li_{3.5}Si$) are consistently present, while Si#d4 ($a$-$Li_{3.5\text{-}3.7}Si$→ $c$-$Li_{3.75(+\delta)}Si$) intensity gradually decreases and widens. On delithiation, no change in Si#c1 ($c$-$Li_{3.75(+\delta)}Si$→ $c$-$Li_{3.75(-\delta)}Si$), followed by a significant decrease of Si#c3 ($c$-$Li_{3.75(-\delta)}Si$ → $c$-$Li_{\sim 1.1}Si$) and concurrent increase of Si#c2 ($a$-$Li_{3.5\text{-}3.75}Si$→$a$-$Li_{2.0}Si$). Eventually, Si#c3 is overshadowed by Si#c2 and Si#c4 ($a$-$Li_{2.0}Si$→$a$-$Li_{x<1.1}Si$) after the 65$^{th}$ cycle and disappeared after the 86$^{th}$ cycle. Importantly, this alteration is consistent even under quasi-thermodynamic cycling conditions (Supplementary Fig.S 6, 7). Thus, iterative DOD100 % cycling protocol self-alters the Li-Si reaction path over cycling with ~20–23 % capacity loss. Note that the alteration is not due to increasing resistance in Li-metal in half-cells (see Methods and Supplementary Fig.S 8, 9, 10). This trend applies to the other $c$-Si nano-powder sources (Supplementary Fig.S 11). On the other hand, when the DOD is controlled from 70-90 %, the alteration is significantly delayed by 40–70 cycles (Supplementary Fig.S 6, 7), Si#c3 disappearing completely only after ~100–150$^{th}$ cycles. Extra capacity decay due to the iterative $c$-$Li_{3.75(+\delta)}Si$ (de)formation [7,26,30,41] is shown in (Supplementary Fig.S 12) and also discussed in Methods part.

***Coulombic reversibility of Li–Si processes for different DOD pathways***

The following shows our new findings on CE alterations by various Li–Si reaction paths over ~190 cycles. Over the first 20 cycles, CE trends under DOD100% cycling protocol show an abrupt decrease down to 96.5% and 97% for type-A and -B electrodes, respectively (Figure 2a,b and Supplementary Fig.S 4a,b). These drops are followed by CE surges and saturations around ~99.9% after ~80 cycles with ~25–35 % capacity loss even without nano-engineered passivation shell/coating around Si. Contribution of Li–Gr processes (Supplementary Fig.S 13) to the overall CE alterations in these electrode systems is negligible and discussed in details in Methods part. In contrast, CE trends under DOD70–90% cycling protocols significantly differ from the one under DOD100%. The initial CE decrease is less severe, while the subsequent increase is more moderate, reaching ~99.7–99.9% only after 140–200 cycles. The initial CE drop for DOD100 and DOD70-90 % differs by ~1.5 %, which shows sacrificial CE drop by the iterative $c$-$Li_{3.75(+\delta)}Si$ (de)formation. As DOD decreases from 90 to 70 %, the initial CE decrease and subsequent gradually increase are somewhat delayed at a constant manner. This clearly shows that a CE profile non-linearly changes between DOD70–90 and DOD100%. Interestingly, electrode thickness trends (Supplementary Fig.S 14) are in line with such non-linearity. This indicates that irreversible Li consumption in the electrodes is correlated with the iterative $c$-$Li_{3.75(+\delta)}Si$ (de)formation.

One issue to be carefully considered is the difference in Si-to-electrolyte exposure-time for different DOD controls during CCCV on lithiation. This is because, under the higher current-rate (e.g. 1C), CCCV duration becomes not proportional to DOD X% owing to exponential current decay in the CV domain. Consequently, the duration can be much longer under DOD100% compared with other DODs (Supplementary Fig.S 15, 16a,b). Therefore, separate experiments at 0.1C are also conducted under the same DOD controls such that the CCCV duration can be controlled to be nearly proportional to DOD% at this rate (Supplementary Fig.S 15, 16c,d). The result show that CE profiles at 0.1 C are similar to those at 1 C (Supplementary Fig.S 15,16e–g). These results suggest that the non-linear CE transition between DOD70–90 and DOD100 % is more prominently triggered by iterative *c*-$Li_{3.75(+\delta)}Si$ (de)formations rather than by incremental volume changes nor by different Si-to-electrolyte exposure times.

To quantitatively and qualitatively explore this in more details, the electrodes under DOD90% are cycled for X cycles (X=22, 43, 64, 85, and 106) and then abruptly switched to DOD100 % from cycle X+1 onward to iterate *c*-$Li_{3.75(+\delta)}Si$ on purpose. Figure 2g,h and Supplementary Fig.S 4g,h show that following the switch to DOD100%, the CE undergoes a sudden decrease followed by a rapid increase for all X. The depth and width of these CE *valleys* at the switching points become shallower and narrower as X increases, and number of cycles that is required to saturate CE decreases as cycling progresses. In contrast, when the cycling pathway under DOD100% in the symmetric reaction regime is switched to the one under lower DOD controls, CE barely changes. These results indicate that duration of inherently remaining asymmetric regime in the given electrode systems determines CE projections in the following cycles under given DOD controls. In other words, CE gets higher and becomes less susceptible to *c*-$Li_{3.75(+\delta)}Si$ presence as the inherent regime gets closer to the symmetric one.

Figure 2i and Supplementary Fig.S 4i show the CE profiles plotted against reversible capacity for type-A and B electrodes, respectively. These plots show that when compared under the same residual reversible capacities, CE can be significantly altered by the ever-cycled phase-transformation pathways. The pathway under DOD100% exhibits the highest reversibility of all the given DODs when the residual capacity gets below ~77–80%. Further, Supplementary Fig.S 17 provides CE profiles plotted over irreversibly consumed accumulative Li capacities, showing that DOD100 % cycling protocol can reach the highest CE of all the given DOD controls after a certain irreversible Li consumption. These findings show that Li–Si phase transformation histories and consequent residual periods of the inherent asymmetric regime quantitatively/qualitatively govern Li–Si irreversible nature. Most importantly, we for the first time provide evidence on a positive aspect of *c*-$Li_{3.75(+\delta)}Si$ (de)formation, which has a potential to evolutionally upheave CE, minimising accumulative Li loss with the manageable capacity losses. These findings are summarised in Supplementary Table S **4**.

***Morphological analysis via electron microscopies***

*Ex situ* TEM imaging is conducted to investigate Si morphological change at every recovery cycling point for under different DOD controls (see Methods). Through the initial amorphisation process, Si transforms from poly-crystalline (Figure 3a) into fully porous amorphised structures (Figure 3b, also see Methods). Selective area diffraction patterns (SADPs) at 10 mV in the 3rd cycle clearly indexes *c*-$Li_{3.75(+\delta)}Si$ (Figure 4a–2). Dark-field TEM images (Figure 4a–2) show that the size of *c*-$Li_{3.75(+\delta)}Si$ grains are on the order of few dozens of nm. Elemental mapping of fully-delithiated Si at recovery cycling points via electron energy loss spectroscopy (EELS, Figure 3a3–5,b4–7) and energy dispersion spectroscopy (EDS, Supplementary Fig.S 18-22) is given in Methods.

Over the initial 23 cycles, for both DOD100% and 90 %, the amorphised spherical structures drastically change morphology, by expanding, and merging with neighbors, resulting in widespread three-dimensional networked structures (Supplementary Fig.S 18–21). After the 23rd cycle, there is more bulky core remained in the structures for DOD 90% compared with that for DOD100% (Figure 3c,d). Supplementary Fig.S 23 shows the average thicknesses (*d*) of 100 randomly picked delithiated Si feature size for DOD90 and 100 %; *d* and standard deviation (*stdev*) after 107th cycle for DOD80%, 90%, and 100 % show *d*~5.9 (*stdev*~2.6), 5.8 (~2.7), and 4.8 nm (~1.0 nm), respectively. This indicates that iterative *c*-$Li_{3.75(+\delta)}Si$ (de)formation leads to a slightly decreased thickness with more clearly defined distribution. At 10 mV after the 107th cycle, *c*-$Li_{3.75(+\delta)}Si$ crystal sizes get much smaller than those in the 3rd cycle (Figure 4b,c). The elemental mappings via EDS and EELS over the first 107 cycles (Figure 3, Supplementary Fig.S 18 and 21) show that SEI distribution gets more uniform along with the morphological changes, indicating that Li-ions can more uniformly access to the complex structures. *Ex situ* XPS analysis under DOD80–100% is presented in Methods and Supplementary Fig.S 24.

***c-$Li_{3.75(+\delta)}Si$ analysis via ex situ XRD***

Figure 4d shows *ex situ* XRD profiles at 0.01 V (see Methods) under DOD 80–100 % over 190 cycles. For all DODs, non-negligible *c*-$Li_{3.75(+\delta)}Si$ reflection profiles are continuously confirmed. The *c*-$Li_{3.75(+\delta)}Si$ presence at 10 mV and the following symmetric delithiation processes contradict previous works,[25-27,30] in which the phase primarily controls the energetics on delithiation, resulting in the hysteretic plateau at 430 mV. To explore this inconsistency, XRD at different potentials on delithiation for DOD100% is conducted (Figure 4e,f) in asymmetric (23rd cycle) and symmetric (86th cycle) regimes for DOD100% cycling protocol. On the 23rd cycle, the reflection corresponding to *c*-$Li_{3.75(+\delta)}Si$ is still present at 250 mV on delithiation and only disappears at 550 mV.[25-28] In contrast, after the 86th cycle, $c$-$Li_{3.75(+\delta)}Si$ reflection at 10 mV significantly diminishes at 150 mV on delithiation, the empirical stoichiometry still being x > 3.2 in $a$-$Li_xSi$. These results show that Li ions enables de-coupling from $c$-$Li_{3.75(+\delta)}Si$ to form $a$-$Li_xSi$ (3.2 <x< 3.75) prior to the Si#c2 process. This is probably due to different Li–Si delithiation energetics altered by the significant morphological change from bulky to more surface dominated system and to different Si interfacial properties. Hereafter, the new process is named Si#c1' ($c$-$Li_{3.75(+\delta)}Si$ → $a$-$Li_xSi$, 3.2 <x< 3.75). These results exclude a possibility that Si#c3 absence is simply due to electrically/ionically isolated Li-ions in $c$-$Li_{3.75(+\delta)}Si$ during delithiation.

Figure 4g shows the full width of half maximum (FWHM) of $c$-$Li_{3.75(+\delta)}Si$ (332) reflection over 190 cycles. For the first 23 cycles, FWHM does not significantly change for any of the DODs, which indicates that average Li–Si feature size is large enough to accommodate a $c$-$Li_{3.75(+\delta)}Si$ grain. From the 23rd cycle onward, FWHM for DOD100 % shows a steeper increase than those of DOD90% and 80 %. The FWHM increase for DOD100 % agrees with the decrease of $c$-$Li_{3.75(+\delta)}Si$ crystal size observed in Figure 4a,b. In the course of iterative $c$-$Li_{3.75(+\delta)}Si$ (de)formation, FWHM is nearly saturated after the 86th cycle. The suppressed increase of FWHM after 86th cycle (Figure 2a and Supplementary Fig.S 4a) is probably due to more effective stress release and consequent more non-destructive structural changes in the sub-5 nm surface-dominated structures. In contrast, FWHM for DOD 90-80 % increases less steeper, reaching the same FWHM level as that for DOD100% only after 100th–150th cycle. Interestingly, this is reminiscent of the CE and electrode thickness trends. This is probably because lower DOD controls can leave chunky Si remained in $a$-Si for longer-cycling (Figure 3d and Supplementary Fig.S 20, 21) which may act as nucleation sites of larger $c$-$Li_{3.75(+\delta)}Si$.

***<u>Local structure probing via ex situ $^7$Li ss-NMR spectroscopy</u>***

In the *ex situ* $^7$Li *ss*-NMR analysis (see Methods), the $^7$Li resonance notation used in the previous study is used:[25] resonances at around 20–10, 10–0, and –10 ppm are labeled respectively as P1 (Li near small Si clusters, correlated with Si#d3 and Si#c2 processes), P2 (isolated Si anions, $c$-$Li_{3.75}Si$ correlated with Si#d4, Si#c3, and extended Si networks correlated with Si#d2 and Si#c4), and P3 (overlithiated crystalline phase, $c$-$Li_{3.75+\delta}Si$, correlated with Si#d4, Si#c1). Summary of these notations on (de)lithiation are summarised in Supplementary Table S **2**. Spectra at different potentials are recorded at the recovery points for DOD100% and 90% over 107 cycles.

In the 23rd cycle, the profile shows signatures of the asymmetric reactions (Figure 5) both for DOD100% and 90% as seen in the previous work (see Methods).[25] However, after the 65th cycle, significantly different behaviour is observed between DOD100 and 90 %. For DOD100%, while the spectra until 80 mV are similar to those at the earlier cycling stages, the spectrum at 10 mV is dominated by P2 *without* P3. In combination with the XRD results, this finding indicates that a non-negligible amount of isolated Si anions remains in the $c$-$Li_{3.75}Si$ state without forming $+\delta$ environments in $c$-$Li_{3.75(+\delta)}Si$. On delithiation, most isolated Si anions in $c$-$Li_{3.75}Si$ can reform Si small clusters with P1 increase at 150 mV. The profile well overlaps with that at 80 mV on lithiation, indicative of symmetric interplays between Si small clusters and isolated Si anions. This agrees with the $c$-$Li_{3.75(+\delta)}Si$ reflection disappearance at 150 mV on delithiation in XRD and newly assigned process, Si#c1'. This is in contrast to the complex hysteretic interactions in the asymmetric regime among Si small clusters, isolated Si anions, overlithiated Si and larger Si clusters. Importantly, such NMR spectra evolution well agrees with the electrochemical signatures' change by the regime shift. It is to be noted that $c$-$Li_{3.75}Si$ nuclei is not further lithiated to form $+\delta$ in $c$-$Li_{3.75+\delta}Si$ despite more favorable energetics of forming $c$-$Li_{3.75+\delta}Si$ than breaking residual Si–Si bonding.[25] In contrast, the spectra for DOD90% in the 65th cycle show mixed interplays of asymmetric and symmetric local environments (see Methods). The sequence only becomes more symmetric after the 107th cycle, which is also in accordance with the delayed shift from asymmetric to symmetric profiles in *dQ/dV* for DOD90% (Supplementary Fig.S 6, 7).

***Local structure probing via ex situ XAFS***

*Ex situ* XAFS (X-ray absorption near-edge structure, XANES, and extended X-ray absorption fine structure, EXAFS) at the Si K-edge is conducted to analyse Si local structure in fully delithiated type-A electrodes cycled at the recovery cycling points under different DOD controls (see Methods). The XANES profiles are summarised in Supplementary Fig.S 25a. The EXAFS profiles are extracted (Supplementary Fig.S 25b) from the XAFS data and converted into Fourier transformed profiles, which are relevant to radial distribution function (RDF). Figure 6a clearly shows that Si-Si correlations longer than 3.0 Å almost disappear after the 2nd cycle, indicating that Si is fully amorphised only leaving 2 Å Si–Si tetrahedral correlations. Normalized integrations of 2 Å Si–Si correlations peaks, thereafter named as $A_{(2Å\ Si–Si)}$, are used to index de-lithiated $a$-Si local environments. $A_{(2Å\ Si–Si)}$ for various cycles is shown in Figure 6b. After the 2nd cycle, $A_{(2Å\ Si–Si)}$ suddenly decreases to ~0.81, which is in line with the initial huge structural change into a complex porous sphere seen in the TEM images (Figure 3b). Over the next dozens of cycles, $A_{(2Å\ Si–Si)}$ increases for all DODs, reaching a local maximum with different timings: at the 23rd cycle for DOD100% and at the 65th cycle for DOD90–80%. After reaching the local maxima with different timings for different DOD controls, $A_{(2Å\ Si–Si)}$ linearly decreases with different gradients; for DOD100%, $A_{(2Å\ Si–Si)}$ starts to decrease after 23rd reaching 0.61 after the 149th cycle, and for DOD90% and 80 %, it does after the 65th or later cycle, reaching 0.78 and 0.87 after the 149th cycle, respectively. The $A_{(2Å\ Si–Si)}$ increase and the following decrease corresponds to the temporal $a$-Si agglomeration with neighbors and the following Si size decrease with the narrower size distribution. The iterative $c$-$Li_{3.75(+\delta)}Si$ (de)formation accelerates the $A_{(2Å\ Si–Si)}$ changes over cycling. Such behaviour also supports the accelerated structural shift from bulk to surface dominated systems in TEM imaging, the FWHM trends (Figure 4g), and the local environments' interplays in NMR (Figure 5).

When CE is plotted as a function of $A_{(2Å\ Si–Si)}$, interestingly, the curves in Figure 6c are constrained by different DOD protocols. Regardless of the DOD conditions, CE from $A_{(2Å\ Si–Si)}$ ~0.81 to 0.95 is significantly susceptible to presence of $c$-$Li_{3.75(+\delta)}Si$ on (de)lithiation. The CE for DOD100% can further drop by ~1.5 % compared to that for DOD80-90 % at the same $A_{(2Å\ Si–Si)}$ values, which corresponds to the initial sacrificial CE drop by the $c$-$Li_{3.75(+\delta)}Si$ iteration. Once $A_{(2Å\ Si–Si)}$ reaches ~0.95, the course is reversed for all DODs and presence/absence of $Li_{3.75(+\delta)}Si$ on (de)formation significantly controls CE increase per unit $A_{(2Å\ Si–Si)}$. CE for DOD 100% quickly increases from 96.4% to 99.5 % as $A_{(2Å\ Si–Si)}$ decreases from 0.94 to 0.86 after the 65$^{th}$ cycle. In contrast, the CE that for DOD80-90 % shows a gentler increase, reaching ~99.5 % after 149 cycles. Interestingly, when $A_{(2Å\ Si–Si)}$ <~0.8–0.85, CE is insusceptible to DOD controls, in contrast, when $A_{(2Å\ Si–Si)}$ >~0.8–0.85, CE is significantly susceptible to $c$-$Li_{3.75(+\delta)}Si$ presence. This indicates that +δ presence ($A_{(2Å\ Si–Si)}$ >~0.8–0.85) and absence ($A_{(2Å\ Si–Si)}$ <~0.8–0.85) in $c$-$Li_{3.75(+\delta)}Si$ may at least partly affect the susceptibility due to its reducing nature.[23] It is to be noted that the turning point at $A_{(2Å\ Si–Si)}$~0.8–0.85 is associated with the transition from the asymmetric to symmetric regime. Such CE behaviours also support that the $A_{(2Å\ Si–Si)}$ shift through 0.8–0.85 is closely related to the shift from bulk to more surface dominated systems as well as that from the asymmetric to the symmetric regime.

## ***Discussion***

The structural probing analyses over 190 cycles are schematically provided in Figure 7a-c, combined with CE profiles under the different DOD protocols. Figure 7a shows a flow chart of overall structural probing scheme (Supplementary Table S **3**) at the recovery points in this study. In Figure 7b, local Li-Si environments and $a$-Si interfacial properties are classified into different colour schemes, which are complementarily interpreted from our mechanistic findings summarised in Supplementary Table S 5. The relative amounts of the local Li–Si environments and their portfolio at each potential are represented by pie charts in Figure 7c at the target cycles.

Over cycling, electrochemical Li–Si reaction sequence on (de)lithiation is altered from the asymmetric to the symmetric regime. As shown in the previous sections, the iterative $c$-$Li_{3.75(+\delta)}Si$ (de)formation drastically accelerates a timing of the regime shift; around the 65$^{th}$ cycle for DOD100%, in contrast, the 100-140th cycle for DOD70-90% cycling protocols. In the course of such regime shift, structural/interfacial characteristics are concurrently changed. In the asymmetric regime, on lithiation, *a*-Si gradually turns into large Si clusters/extended Si networks, then into small Si clusters, which finally transform into *c*-$Li_{3.75(+\delta)}Si$ (isolated Si and over-lithiated Si anions) at 10 mV. On delithiation, the isolated and overlithiated anions do not reform small Si clusters, instead, asymmetrically transform back into larger Si clusters and extended Si networks. In the symmetric regime, lithiation processes are almost identical to the asymmetric ones, but +δ components (potentially being reduction agents) in *c*-$Li_{3.75(+\delta)}Si$ are not formed at 10 mV. In contrast, on delithiation, isolated Si anions in *c*-$Li_{3.75}Si$ symmetrically reform small Si clusters at 150 mV. Then, small Si clusters turn into large Si clusters/extended Si networks at 300 mV, followed by *a*-Si formation at 550 mV. Thus, the overall processes consequently become more symmetric. These flows are summarised in Supplementary Table S 2. When population index of Si–Si tetrahedral correlation, $A_{(2Å\ Si–Si)}$, decreases near 0.8–0.85, the regime shift consistently occurs regardless of ever-cycled DOD protocols. This agrees with a morphological transition from bulk to more surface dominated Si structures, being sub-5-nm structures. Incremental increase of electrode thickness over cycling is nearly saturated when the regime shifts. This saturation trend is also observed in FWHM in *c*-$Li_{3.75(+\delta)}Si$ XRD reflection. These trends indicate that there are more non-destructive or efficient stress release processes with much less irreversible reactions at Si/SEI interface in the symmetric regime. Further, our preliminary DFT calculation (see Methods and Supplementary Fig.S-26) shows that a driving force for lithiating *a*-$Li_xSi$ beyond x=3.25 in the sub-5-nm-structure is significantly lowered, owing to increased surface energy, compared with that in bulk Si. In such system, i.e. the symmetric regime, Li atoms might be less motivated to inhomogeneously over-lithiate *c*-$Li_{3.75}Si$ nuclei. Instead, breaking residual Si–Si bonds is more preferred, resulting in more uniform lithiation and +δ absence in *c*-$Li_{3.75(+\delta)}Si$. This may partly explain the CE susceptibility change in asymmetric/symmetric regimes.

All of these changes in structural/interfacial characteristics are primarily altered by the accumulative iteration of *c*-$Li_{3.75(+\delta)}Si$ (de)formation, which is clearly distinguished from incremental *amorphous* Li–Si volume changes by the DOD controls. As summarised in Supplementary Table S **4**, electrochemical quasi-thermodynamic reaction pathways are also prominently altered from the asymmetric to symmetric regime by the iteration. The CE behaviours show strong correlations with the affiliated regimes and with subjected DOD controls on (de)lithiation. In other words, the CE behaviours are significantly governed by accumulative *c*-$Li_{3.75(+\delta)}Si$ iteration. More specifically, inherently remaining asymmetric regime can project how CE can develop in the following cycles. Complimentarily combined all these together, the regime shift in the structural/interfacial characteristics and in the consequent Li–Si reaction mechanism is clearly associated with the CE behaviours. Importantly, iterating *c*-$Li_{3.75(+\delta)}Si$ (de)formation, usually featured as capacity degradation factors, can boost CE in the most efficient manners among the given Li–Si reaction pathways, minimising accumulative irreversible Li consumption.

## Conclusion

In this study, we investigated how Coulombic efficiency (CE) can be quantitatively and qualitatively altered over longer cycling by different electrochemical Li–Si reaction paths. The alteration is further associated with various atomistic probing methodologies. CE precision is delicately managed to argue subtle CE alterations. Li–Si reaction paths are quantitatively and qualitatively controlled by capacity-cutoff depth of discharge (DOD) from 70 to 100 %. When cycled under DOD100% protocols with iterative *c*-$Li_{3.75(+\delta)}Si$ (de)formation, quasi-thermodynamic Li-Si processes are discontinuously altered from an asymmetric to symmetric sequence regime. We showed that the CE behaviours can be significantly controlled by iterating *c*-$Li_{3.75(+\delta)}Si$ (de)formation. Particularly, the iteration, usually featured as a capacity degradation factor, can upheave CE up to ~99.9 % in the most efficient manners among various Li–Si reaction routes, minimising accumulative irreversible Li consumption. This is clearly distinguished from incremental *amorphous* Li-Si volume changes cycled under DOD70–90%. Atomistic probing methodologies also show that the iterative *c*-$Li_{3.75(+\delta)}Si$ (de)formation prominently alters structural/interfacial characteristics and Li–Si reaction mechanism on (de)lithiation. The alteration is closely associated with a shift from bulk to more surface dominated systems. Also, local Li–Si reaction sequence shifts from asymmetric to symmetric one. Complimentarily combined all these together, the regime shift in the structural/interfacial characteristics and in the consequent Li–Si reaction mechanism is clearly associated with the CE behaviours, both of which are triggered by the iterative *c*-$Li_{3.75(+\delta)}Si$ (de)formation. The insights in this study can open up new possibilities for designing cells to undergo desirable Li–Si structural transformations that accumulatively minimise Li loss. For example, re-defining anode/cathode capacity ratio and pre-lithiation loading may lead to pushing the anodes one-step-forward not only for advanced Li-ion batteries but also all-solid-state, Li-sulfer, and Li-metal batteries.

**Main figures**

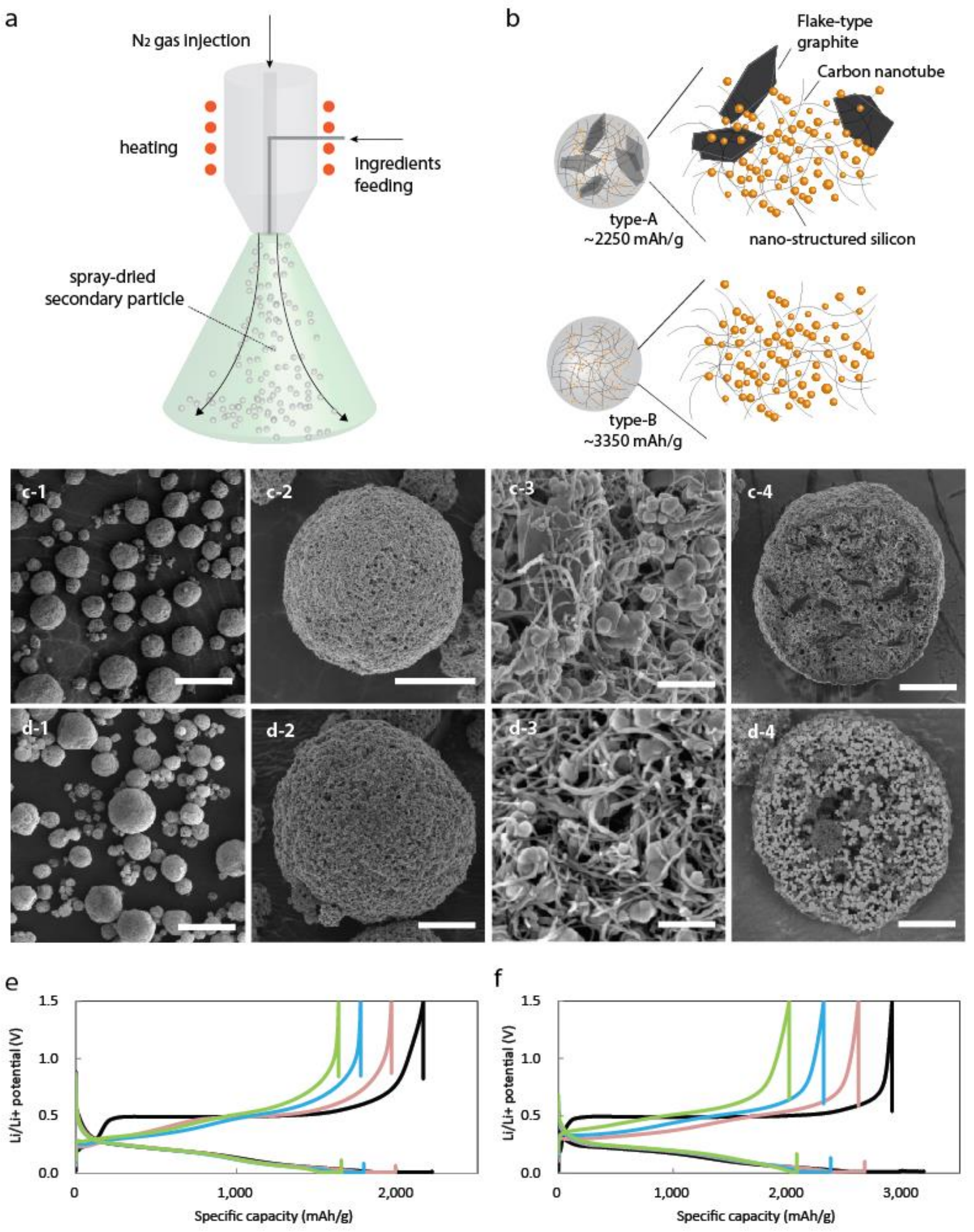

**Figure 1 Schematics and secondary electron images of baseline active materials with potential-capacity profiles under different lithiation depth controls.**

(a) Schematic of spray-drying process for secondary particle fabrication. (b) Shematic of spray-dried secondary particles for type-A (the first reversible capacity being, 2250 mAh/g) and type-B (3350 mAh/g) active materials. (c,d1–4) SEM images of (c) type-A and (d) type-B secondary particles. (c,d1) planar zoomout, (c2–3,d2–3) planar views, (c4,d4) FIB cross-sectional views. The scale bars for (c,d1–4) are 20 µm, 5 µm, 200 nm, and 5 µm, respectively. (e,f) $Li^{/}Li^{+}$ potential as a function of specific capacity on the 10$^{th}$ cycle for different capacity cut-off depth of discharge X% (DOD X%) for (e) type-A and (f) type-B electrodes. The black, pink, blue, and green profiles correspond to cycling under DOD100%, 90%, 80%, and 70%, respectively.

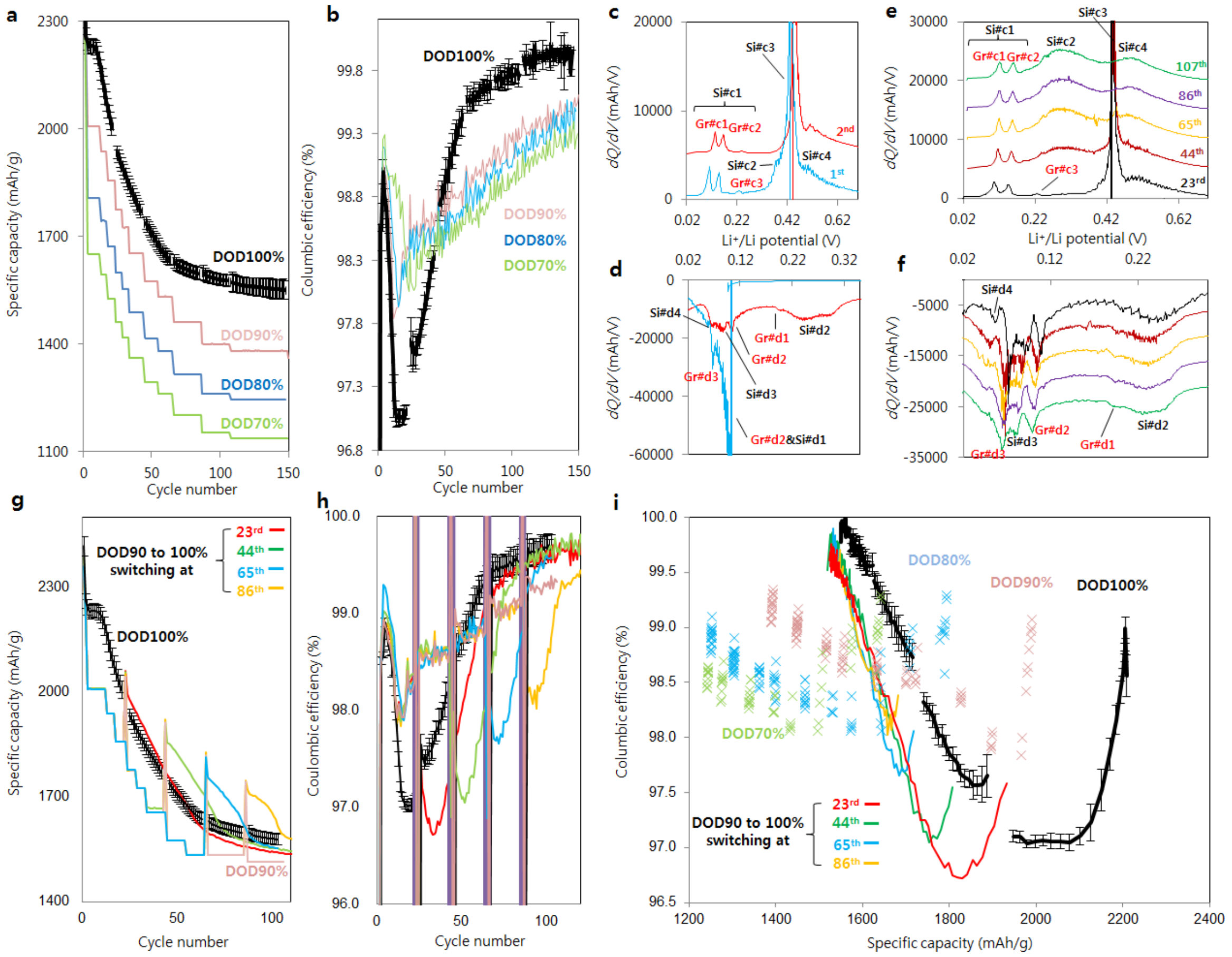

**Figure 2 Baseline electrochemical properties under different lithiation depth controls.**

Electrochemical properties for type-A electrodes in 2032-type coin half cells under different depth of discharge (DOD) controls under CCCV-CC cycling conditions (current cutoff at 0.01 C in the CV domain). (a) Specific capacity and (b) Coulombic efficiency (CE) for different DOD controls over 150 cycles. *dQ/dV* plots cycled under DOD100% during amorphising *c*-Si on (c) delithiation (charge) and (d) lithiation (discharge) at 0.1 C ($1^{st}$ cycle) and 0.2 C ($2^{nd}$ cycle), and at recovery points at 0.1 C on (e) delithiation and (f) lithiation. Note that at the recovery points, all the electrodes are cycled at 0.1 C regardless of the ever-cycled DOD pathways. The different Li–Si processes are labeled as Si#d X/Si#c X and Gr#d X/Gr#c X for Si and Gr for the #$X^{th}$ discharge/charge process; the *dQ/dV* profiles are stacked with a constant pitch to show the different processes more clearly. The notations for Li–Si processes are summarised in Supplementary Table S 2. (g) Specific capacity and (h) CE for DOD90% switched to DOD100% at different cycle numbers; the switching points for red, green, blue, and yellow solid lines are the $23^{rd}$, $44^{th}$, $65^{th}$, and $86^{th}$, respectively. (i) CE as a function of the specific capacity for different DOD controls. Coloured solid lines show the profiles for switching from DOD90% to DOD100% at different cycle numbers in (g,h). CE error bars for DOD70-90% are all within a range of ±0.1 % and omitted in (b,h,i).

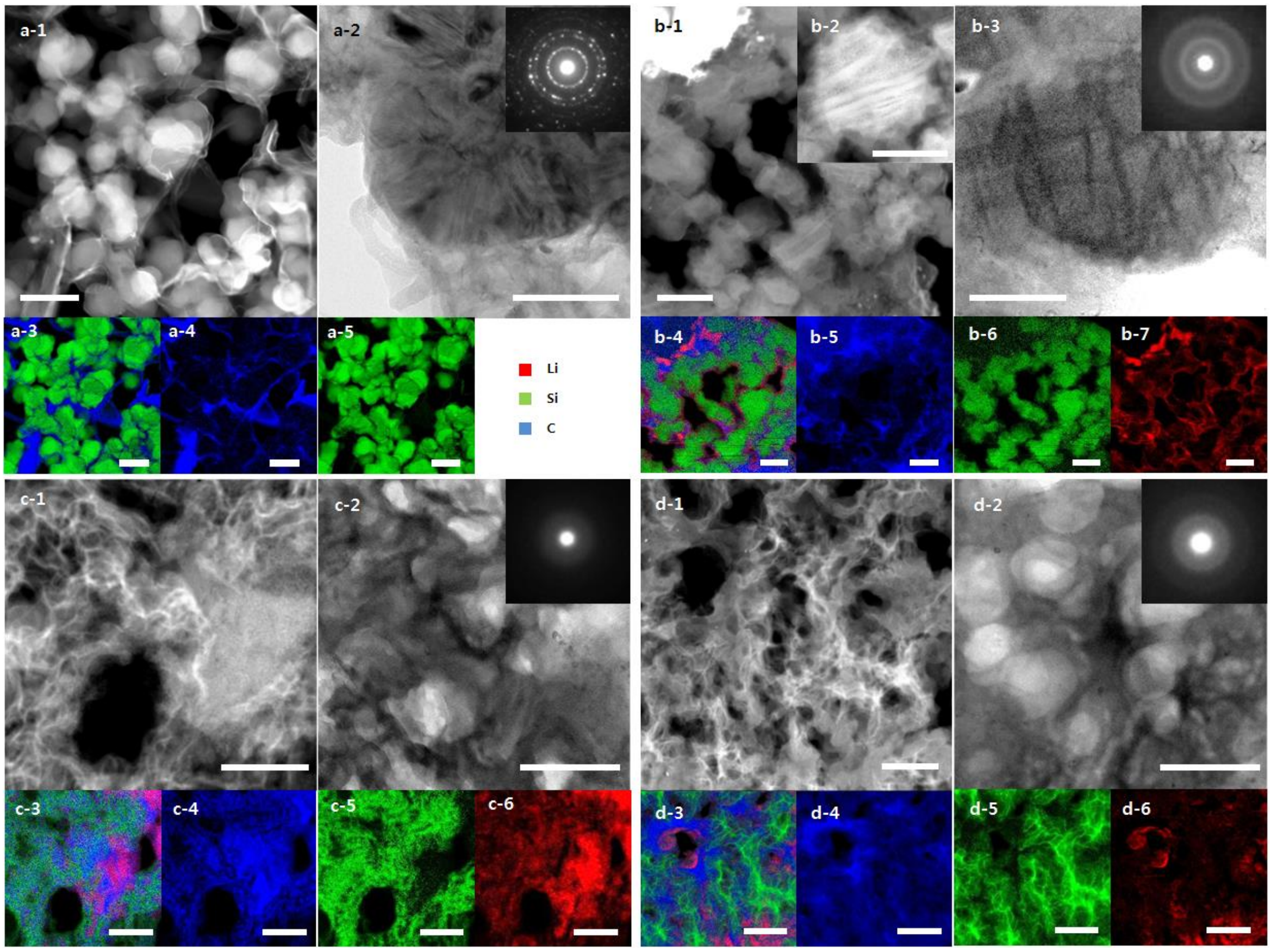

**Figure 3 Electron microscopy images of delithiated electrodes.**

(a1,b1,2,c1,d1) High-angle annular dark field (HAADF), (a2,b3,c2,d2) bright-field (BF) TEM images, and (a3–5,b4–7,c3–6,d3–6) electron energy loss spectroscopy (EELS) spectrum image (SI) elemental mapping. Electrodes (a) before cycling, (b) after the amorphisation of *c*-Si, and after the 65$^{th}$ cycle under (c) depth of discharge (DOD)100%, and (d) DOD90%. Inset selected area diffraction pattern (SADP) in same area as the BF-TEM image. Scale bars are (a1,b1) 200, (a2,b2) 100, (a3–6,b4–7) 200, (b3) 80, (c1–2,d1–2, c3–6, and d3–6) 100 nm. EELS SI elemental mapping is (c4,d4) blue for C, (c5,d5) green for Si, and (c6,d6) red for Li, (a3,b4,c3,d3) EELS SI elements overlaid.

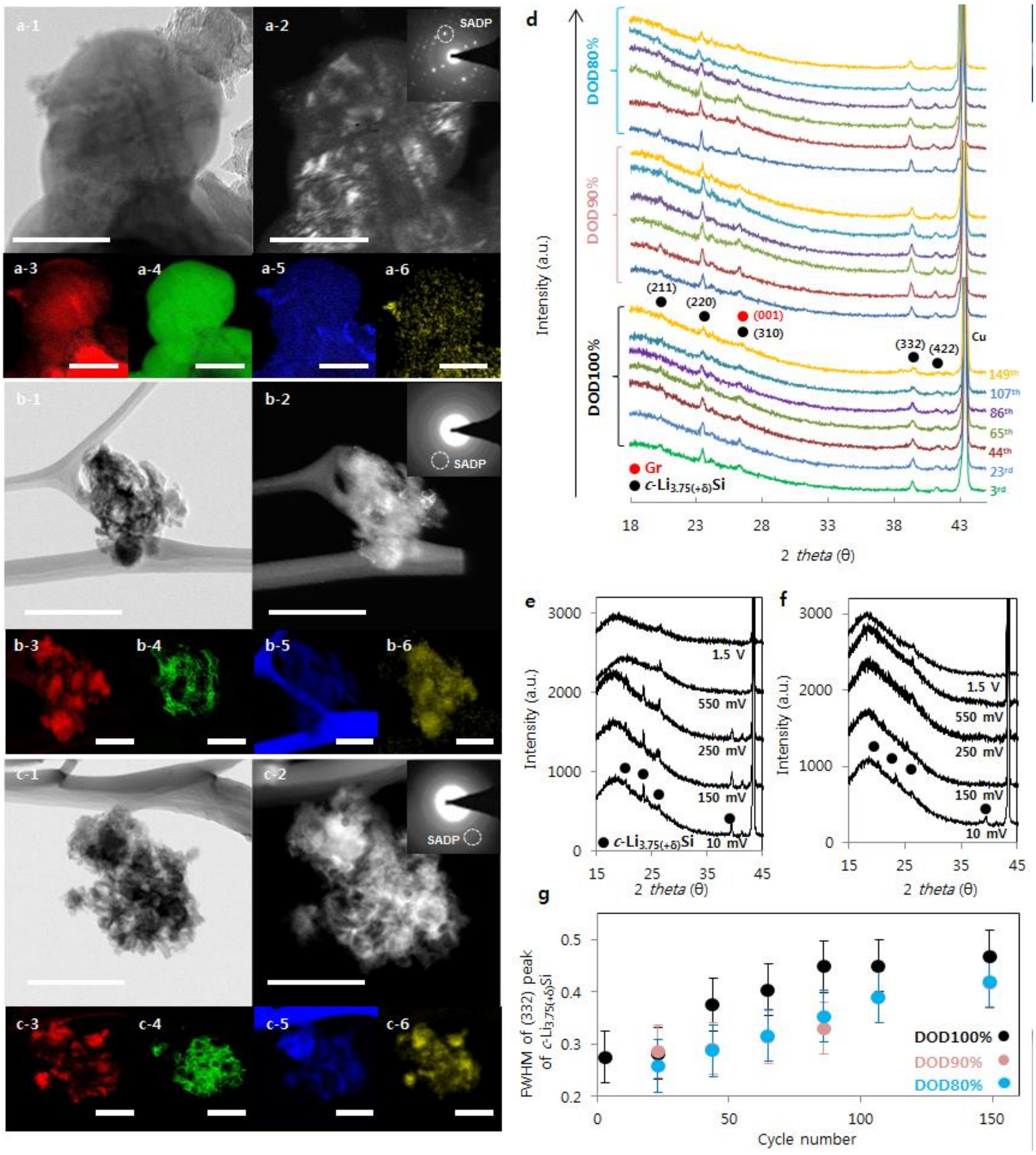

**Figure 4 Electron microscopy images and XRD profiles of lithiated (at 10 mV) electrodes over 150 cycles.**

Fully lithiated type-A electrodes at 10 mV in a different cycle. (a1–c1) TEM images. (a2–c2) Dark-field TEM images with selected area diffraction pattern (SADP) shown in inset, the white dotted circle showing the selected diffraction spot for DFTEM. (a3–6 to c3–6) electron energy loss spectroscopy (EELS) spectrum image (SI) elemental mapping. (a3–c3) red for Li, (a4–c4) green for Si, (a5–c5) blue for C, and (a6–c6) yellow for F. (a) After the amorphisation of *c*-Si, (b) depth of discharge (DOD)100% after the 65$^{th}$ cycle, (c) DOD90% after the 65$^{th}$ cycle. Scale bars are (a1–6) 200 and (b1–6,c1–6) 500 nm. XRD spectra at 10 mV (d) for DOD100%, 90%, and 80% over 190 cycles. XRD spectra at various delithiation potentials for DOD100% after (e) the 23$^{rd}$ cycle and (f) the 86$^{th}$ cycle. (g) Full width of half maximum (FWHM) over 150 cycles for DOD100%, 90%, and 80% for the (332) diffraction of *c*-$Li_{3.75(+\delta)}Si$.

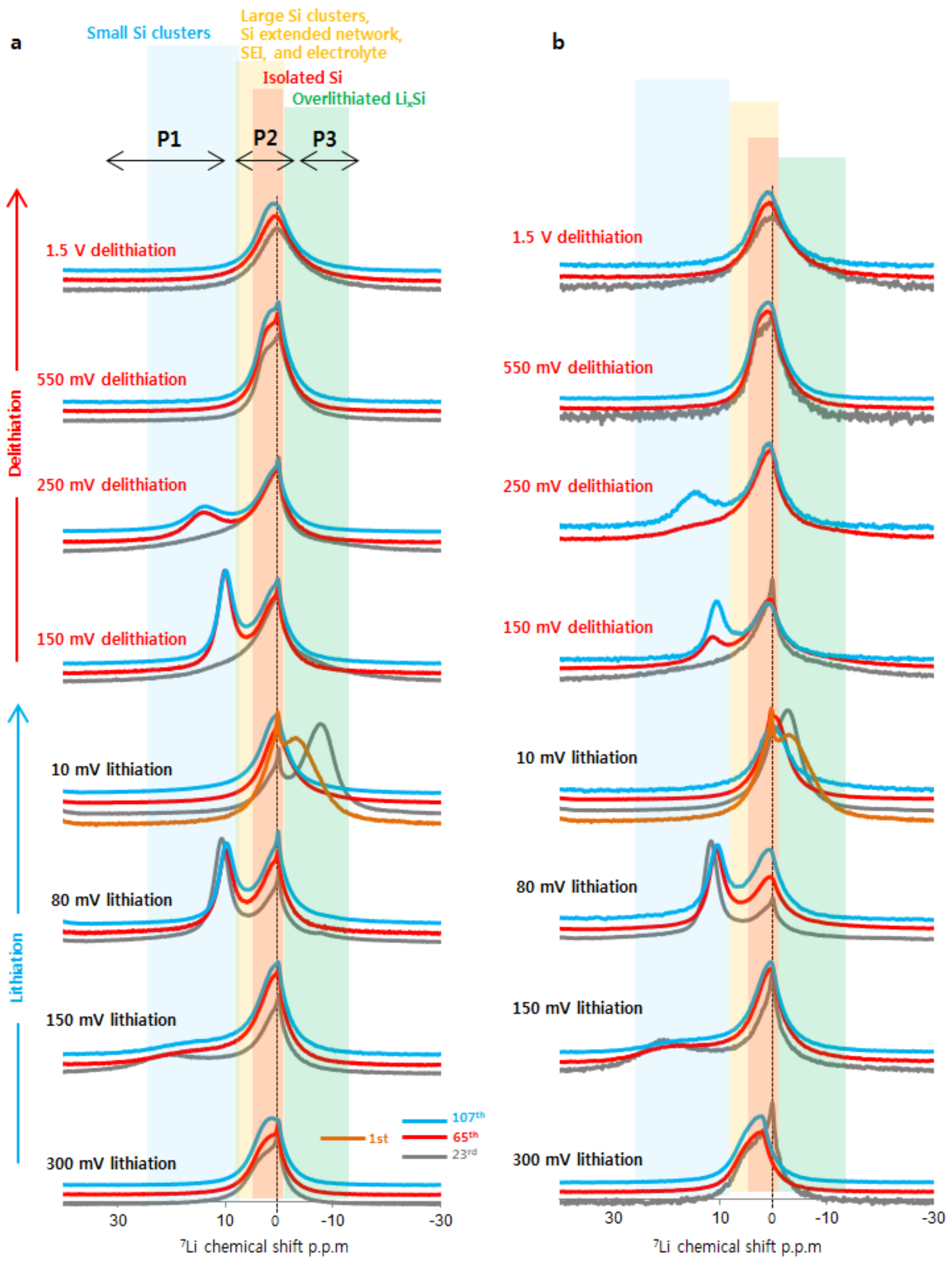

**Figure 5 Stacked *ex situ* $^{7}$Li *ss*-NMR spectra.**

The recorded spectra at various $Li/Li^{+}$ potentials over 107 cycles at the recovery points under (a) depth of discharge 100 % (DOD100%) and (b) DOD90%. The $^{7}$Li resonances at around 20–10 ppm, 0–10 ppm, and –10 ppm correspond to P1 (Li near small Si clusters), P2 (electrolyte, solid electrolyte interphase components, extended Si networks and isolated Si anions including *c*-$Li_{3.75}Si$), and P3 (overlithiated crystalline phase, *c*-$Li_{3.75+\delta}Si$), respectively. The notations for $^{7}$Li *ss*-NMR Li–Si environments are also summarised in Supplementary Table S 2. Orange, grey, red, and blue solid lines correspond to the spectra after the 1$^{st}$, 23$^{rd}$, 65$^{th}$, and 107$^{th}$ cycles, respectively. The spectra for different cycle numbers and different potentials are stacked with constant pitch to show the different processes more clearly.

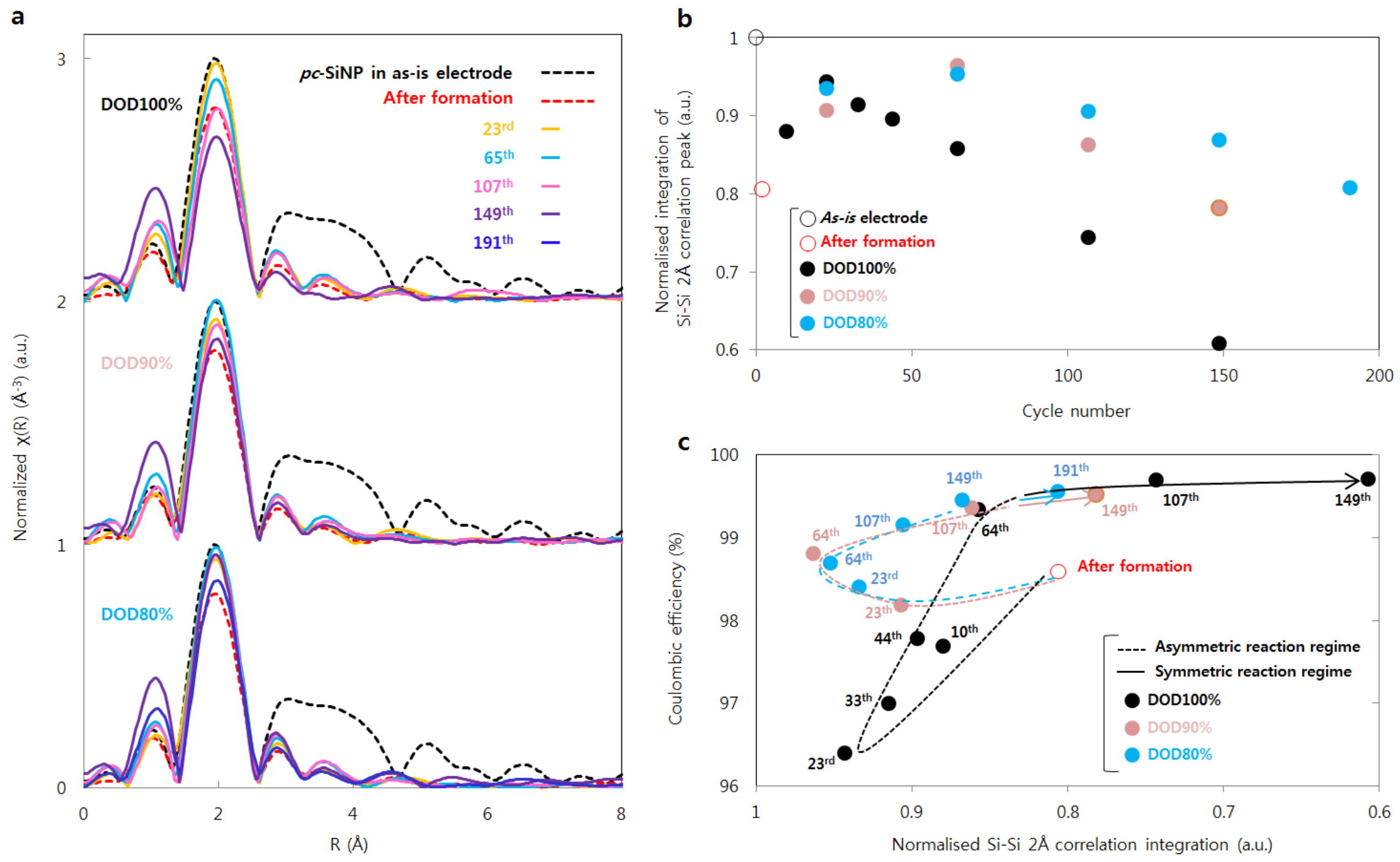

**Figure 6** ***Ex situ*** **XAFS analyses for delithiated Si.**

(a) Stacked Fourier transformed profiles using EXAFS profiles for Si K-edge for fully delithiated type-A electrodes at 1.5 V over 190 cycles under depth of discharge 100% (DOD100%), 90%, and 80%. Analysis is done at the recovery points, in which a cell is cycled at lower current rates (0.02–0.1C). Black and red dotted lines, and yellow, blue, magenta, purple, and blue solid lines show profiles for the as-is electrodes, the electrodes after the amorphisation of *c*-Si, and after the $23^{rd}$, $65^{th}$, $107^{th}$, $149^{th}$, and $191^{st}$ cycles, respectively. The profiles for the same DOD control are overlaid, whereas those under different DOD controls are stacked with a constant pitch to show the intensity changes more clearly. (b) Normalised integration of 2 Å Si–Si correlation peak in (a), named $A_{(2Å\ Si\text{-}Si)}$, over 190 cycles for different DOD controls. The as-is type-A electrode and the electrode after amorphising *c*-Si are shown by black and red empty circles, respectively. (c) Coulombic efficiency as a function of $A_{(2Å\ Si\text{-}Si)}$ over 190 cycles for different DOD controls. The dashed and solid lines indicate asymmetric and symmetric reaction regimes, respectively.

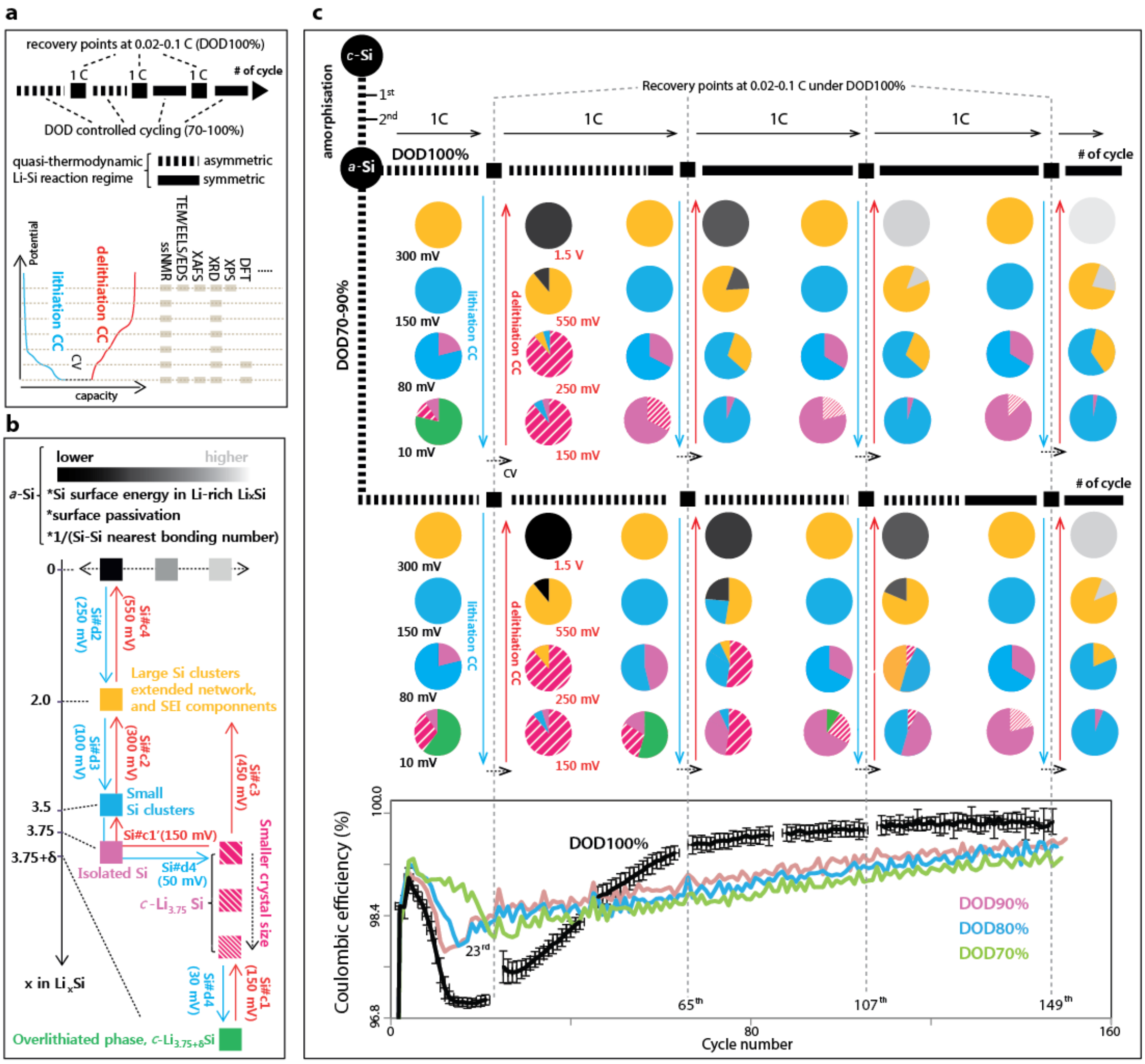

**Figure 7 Schematic of links between *a*-Si/Li$_x$Si structural characteristics and CE evolution.**

(a) Schematics of analytical scheme. Atomistic probing methodologies are complementarily combined (*ex situ* NMR, XAFS, XRD, TEM/EELS/EDS, and DFD calculation). The analysis is done at the recovery points, in which a cell is cycled at lower current rates (0.02–0.1C). (b) Various Li–Si structural environments, present during cycling over 150 cycles, are categorised into different colour schemes, in which fully delithiated *amorphous*-Si and *c*-Li$_{3.75}$Si are further sub-categorised. The classification is based on comprehensive interpretation from the complimentarily combined atomistic probing methodologies. These classifications are used in the right schematic flow. The notations for Li–Si processes (Si#d-X/Si#c-X represents the X$^{th}$ Li–Si discharge/charge process) are summarised in Supplementary Table S 2.(c) The phases formed on lithiation and delithiation are shown on the left and right hand side along blue and red arrows, respectively. Constant voltage domain at 10 mV is illustrated by dotted arrows. Each pie along the arrows shows the *a*-Si/Li$_x$Si phases present at each potential. The sizes of the pie segments indicate the relative proportion of each phase. The figure illustrates that the local environments and their profile on (de)lithiation depends significantly on the *a*-*c* phase transformation pathways controlled by different (DOD) cycling protocols, which is further linked to the CE profiles over 150 cycles.

## Methods

*Baseline active materials*

The spray-drying method (B290 Mini Spray-dryer, Buchi) enables synthesis of a well-controlled slurry for subsequent electrode fabrication, leading to reproducible electrochemical behaviour. Active materials in the form of secondary particles are designed to have porous open structures that can be easily wet with the electrolyte and undergo relatively abrupt phase transformations, even at higher current rates. The secondary particle composites composed of Si nano-powder (SiNP, Stream-Si, ~120 nm), multi-wall carbon nanotube (MWCNT, 15 nm, CNT Co. Ltd.) with/without flake-type Gr (FT-Gr, SPG1, SEC carbon), and polyvinyl alcohol (PVA, Sigma Aldrich, MW ~ 50k) are fabricated via spray drying. First, the components (with SiNP/MWCNT/FT-Gr/PVA ratios of 55.9/7.8/34.3/2.0 and 87.0/10.8/0/2.2 (wt %) for type-A and -B secondary particles, respectively, Supplementary Table 1) are dispersed in DI water, followed by 2 h of ultrasonication. The dispersed slurry is then spray-dried with a two-fluid-type nozzle at an inlet temperature of 220 °C in 60 slm $N_2$ flow, with subsequent thermal treatment at 900 °C for 5 h in a $N_2$ atmosphere (100 °C/h ramping rate), followed by sieving (< 32 μm) to remove larger secondary particles. The average secondary particle diameter is ~10 μm with specific surface areas (SSAs) of 29.5 and 39.5 $m^2/g$ for type-A and -B, respectively. SiNP and MWCNT are well entangled in the secondary particles, securing good electrical connections and buffer space to accommodate the volume expansion of Si (Figure 1c,d). Gr in type-A secondary particles acts as a scaffold to maintain the spherical shape and electrical contacts (Figure 1c). The wt% of Si in type-A and -B secondary particles is quantified by inductively coupled plasma spectroscopy (ICP, Shimadzu quartz torch, Nebulizer-Meinhard-type glass), resulting in ~54–55 and ~86–87 wt %, respectively, which is in a good agreement with tabulated wt % of Si in the secondary particles after calcination, assuming that the wt% of residual carbon from PVA after the calcination is about 20%.

The initial reversible theoretical capacities of Si, MWCNT, and Gr are supposed to be 3818 (3579 + 239, assuming $\delta$ in $c$-$Li_{3.75+\delta}Si$ is about ~0.2–0.3),[25] 200, and 350 mAh/g,[42] respectively, giving theoretical capacities of 2273 and 3342 mAh/g for type-A and -B, respectively. These values are in good agreement with the experimentally determined initial reversible capacities (on delithiation) of ~2250 and ~3350 mAh/g for type-A and -B, respectively (Supplementary Table 1). It should be noted that the portion of the theoretical capacity that results from Li–Si processes in type-A and -B active materials are calculated to be larger than 94.7% and 99.5%, respectively, with minor capacity contributions from Gr, MWCNT, and/or others. The presence of $c$-$Li_{3.75(+\delta)}Si$ in the early cycle stage is clearly confirmed by the characteristic sharp peak at 430 mV in *dQ/dV* profiles and capacity-voltage profiles (Figure 1e,f, Supplementary Fig.S 6-7). This observation is in contrast to that for other types of Si-based anodes, in which Li incorporation into Si is limited by a protective shell [2,3,15] and/or a buffer medium [13,16] to minimize irreversible side reactions. In such anodes, Li–Si incorporation can also be kinetically limited, with x in $Li_xSi$ usually less than 3.75 at the end of lithiation, and consequently, these systems are dominated by *a-a* phase transformations throughout cycling. A signature of this in half cells is the plateau at 300 and 550 mV in *dQ/dV* profiles on delithiation corresponding to $a$-$Li_{3.5}Si$ → $a$-$Li_{2.0}Si$ and $a$-$Li_{2.0}Si$ → $a$-$Li_{x<\sim1.0}Si$, respectively.[26,30,41] In such anodes, it is hard to define the initial cycling points as DOD100% ($x = 3.75 + \delta$) because x is typically less than 3.75, and consequently the definition of the DOD can be imprecise. Thus, it is also difficult to interpret whether the retention in such anodes originates from capacity sustenance or merely from balancing active material loss and gradual activation of the kinetically unused capacity.[3,13,15]

*Electrode fabrication and cycling conditions*

The electrodes are made of 79 wt% secondary particle (type-A or -B), 20 wt% polyacrylic acid (Li-PAA, Hwagyong Chemical) as a binder, and 1 wt% Kechen Black as a conductive additive. The components are mixed in a planetary mixer (Awatorirentaro, Thinky) for 15 min at 1000 rpm. The slurry is pasted onto a 10 µm thick Cu foil, and the mass loading level (weighed by Mettler Toledo XP26, ±1 µg accuracy) for type-A and -B is typically 1.8 and 1.2 mg/cm$^2$ (~3.0–3.5 mAh/cm$^2$), respectively, with an electrode density of ~0.4 g/cc for both types, resulting in Si loading of ~0.7–0.8 mg/cm$^2$. 2032-type coin cells (Hohsen Corp.) are used in all the following experiments. The electrolyte is 1 M $LiPF_6$ in a 25/5/70 (vol%) mixture of fluoroethylene carbonate (FEC), dimethyl carbonate (EC), and dimethyl carbonate (DEC) (LP 30 Selectilyte, Merck). A 10-µm-thick separator (Asahi, Celguard, 1-µm-thick $Al_2O_3$ coated on both sides) is used.

In this study, we defined the electrode specific capacity (mAh/g) by dividing the total capacity (mAh) by the weight of spray-dried secondary particles on the electrode, i.e., divided by 79% of the total mass loading. DOD100% is defined such that the reference electrode reached a cut-off current of 0.01 C at 10 mV at 1 C on lithiation in every cycle. With reference to the DOD100% capacity trend during cycling, the capacity is controlled between 70% and 100% in a staircase manner, as shown in Figure 1a and Supplementary Fig.S 4a. For all experiments, the first two cycles are carried out under DOD100% at 0.1 and 0.2 C to fully amorphise $c$-Si, followed by subsequent cycles under different DOD controls at 1 C. Slower cycles at 0.02–0.1 C under DOD100% are inserted every 20 cycles, regardless of the past DOD history to screen nature of quasi-thermodynamic reactions and corresponding local structural/interfacial changes.

*Interpretation of cycler's accuracy*

The 2032-type coin cells are cycled in a commercial cycler (TOYO system, TOSCAT-3100 series). The internal temperature of the cycler is maintained at ~23 °C (±1 °C) during the measurements. The internal system set a current acquisition pitch of about ~1 s. As 50 identical channels are used in this study, the instrument can measure the current with an accuracy of ±0.0167% (167 ppm) at a 2–10 mA range. To confirm the reproducibility of our samples, six identical control electrodes are cycled. The difference in mass loading of the samples on Cu foil is kept within ~5%. The average standard deviation of Coulombic efficiency (CE) over 107 cycles is ±~0.07% (Supplementary Fig.S 3). As the change in CE over the iterative phase transformations is as much as ~2 % over 150 cycles, these changes can be detected with our instrumental setup and electrode reproducibility. Despite this capability, the implementation of instruments with one-order-higher accuracy [34-36] is desirable, which we will pursue in future.

*Reference electrochemistry*

To obtain the background electrochemical signal for type-A electrodes, Gr-based electrodes are separately prepared using the same electrode fabrication method. Three distinct processes are observed on both discharge and charge with a capacity of ~300 mAh/g: *dQ/dV* peaks at 200 mV (namely C#d1, 84 mAh/g ≡ $LiC_{27}$), 110 mV (C#d2, 171 mAh/g ≡ $LiC_{13}$), and 80 mV (C#d3, 300 mAh/g ≡,$LiC_{7.4}$) on discharge, and at 90 mV (C#c1, 171 mAh/g ≡ $LiC_{13}$), 140 mV (C#c2, 80 mAh/g ≡ $LiC_{28}$), and 230 mV (C#c3) on charge.[42] These *dQ/dV* processes are in good agreement with those previously reported,[42,43] even though our Gr is partially disordered. Li–Gr processes can be separated from Li–Si processes, as shown in Figure 2c–f. The CE originated from Li–Gr processes is above 99.5% after 5 cycles (Supplementary Fig.S 13) and saturated above 99.9% after different number of cycling for the different DOD controls from 70 to 100%.

For type-A and -B electrodes, upon the 1st lithiation, the process is dominated by a sharp peak at 100 mV (Si#d1, *c*-Si → *c*-$Li_{3.75(+\delta)}Si$; gradual lithiation of the *c*-Si lattice into *a*-$Li_xSi$, with further transformations into *c*-$Li_{3.75}Si$ and *c*-$Li_{3.75+\delta}Si$).[25-27] On delithiation, which is initiated by a rather flat process up to 300 mV (Si#c1, *c*-$Li_{3.75(+\delta)}Si$ → *c*-$Li_{3.75(-\delta)}Si$), the characteristic plateau at

~430 mV dominated (Si#c3, $c$-$Li_{3.75(-\delta)}Si$ → $a$-$Li_{\sim1.1}Si$; a signature of converting $c$-$Li_{3.75(-\delta)}Si$ into a Li-substituted amorphous phase).[25] A small 300 mV peak (Si#c2, $a$-$Li_{\sim3.5}Si$→$a$-$Li_{\sim2.0}Si$) is also seen, which originates from residual $a$-$Li_xSi$ at the end of lithiation.[25] On the 2nd lithiation, at least three different processes are observed (Figure 2c–f and Supplementary Fig.S 4c–f); note that the 30 mV peak [25] (Si#d5, $c$-$Li_{3.75}Si$ → $c$-$Li_{3.75+\delta}Si$; $\delta$ = ~0.2–0.3) is probably overshadowed by signals from the other components. Hence Si#d5 is merged in Si#d4 process in this study, i.e. Si#d4 ≡ $a$-$Li_{\sim3.5\text{-}3.75}Si$→ $c$-$Li_{3.75(+\delta)}Si$. The 2nd delithiation is almost identical to that in the 1st cycle. These classifications are used in the right schematic flow. The notations (Si#dX/Si#cX) for Li–Si processes are summarised in Supplementary Table S 2

The negative/positive electrode capacity loading ratio (N/P ratio) in commercial full cells is typically designed to be 1.05–1.10 to satisfy safety constraints, i.e. avoiding Li dendrite formation on higher-electrode-density Gr-dominant anodes. In other words, the state of charge (SOC) in the anodes in such full cells at an initial cycling state becomes around ~90–95%, which is equivalent to $x < 3.5$ in $Li_xSi$ if Si is present in the anodes and x is proportional to SOC. However, $c$-$Li_{3.75(+\delta)}Si$ can be still present at least within the following two representative anode strategies. The first strategy is to use Si/Gr composite electrodes, in which the good cyclability and lower volume changes of Gr are coupled with the poor cyclability and large volume changes of Si. This strategy necessitates that electrode are cycled at lower voltages (typically less than 60 mV) [42,43] to access the full Gr capacity. The decrease in potential is even more significant when cycled at higher current rates (e.g. >1 C) with higher energy densities (mAh/cc) owing to an increase of kinetics and consequent Li-concentration inhomogeneity, particularly along the direction perpendicular to the electrode.[29,38] The second contrasting strategy is to use an electrode that contains a much higher proportion of Si and to limit the capacity to e.g. ~1,500–2500 mAh/g, cutting-off higher potentials and cycling between different $a$-$Li_xSi$ phases.[7,41,44] However, this strategy can still involve phase transformations due to potential drift-down owing to capacity loss and inhomogeneity issues with cycling. Hence, $c$-$Li_{3.75(+\delta)}Si$ is inevitable in either case.

Type-A and -B electrodes are cycled at slower current rates (0.02–0.1 C) under DOD100% at the recovery points to investigate capacity recovery and potential alteration of quasi-thermodynamic reaction nature depending on ever-cycled DOD cycling protocols. CRR at the recovery points is defined with respect to the 1st delithiation capacity. As shown in Supplementary Fig.S 12, the CRR values at the recovery points decrease suddenly from DOD90% to 100% by ~7.3% for type-A and –B, while the difference between DOD90% and 80% is around 1%. This observation indicates that the $a$-$Li_xSi$ → $c$-$Li_{3.75(+\delta)}Si$ phase transformation has a greater impact on the degradation compared with incremental $Li_xSi$ volume changes under DOD80–90% cycling protocols (corresponding to x = ~3.0–3.56). The difference in CRR between DOD100% and the other DOD controls gradually increases as cycling proceeds, reaching a local maximum of 7–8% around the 65$^{th}$ cycle, and then decreasing to 3–4% after the 107$^{th}$ cycle. Owing to such minor difference, DOD in the range of X=70–100 still allows us to separate CE alterations by the volume changes and by the *a*-*c* phase transformations.

The similarity of CE trends for type-A and -B electrodes under the given DOD controls rationalises the followings; firstly, the rapid CE increase/saturation in type-A electrodes is not due to Si capacity decay with a concurrent increase of Li–Gr processes that have relatively higher CE (>99.5 % after 5$^{th}$ cycle, Supplementary Fig.S 13). That is also not due to changes in the portion of capacity resulting from Li–Si and Li–Gr processes under kinetics cycling conditions.

*TEM*

The control secondary particles and electrodes are characterized using SEM by slicing with a Ga focused ion beam (FIB, 5 keV acceleration, Helios Nanolab 450F1, FEI). After slicing the electrodes by FIB, *ex situ* TEM analyses are carried out using a double-Cs-corrected Titan Cubed microscope (FEI) at 300 kV with a Quantum 966 energy filter (Gatan Inc.) and a probe Cs-corrected Titan 80-200 microscope (FEI) at 200 kV with a Super-X EDS detector. To avoid contamination and reaction on exposure of the TEM sample to air, a vacuum transfer TEM holder (Model 648, Gatan, Inc.) and transfer vessel for FIB (hand-made) are used. All samples are transferred from the FIB transfer vessel to the vacuum transfer TEM holder in a glovebox filled with Ar. Coin half-cells with type-A electrodes are cycled at 1 C over 107 cycles under different DOD controls. To fully delithiate the electrodes, the half-cell potential is increased to 1.5 V and held for at least 24 h until the current is less than 0.001 C. To avoid potential structural differences with respect to the position on the perpendicular electrode due to Li concentration differences on (de)lithiation at higher current rates, the section to be analysed is always taken from 0 to 5 µm from the surface (separator side). The amorphised structure consists of a frame of extremely-fine pores (~3.5 nm with *stdev*~0.8 nm) and quantum size frames (~1.9 nm with *stdev*~0.5 nm) with 10 nm-thick chunky stripes, all of which are confined within the sphere (Figure 3b), still maintaining the original spherical frame. It seems that the amorphous plane observed to fill the porous Si frame after the 3$^{rd}$ cycle acts as a building block for the overall structure (Figure 3c,d), altering its feature size over cycling. Supplementary Fig.S 23 shows size and its standard deviation of the delithiated porous Si frame (100 features randomly picked) for DOD80%, 90%, and 100% after the 107$^{th}$ cycle, being *d* ~ 5.9, 5.8, and 4.8 nm with *stdev* ~ 2.6, 2.7, and 1.0 nm, respectively.

For observations on lithiated samples, the electrode is cycled down to 10 mV at 0.05 C and held there until the current decayed to less than 0.001 C. The disassembled sample is washed with DMC for 5 min, and then dried for 30 min. The electrodes are then scraped from the Cu collector onto a lacey-carbon TEM grid (Sigma Aldrich) rather than using the FIB process. This method is adopted because self-discharge can occur during the process, which would accelerate relaxation of the metastable crystalline structure. The grid is transferred to the TEM holder without exposure to ambient air. TEM imaging is promptly conducted, as exposure to the high-energy electron beam can easily relax the phases.

*XRD*

For *ex situ* XRD measurements, the coin half-cells are cycled at 1 C under different DOD controls, subsequently at the recovery points, the electrodes are slowly lithiated at 0.05 C under the CCCV condition. The CV regime is maintained for at least for 24 h until the current decayed to less than 0.001 C to stabilize metastable *c*-$Li_{3.75(+\delta)}Si$. In the previous study,[25] we showed that relaxation of the metastable phase is sluggish, with the phase held for 10–20 h, if cycled in this manner. The coin cells are disassembled in an Ar-filled glovebox, sealed with airtight Kapton tape, and immediately transferred to the XRD instrument (Bruker, D8 Advance). Conventional XRD measurements are performed using Cu Kα radiation (1.54 Å). Each spectrum is acquired in the range of 5–80° (2θ) for ~50 min. The *c*-$Li_{3.75(+\delta)}Si$ (332) reflection peak is fit by the Voigt function using free software (Fytik) to determine the FWHM. The error in the FWHM is considered to be 0.05°, considering the instrumental data acquisition pitch.

*XPS*

For *ex situ* XPS measurements (PHI Quantera-II), the core-level spectra are measured using Al Kα as the excitation source (1486.6 eV) at an acceleration voltage of 1 kV. The atomic concentrations are determined and curve fitting carried out after Shirley background subtraction. All the spectra are referenced to the C 1s peak at 284.8 eV. Coin cells cycled under different DOD controls at 1 C are disassembled in an Ar-filled glovebox and washed with DMC for 5 min, followed by 30 min drying under vacuum. Subsequently, the electrodes are loaded into an in-house airtight vessel and transferred to the instrument without exposure to ambient air. Spectra are recorded for the electrodes before cycling, after the amorphisation of *c*-Si, and at recovery points over 107 cycles for DOD80%, 90%, and 100%. The as-is electrode is soaked into the electrolyte prior to the DMC washing to see signal from residual $LiPF_6$. Each electrode is analysed after sputtering with Ar ions for different amounts of time (t = 0–5 min) to remove potential contamination and/oxidation, the etching rate being ~6 nm/min. The $SiO_x/Li_xSiO_y$ peak always accompanies the Si signal, both of which only start after 1 min etching. This observation indicates that $SiO_x/Li_xSiO_y$ is present in close proximity to Si. As shown in Supplementary Fig.S 24, the Li 1s, F 1s, and P 2p depth profiles show that the majority of F-related composites are made of LiF with fractions of residual $LiPF_6$ (in the surface region), $PO_yF_z$, and/or $LiP_yO_zF_z$,; note, we only observed crystalline LiF by TEM imaging and the amount of P is less than 0.5 at % for all the samples, regardless of the number of cycles and etching depth range. LiF may mostly originate from defluorination of FEC.[45,46] Li–X (X = $O_{0.5}$, O, OH) is also found in the profiles; however, it is difficult to clearly distinguish between oxide and non-oxide components in the Li-1s spectra.

### *$^7$Li ss-NMR spectroscopy*

$^7$Li NMR spectra are acquired using *ex situ* MAS *ss*-NMR at 233.2 MHz (14.1 T magnet and Avance III spectrometer, Bruker) with a 2.5 mm MAS probe (Bruker) at a spinning rate of 15 kHz with π/2-(one-pulse) measurements with a 2.0 s last-delay duration over 64 scans. After the coin half-cells reach the recovery points at 1 C under different DODs, the cells are cycled at 0.05 C until reaching the target potential and held there for at least for 24 h until the current decayed to less than 0.001 C. The cell is then is immediately disassembled in an Ar-filled glovebox, dried for at least 30 min under vacuum, and packed in the rotor for the NMR measurements. All the $^7$Li NMR chemical shifts are referenced to 1 M LiCl (sol.) at 0 ppm as an external reference. The notations for $^7$Li *ss*-NMR Li–Si environments (P1, P2, and P3) are summarised in Supplementary Table S 2. Spectra are recorded at 300, 150, 80, and 10 mV on lithiation and 150, 250, and 550 mV, and 1.5 V on delithiation at the recovery points for DOD100% and 90% over 107 cycles. In the 23$^{rd}$ cycle, on lithiation, P2 appears at 300 mV (Si#d2 ≡ *a*-Si→*a*-$Li_{<2.0}$Si) [23-25] P1 gradual increase and shift to lower resonances at 150 and 80 mV, respectively (Si#d3 ≡ *a*-$Li_{\sim 2.0}$Si → *a*-$Li_{\sim 3.5}$Si), and P2 and P3 prominence at 10 mV (*c*-$Li_{3.75}$Si and *c*-$Li_{3.75(+\delta)}$Si after Si#d4). [7,23,25,27,30] On delithiation, sudden P3 disappearance at 150 mV (after Si#c1 a*c*-$Li_{3.75(+\delta)}$Si→*c*-$Li_{3.75(-\delta)}$Si),[25] no profile change from 150 to 250 mV, P2 increase and widening at 550 mV (Si#c3 ≡ *c*-$Li_{3.75(-\delta)}$Si → *a*-$Li_{x<1.1}$Si).[25] The isolated Si at 10 mV does not symmetrically turn back into Si-Si small clusters, instead, into large Si clusters and extended networks due to hysteretic energetics. For DOD90% in the 65$^{th}$ cycle, on lithiation, while a P2 → P1 sequence similar to that for DOD100% is observed from 300–80 mV, there is small P3 shoulder (–1.5 ppm) seen at 10 mV in the absence

of P1. On delithiation, the P1 to P2 intensity ratio at 150 mV is much smaller than that for DOD100%, which indicates that some Si anions remain in the *c*-$Li_{3.75(+\delta)}Si$ state at 10 mV and do not reform small clusters (P1).[25] The subsequent behaviour on delithiation is similar to that for DOD100%. Such differences in the Li-rich domain indicate that the electrode is in a transient state between the asymmetric and the symmetric regimes in the 65$^{th}$ cycle under DOD90% cycling protocol.

*XAFS*

*Ex situ* XAFS at the Si K-edge is measured at BL-10, Synchrotron Radiation (SR) Center, Ritsumeikan University. The photon beam energy delivered to the samples ranged from 1000 to 2500 eV with a resolution of 0.5 eV or less. 2032-type coin half-cells are cycled at 1 C under the designated DOD controls until reaching the target number of cycles. To fully delithiate the electrode, the half-cell potential is held at 1.5 V for at least 24 h until the current decayed to 0.001 C. The cells are then disassembled in an Ar-filled glovebox. The electrode is rinsed with DMC for 5 min, set on carbon-taped sample holders, loaded into an airtight vessel, and then transferred to the BL-10 chamber without exposure to ambient air. The vessel is immediately evacuated and the samples are loaded into the measurement chamber with a vacuum level of $5 \times 10^{-8}$ Pa. Partial fluorescence yield (PFY) mode is adopted to measure XAFS over the EXAFS range for the Si K-edge, which enables effective elimination of the P K-edge absorption signal by energy selected fluorescence detection with a Si drift detector (SSD). Small amounts of residual P on the surface of the Si anode could not be completely removed, even after rinsing, which agrees with the XPS results, which show 0.5 at % or less P in all the samples at all etching durations. The total electron yield (TEY) is also simultaneously measured and P is detected in the EXAFS region of Si. For XANES (Supplementary Fig.S 25a), after the initial amorphisation, the $SiO_2$ peak shifts to a lower energy at 1843–1846 eV ($SiO_x$ and/or $Li_ySiO_z$), indicative of native oxides on Si forming Li silicate or being further oxidized. As cycling progressed, the absorption at 1843–1846 eV increased for all DODs, with this increase more prominent at higher DOD controls (Supplementary Fig.S 25a). Using open analysis software (Athena), EXAFS (Supplementary Fig.S 25b) is extracted from XAFS and converted to RDF profiles. Tabulating coordination number of Si involves a number of uncertainties, such as statistical EXAFS fitting errors, sample preparation reproducibility, and deviations from the assumptions made during data analysis for the physical structures surrounding the absorber.[47] Hence, in this study, we integrated the 2 Å Si–Si correlation peaks to index the Si local environments in the delithiated states. It seems that oxidation of the anodes in the ambient environments after cycling is minimised, as the amplitude of the 1A coordination peaks in the Fourier transformed profiles is lower than ~0.3–0.4. The error in $A_{(2Å\ Si\text{-}Si)}$, originating from sample reproducibility and handling issues, is not large (mostly around ~10 %) to invalidate the overall trend observed in Figure 6b,c. It should be also noted that when $A_{(2Å\ Si–Si)}$ increase is reversed around ~0.95, the same track is not followed, instead, higher CEs are observed at the same $A_{(2Å\ Si\text{-}Si)}$. This change is probably caused by the evolving morphological changes in Si from more entangled/agglomerated to disengaged structures and by the changes in Si interfacial property/energetics (Supplementary Fig.S 14 and 24).

*Electrochemical impedance spectroscopy*

The possibility that an increase in Li-metal resistance in half-cells limits Li–Si incorporation is examined to account for the absence of Si#c3 in the symmetric regime. *Ex situ* electrochemical impedance spectroscopy (EIS) is conducted over 150 cycles for different DODs. The frequency is swept from 1 MHz to 0.1 Hz with a fluctuating voltage of ±5 mV. The cell is cycled at 1 C until the recovery point, and then switched to CCCV(10mV) at 0.05 C, holding the voltage for at least for 24 h to stabilize the metastable phase. As shown in Supplementary Fig.S 8, the semicircle at mid-range frequencies (10–10000 Hz) for half-cells (corresponding to charge-transfer components) is much smaller when EIS is instead measured for symmetric cells (Supplementary Fig.S 9). This observation indicates that the resistance increases on the counter electrode in the half-cells, which may limit the lithiation process. To investigate this phenomenon further, a coin half-cell with a type-A electrode cycled under DOD100% is reassembled after 107 cycles in a coin cell with *fresh* Li-metal, separator, and electrolyte without exposure to air, and then cycled at 0.02 and 0.05 C for three cycles each. Si#c3 is still absent in *dQ/dV* profile (Supplementary Fig.S 10), which shows that the resistance increase in Li-metal over cycling does not contribute to Si#c3 absence.

*Computational simulations*

Bulk amorphous $Li_xSi$ structures are constructed using a series of melting, quenching, and relaxation processes of thermodynamically stable crystalline $Li_xSi$ structures (x = 2.33, 3.25, and 3.75).[48] The structures are obtained using *ab initio* calculations based on DFT implemented in the Vienna *ab initio* simulation package (VASP),[49] in which the generalized gradient approximation (GGA) suggested by Perdew, Burke, and Ernzerhof (PBE)[50] is adopted for the exchange-correlation functional, and the projector augmented wave (PAW) method[50] is used for the atomic quasi-potentials of all elements. To ensure amorphism of the generated structures, a sufficiently large number of atoms is included in the supercells: 120, 136, and 152 atoms for x = 2.33, 3.25, and 3.75, respectively. By using *ab initio* molecular dynamics, *c*-$Li_xSi$ structures are melted at 4000 K for 5 ps with 1 fs time steps, and then quenched at 300 K, by assuming the canonical ensemble based on the Nosé algorithm. Here, k-points of 1×1×1 with Γ symmetry-point-centered sampling and a cutoff energy of 300 eV for the plane-wave basis are used. Subsequently, full structural relaxations of the quenched structures by DFT led to the final bulk *a*-$Li_xSi$ structures, in which we used an atomic force tolerance of 0.02 eV/Å, electronic energy tolerance of $10^{-6}$ eV, energy cutoff of 500 eV, and Γ-centered k-point sampling of 2×4×2, 2×2×4, and 3×3×2 for x = 2.33, 3.25, and 3.75, respectively. The density of the amorphous structures is determined by thermodynamic evolution of crystalline structures with well-defined density.

Initial spherical *a*-$Li_xSi$ nanoclusters are created with the same 84 Si atoms for all Li fractions by using bulk amorphous $Li_xSi$ structures previously obtained by DFT and preserving the relative atomic coordinates. The diameters of *a*-$Li_xSi$ are 20.83, 22.41, and 23.42 Å, and the total numbers of atoms contained in the clusters are 280, 357, and 399 for x = 2.33, 3.25, and 3.75, respectively. In addition, slightly larger amorphous bulk structures are regenerated with the amorphous bulk structures obtained by DFT. The final spherical *a*-$Li_xSi$ clusters and bulk structures are obtained by performing a classical molecular dynamics simulation implemented in the Large-scale Atomic/Molecular Massively Parallel Simulator (Lammps) package [51] with the reactive force field (ReaxFF),[52] as shown in Supplementary Fig.S 26. The structures are relaxed at 300 K under a Nosé-Hoover thermostat for 1 ns with 1 fs time steps in a canonical (NVT) ensemble. The formation energy for a given Li fraction x is calculated as:

$$FE(x) = E_{Li_xSi} - xE_{Li} - E_{Si}$$

where $E_{LixSi}$ is the total energy of a $Li_xSi$ structure divided by the number of Si atoms, and $E_{Li}$ and $E_{Si}$ are energies per atom in the body-centered cubic (bcc) structure of Li and diamond structure of Si, respectively. The surface energy is given by:

$$\sigma = \frac{1}{A}\left(E_{sphere} - E_{bulk}\right)$$

where $E_{sphere}$ is the total energy of the $Li_xSi$ spherical cluster, $E_{bulk}$ is the bulk energy of $Li_xSi$ with the same number of Si atoms as in the spherical cluster with the corresponding x value, and A is the surface area of a spherical cluster.

The key question to be addressed is the reason for the absence of +δ in *c*-$Li_{3.75(+\delta)}Si$ at the recovery in the asymmetric regime despite *c*-$Li_{3.75}Si$ presence. The formation of *c*-$Li_{3.75+\delta}Si$ is energetically favorable in the event of $c$-$Li_{3.75}Si$ formation owing to the lower energy cost of inserting Li atoms into nucleated $c$-$Li_{3.75}Si$ than that of breaking residual Si–Si bonds.[25] Attributing to the increasing resistance from accumulated SEI and degraded electric network over Si expansion/contraction might not be suffice since $+\delta$ is absent under the potentiostatic quasi-thermodynamic cycling conditions (Figure 5). To interpret this phenomenon further, DFT is used to calculate $Li_xSi$ formation energy in the 2 nm spherical Si nanocluster as shown in Supplementary Fig.S 26. In the nanocluster, a gradient for the decrease in formation energies ($\xi$), which is equivalent to the driving force for lithiation, is significantly suppressed around x = 3.25 ($\xi$ = -0.177 and -0.028 eV for x = 2.33–3.25 and x = 3.25–3.75, respectively), whereas that in the bulk barely changes ($\xi$ = -0.23 and -0.297 eV for the same intervals of x). This sudden decrease in the driving force is partly attributed to an increase of the surface energy in the structures. Hence, in such nanoclusters, there is less momentum to reach x = 3.75 ($+\delta$) due to decreased $\xi$ and increased surface energy, which may result in more uniform lithiation by breaking residual Si–Si bonds, rather than locally inserting extra Li atoms into $c$-$Li_{3.75}Si$ nuclei. This may at least in part result in the $+\delta$ absence in $c$-$Li_{3.75(+\delta)}Si$ in the symmetric regime.

*Supplementary information (SI)*

# Revealing evolving affinity between Coulombic reversibility and hysteretic Li-Si phase transformations

**K. Ogata[1,2*§], SH. Joen[1*§], DS. Ko[1*], IS. Jung[1], JH. Kim[1], K. Ito[3], Y. Kubo[3], K. Takei[1], S. Saito[2], YH. Cho[1], HS. Park[1], JH. Jang[1], HG. Kim[1], JH. Kim[1], YS. Kim[1], M. Koh[1], K.Uosaki[3], SG. Doo[1], YI. Hwang[1], SS. Han[1§]**

1. Samsung Advanced Institute of Technology (SAIT), Samsung Electronics, Samsung-ro 130, Suwon, Gyeonggi-do, 16678, Korea

2. Samsung Research Institute of Japan (SRJ), Samsung Electronics, 2-1-11, Senba-nishi, Mino-shi, Osaka-fu, 562-0036, Japan

3. C4GR-GREEN, National Institute for Materials Science (NIMS), 1-1 Namiki, Tsukuba, Ibaraki, 305-0044, Japan

*These authors equally contributed to the work

§Corresponding authors

**Supplementary Table S 1**

Physical parameters and composition ratios of materials in secondary particles (type-A and -B). The composition ratios are tabulated assuming that carbon residue after calcination is about 20 wt%. The theoretically calculated and empirically determined capacities agree well.

| Spray-dried secondary particle's types | type-A | type-B |
|---|---|---|
| $O_2$ content in nano-structured Si particles (at %) | < 0.5 | |
| Specific surface area of nano-structured Si particles ($m^2/g$) | 17 | |
| Nano-structured Si particles in secondary particle (wt %) | 55.9 | 87.0 |
| CNT in secondary particle (wt %) | 7.8 | 10.9 |
| Gr in secondary particle (wt %) | 34.3 | 0 |
| Polyvinyl alcohol in secondary particle (wt %) | 0.39 | 2.17 |
| Theoretical capacity of secondary particle (mAh/g) | 2273 | 3342 |
| Specific surface area secondary particle ($m^2/g$) | 29.5 | 39.5 |
| Experimental first reversible capacity (mAh/g) | 2250 | 3350 |

**Supplementary Table S 2**

Electrochemical Li-Si processes on (de)lithiation linked to the used notations in the main text. Si#d-X and Si#c-X denote the X$^{th}$ Li–Si discharge (lithiation) and charge (delithiation) processes, respectively. The notation is further linked to that of Li-Si local structures interpreted from $^{7}$Li NMR spectrum. Resonances at around 20–10, 10–0, and –10 ppm are labeled respectively as P1 (Li near small Si clusters), P2 (isolated Si anions, $c$-Li$_{3.75}$Si, and extended Si networks), and P3 (overlithiated crystalline phase, $c$-Li$_{3.75+\delta}$Si). Blue and red letters represent lithiation and delithiation processes, respectively.

| Electrochemical Li-Si processes | | | Li-Si processes in $^{7}$Li NMR processes | | | | |
|---|---|---|---|---|---|---|---|
| | | | | Li-Si local structure | | | |
| Li/Li$^{+}$ potential (mV) (Lithiation /Delithiation) | Li-Si processes | Stoichiometry | $^{7}$Li NMR peak flow | Small Si clusters | Large Si clusters and extended Si networks | Isolated Si | Over-lithiated Si |
| | | | | P1 | P2 | | P3 |
| 100 | Si#d1 (only the 1st cycle) | $c$-Si > $c$-Li$_{3.75(+\delta)}$Si | P2→P1→P2→P3 | ✓ | ✓ | ✓ | ✓ |
| 250~300 | Si#d2 | $a$-Si > $a$-Li$_{2.0}$Si | P2 | | ✓ | | |
| 100 | Si#d3 | $a$-Li$_{2.0}$Si > $a$-Li$_{3.5}$Si | P2→P1 | ✓ | | ✓ | |
| <50 | Si#d4 | $a$-Li$_{3.5}$Si > $c$-Li$_{3.75(+\delta)}$Si (only in asymmetric reaction regime) | P2→P3 | | | ✓ | ✓ |
| | | $a$-Li$_{3.5}$Si > $c$-Li$_{3.75}$Si (only in symmetric reaction regime) | P2 | | | ✓ | |
| 0~150 | Si#c1 | $c$-Li$_{3.75(+\delta)}$Si > $c$-Li$_{3.75(-\delta)}$Si | P2 | | | ✓ | |
| 0~150 | Si#c1' (only in symmetric reaction regime) | $c$-Li$_{3.75}$S > $a$-Li$_{3.5}$Si | P2→P1 | ✓ | | ✓ | |
| 300 | Si#c2 | $a$-Li$_{3.5}$Si > $a$-Li$_{2.0}$Si | P2→P1 | ✓ | ✓ | | |
| 430 | Si#c3 | $c$-Li$_{3.75(-\delta)}$Si > $a$-Li$_{1.1}$Si | P2 | | ✓ | | |
| >550 | Si#c4 | $a$-Li$_{2.0}$Si > $a$-Li$_{<1.0}$Si | P2 | | ✓ | | |

**Supplementary Table S 3**

Electrochemical and structural probing schemes. (Blue) Electrochemical control parameters (Input factor, Input#1,2), (Red) Electrochemical output parameters at least discussed in this study (Output factor, Output#1–3), and (Green) Probing methodologies (Material analysis, Material analysis#1–9).

| Input# | Input factor | Input factor category | Parameters |
|---|---|---|---|
| 1 | Incremental volume changes without phase transformations | Electrochemical | DOD 70, 80, 90% |
| 2 | Volume changes with *a-c* phase transformations | Electrochemical | DOD100% |
| **Output#** | **Output factor** | **Output factor category** | **Relevant figures** |
| 1 | CE | Electrochemical | Fig.2(b,h,i), SI-Fig.4(b,h,i) SI-Fig.3, SI-Fig.15-17 |
| 2 | Capacity retention rate (CRR) | Electrochemical | Fig.2(a,g,i), SI-Fig.4(a,g,i) SI-Fig.12 |
| 3 | *dQ/dV* | Electrochemical | Fig. 2(c-f), SI-Fig. 4(c-f) SI-Fig.5-7, SI-Fig.11 |
| **Material analysis#** | **Material analysis** | **Material analysis category** | **Relevant figures** |
| 1 | TEM | Structural | Fig.3-4, SI-Fig.18-23 |
| 2 | EELS/EDS | Elemental | Fig.3-4, SI-Fig.18-23 |
| 3 | Electrode thickness | N/A | SI-Fig.14 |
| 4 | XRD | Structural | Fig.4(d-g) |
| 5 | Symmetric EIS | Impedance | SI-Fig.8.9 |
| 6 | XPS | Interfacial | SI-Fig.24 |
| 7 | NMR | Local structural | Fig.5 |
| 8 | XAFS | Local structural | Fig.6, SI-Fig.25 |
| 9 | DFT | Energetics | SI-Fig.26 |

**Supplementary Table S 4**

Summary on electrochemical new findings and corresponding interpretation sources.

| Electrochemical new finding # | Electrochemical new findings | Interpreted from |
|---|---|---|
| **1** | **<u>#1 Summary: An accelerated shift of electrochemical Li–Si reaction regimes by the iterative *a-c* phase transformations</u>**<br><br>The iterative *a-c* phase transformations accelerate the alteration of quasi-thermodynamic reaction pathways from asymmetric to symmetric reaction sequences; in the former, *c*-$Li_{3.75(+\delta)}Si$ formed around ~50 mV asymmetrically transforms back into *a*-$Li_{\sim1.1}Si$ on delithiation with the well-known large hysteresis at 430 mV, while in the latter *c*-$Li_{3.75(+\delta)}Si$ can reform *a*-$Li_{\sim3.75\text{-}3.2}Si$ at as low as ~150 mV (a newly elucidated quasi-thermodynamic pathway), followed by the subsequent symmetric *a-a* transformations at 300 mV (*a*-$Li_{\sim3.5}Si$→*a*-$Li_{\sim2.0}Si$) and 550 mV (*a*-$Li_{\sim2.0}Si$→*a*-$Li_{x<\sim1.0}Si$). Iterating the *a-c* transformations under DOD100%, the regime shift occurs around $60^{th}$ cycle, while cycled under lower DOD% it is postponed to after $110^{th}$ cycle. | **CE**<br>***dQ/dV***<br>**CRR** |
| **2** | **<u>#2 Summary: To quantify and qualify evolving CE alterations by different Li–Si structural changes</u>**<br><br>Iterating *c*-$Li_{3.75(+\delta)}Si$ (de)formation (usually featured as capacity degradation factors) can upheave CE up to ~99.9 % in the most efficient manners among the given Li–Si reaction sequences when compared under the same reversible capacity and the same accumulative irreversible Li consumption. The CE alteration by the iterating *c*-$Li_{3.75(+\delta)}Si$ (de)formation on is quantitatively distinguished from that by mere *amorphous* Li–Si volume changes under different DOD controls from DOD70 to 90%. | **CE**<br>***dQ/dV*** |
| **3** | **<u>#3 Summary: Inherent CE behaviours in different electrochemical Li–Si process regimes</u>**<br><br>CE has a strong correlation with an affiliated Li–Si process regime and with subjected DOD controls. In other words, inherently remaining asymmetric regime (i.e. duration of residual asymmetric sequence when DOD is set to DOD100%) and subjected DOD% in the following cycles can project how CE profiles can develop. The asymmetric-to-symmetric shift can also alter CE susceptibility; CE becomes more susceptible to *c*-$Li_{3.75(+\delta)}Si$ presence/absence upon (de)lithiation by the DOD controls in the former regime, in contrast, more stabilised/saturated at a higher level regardless of the presence/absence in the latter. | **CE**<br>***dQ/dV*** |

**Supplementary Table S 5**

Summary on mechanistic findings and corresponding interpretation sources.

| Mechanistic new finding # | Mechanistic new findings | Interpreted from |
|---|---|---|
| **1** | **<u>#1 Summary: An accelerated shift of Li–Si local environments' reaction sequences</u>**<br><br>Iterative *c*-$Li_{3.75(+\delta)}$Si (de)formation accelerates a shift of a reaction sequence of Li–Si local environments from an asymmetric to symmetric sequence. In the former, on delithiation, isolated Si and over-lithiated Si in *c*-$Li_{3.75(+\delta)}$Si form larger Si clusters and extended Si networks in *a*-$Li_{\sim 1.1}$Si upon the hysteretic quasi-two-phase process at 430 mV. In contrast, in the latter, +δ components in *c*-$Li_{3.75(+\delta)}$Si are absent and isolated Si anions in *c*-$Li_{3.75}$Si can reform Li-deficient isolated Si and small Si cluster at ~150 mV, followed by double plateaus at 300 and 550 mV, consequently being a symmetric Si clusters' reaction sequence . | ***dQ/dV***<br>**NMR**<br>**XRD** |
| **2** | **<u>#2 Summary: A shift from bulk to surface dominated systems</u>**<br><br>The sequence shift in #1 is correlated with changes in $A_{(2Å\ Si\text{-}Si)}$ *. When $A_{(2Å\ Si\text{-}Si)}$ > ~0.8–0.85, the sequence becomes asymmetric and the Si morphology is more bulky, while $A_{(2Å\ Si\text{-}Si)}$ < ~0.8–0.85 the sequence becomes more symmetric and more surface dominated systems, e.g. sub-5nm scale Si with a narrower feature-size distribution. Also, when $A_{(2Å\ Si\text{-}Si)}$>0.8–0.85, CE alteration is very susceptible to *c*-$Li_{3.75(+\delta)}$Si presence/absence. In contrast, when $A_{(2Å\ Si\text{-}Si)}$<0.8–0.85 CE is more stabilised at higher values (99.5–99.9 %) regardless of *c*-$Li_{3.75(+\delta)}$Si presence/absence.<br><br>* $A_{(2Å\ Si\text{-}Si)}$ is population index of Si-Si tetrahedral correlation obtained from XAFS analysis | **CE**<br>***dQ/dV***<br>**NMR**<br>**Electrode thickness**<br>**EIS**<br>**XPS**<br>**XAFS** |
| **3** | **<u>#3 Summary: Interfacial stabilisation in the surface dominated system</u>**<br><br>In the transition from the asymmetric to symmetric regime at $A_{(2Å\ Si\text{-}Si)}$ ~0.8–0.85, +δ in *c*-$Li_{3.75+\delta}$Si (potential reduction sources) disappears even under quasi-thermodynamic reactions. Also, an increase of the electrode thickness is saturated at the regime shift point. The same trend is observed in FWHM of *c*-$Li_{3.75(+\delta)}$Si XRD reflection. These trends indicate that there are more non-destructive or efficient stress release processes with much less irreversible reactions at Si/SEI interface in the symmetric regime. | **CE**<br>***dQ/dV***<br>**NMR**<br>**Electrode thickness**<br>**EIS**<br>**XPS**<br>**XAFS** |
| **4** | **<u>#4 Summary: Altered surface energetics in sub-5nm Li-Si structures</u>**<br><br>Preliminary DFT calculation indicates that a driving force for lithiating *a*-$Li_x$Si beyond x=3.25 in the sub-5-nm-clusters is significantly lowered compared with that in bulk Si, and that surface formation energy significantly increases. In such surface dominated system, there may be less motivation for Li atoms to over-lithiate *c*-$Li_{3.75}$Si nuclei inhomogeneously, instead breaking residual Si–Si bonds is more preferred, resulting in more uniform lithiation, +δ absence. | ***dQ/dV***<br>**DFT**<br>**XRD**<br>**NMR**<br>**Electrode thickness** |

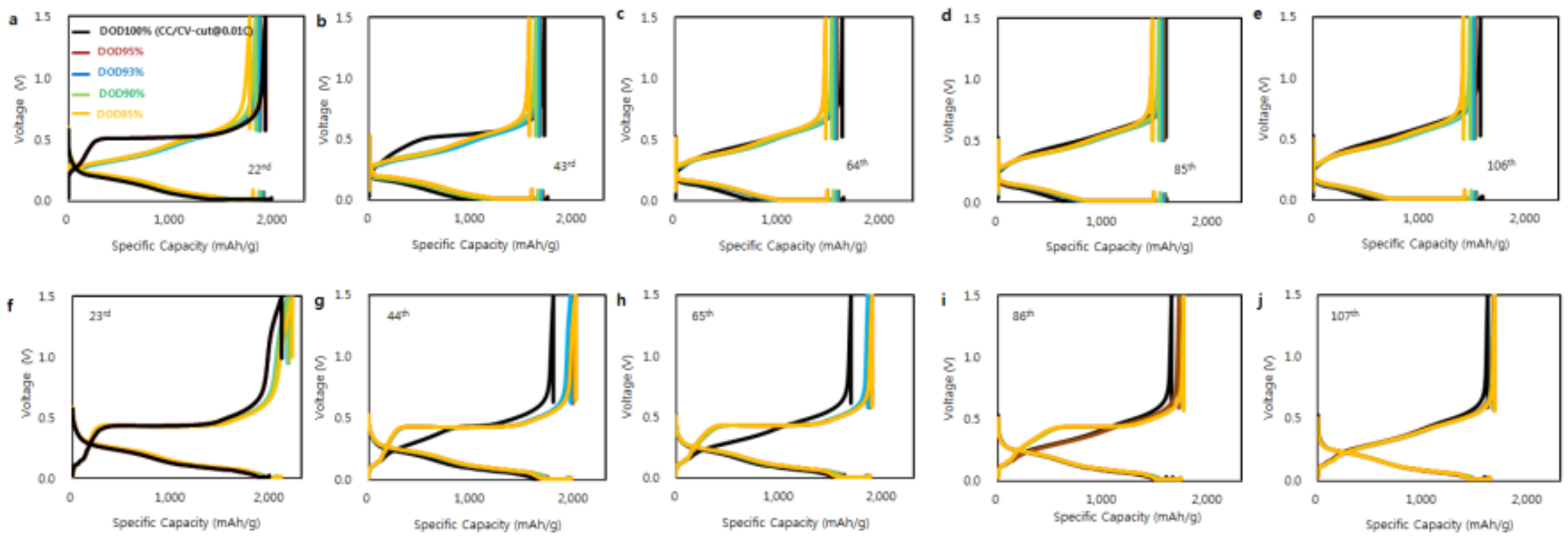

**Supplementary Fig.S 1**

$Li/Li^+$ potential in 2032-type coin half cells as a function of the specific capacity of type-A electrodes at recovery points (0.1 C with depth of discharge 100% (DOD100%), regardless of ever-cycled DOD) and at (recovery point − 1)$^{th}$ cycle under more precise DOD controls from 85-100% compared with Supplementary Fig.S-4. At the recovery points, all the electrodes are cycled at 0.1 C under CCCV-CC with DOD100%, regardless of the DOD history, while cells not at the recovery points are cycled at 1 C under CCCV-CC within the given DOD controls. (a–e) (recovery points − 1) cycle for type-A electrodes, and (f–j) recovery points for type-A electrodes. Black, brown, blue, green, and yellow profiles correspond to cycling under DOD100%, 95%, 93%, 90%, and 85%, respectively.

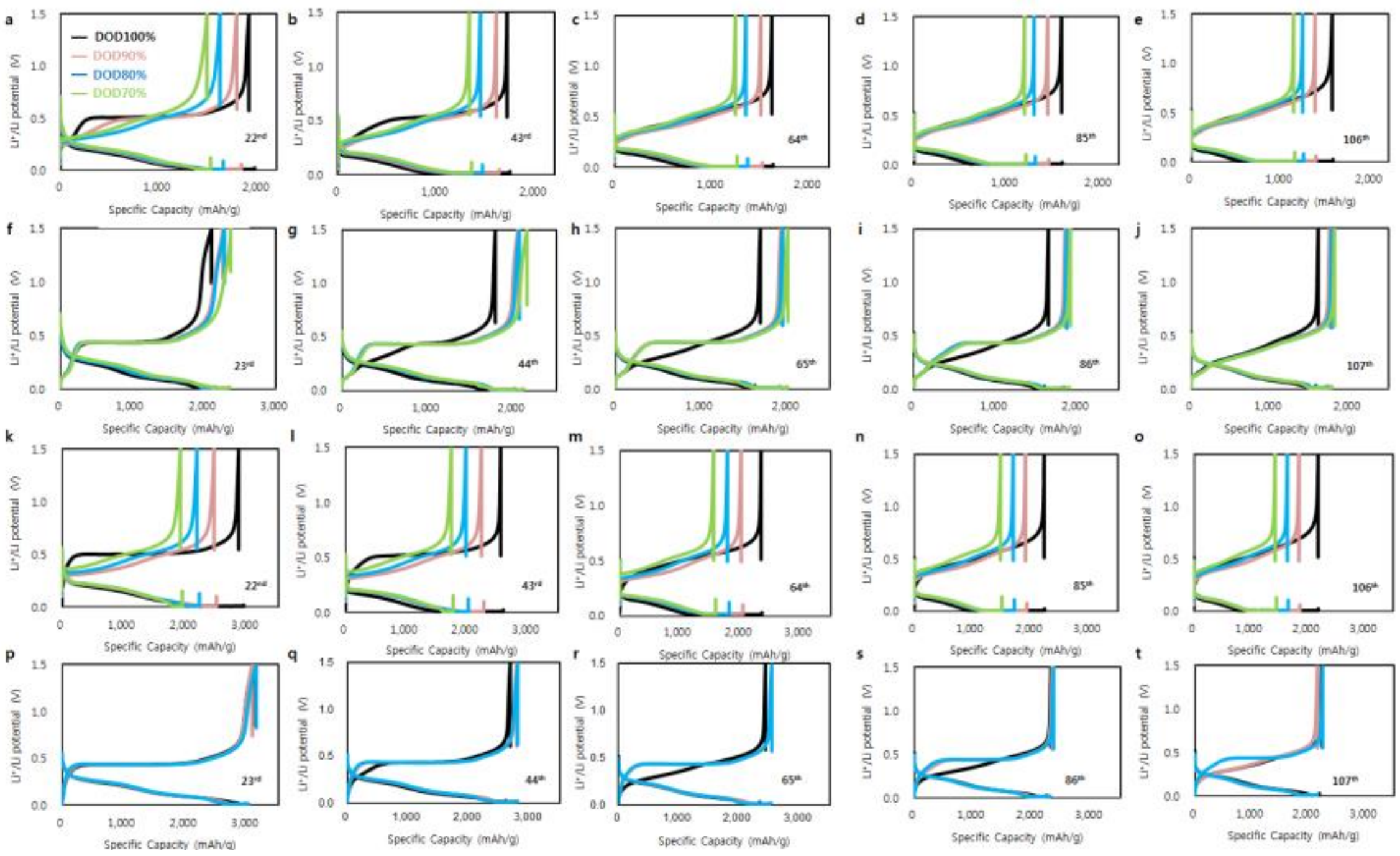

**Supplementary Fig.S 2**

$Li/Li^+$ potential in 2032-type coin half cells as a function of the specific capacity of type-A and -B electrodes at recovery points (0.1 C with DOD100%, regardless of ever-cycled depth of discharge (DOD)) and at (recovery point − 1)$^{th}$ cycle. At the recovery points, all the electrodes are cycled at 0.1 C under CCCV-CC with depth of discharge 100% (DOD100%), regardless of the DOD history, while cells not at the recovery points are cycled at 1 C under CCCV-CC within the given DOD controls. (a–e) (recovery point − 1)$^{th}$ cycle for type-A electrodes, (f–j) recovery points for type-A electrodes, (k–o) (recovery point − 1)$^{th}$ cycle for type-B electrodes, and (p–t) recovery points for type-B electrodes. Black, pink, blue, and green profiles correspond to cycling under DOD100%, 90%, 80%, and 70%, respectively.

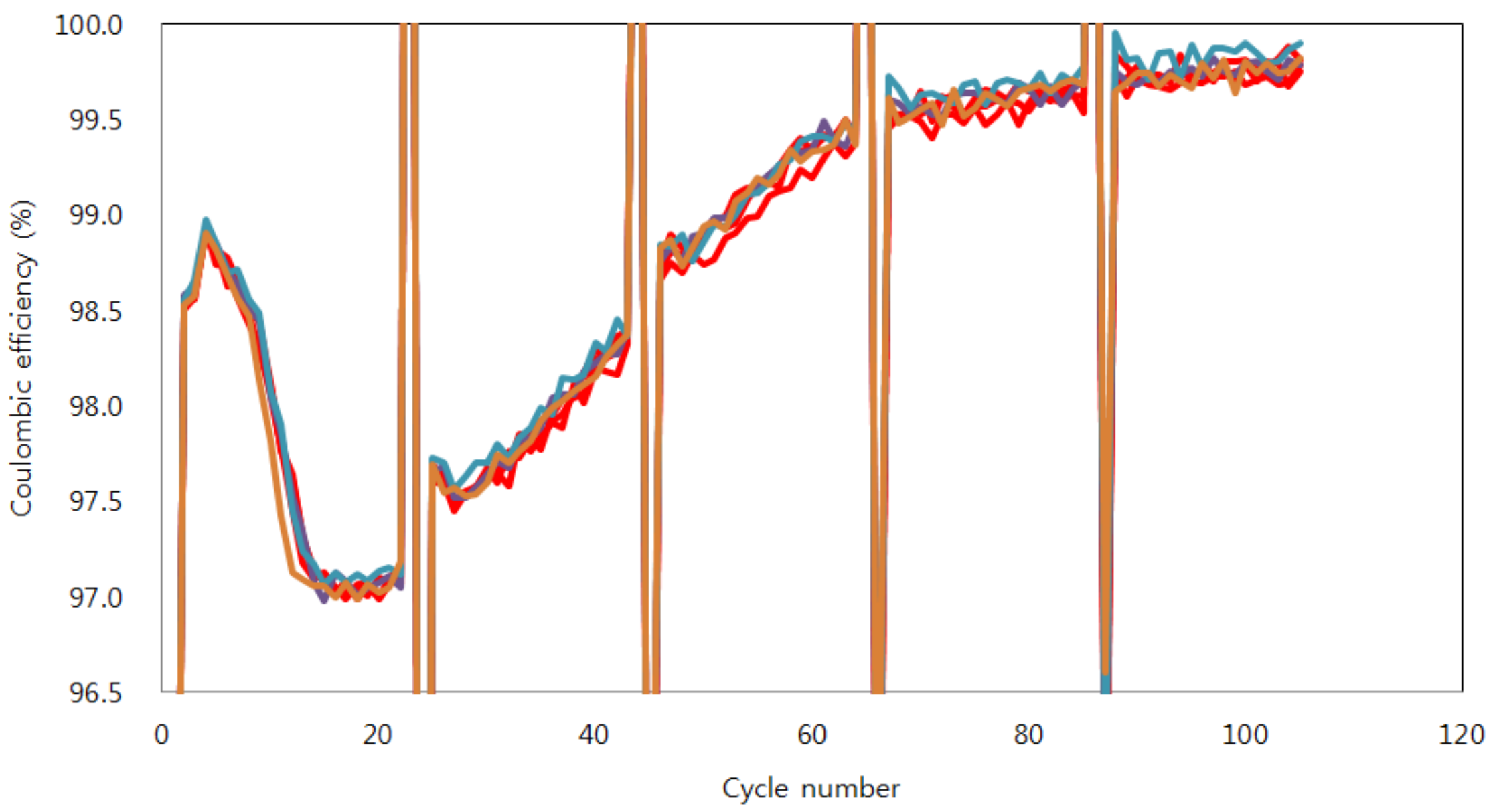

**Supplementary Fig.S 3**

Coulombic efficiency (CE) for six identical type-A electrodes in 2032-type coin half cells with depth of discharge 100% (DOD100%) at 1 C under CCCV–CV cycling conditions (current cutoff at 0.01 C in the CV domain). The standard deviation is ~0.07%.

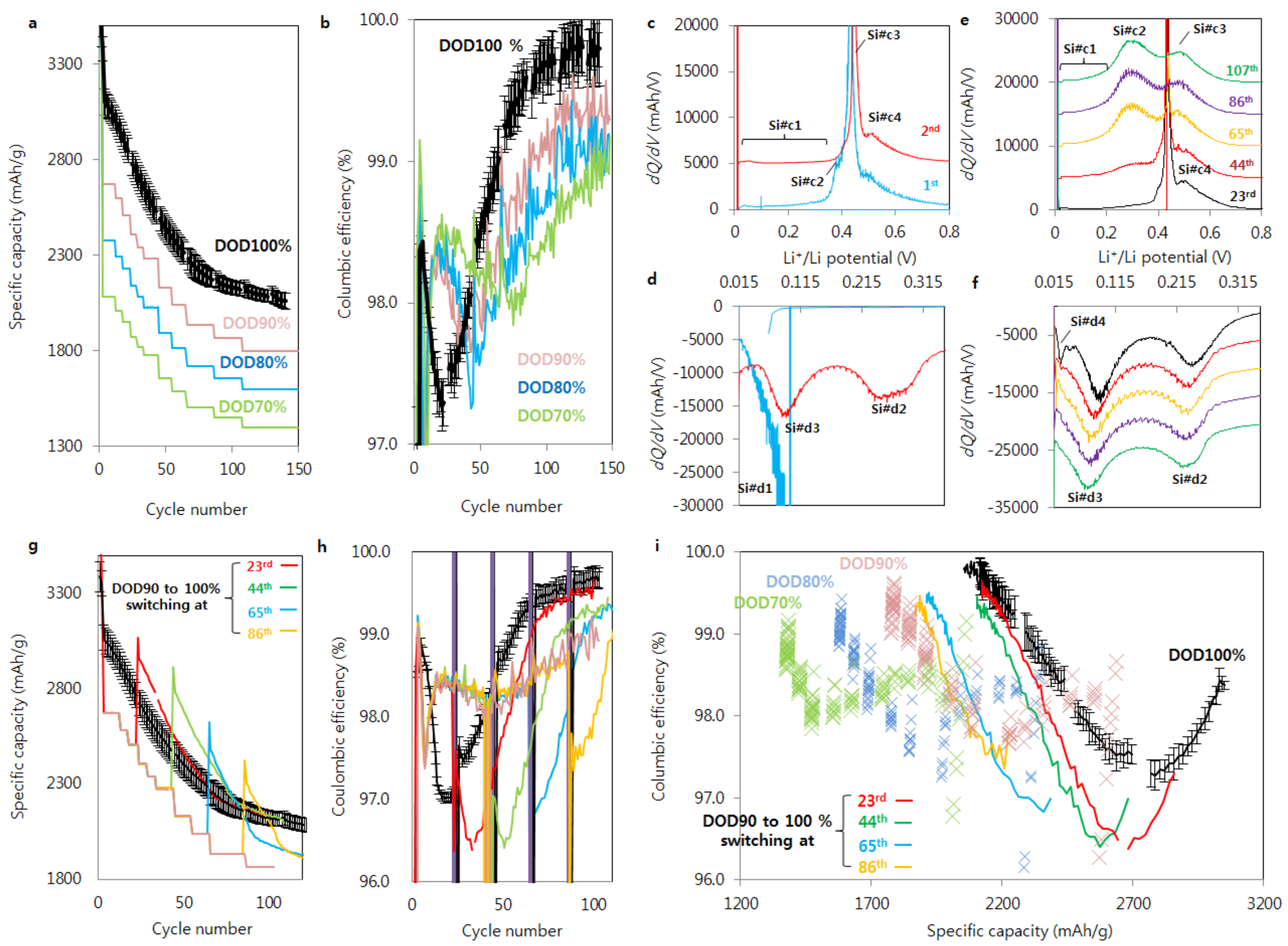

**Supplementary Fig.S 4**

Electrochemical properties for type-B electrodes in 2032-type coin half cells under different depth of discharge (DOD) controls under CCCV-CC cycling conditions (current cutoff at 0.01 C in the CV domain). (a) Specific capacity and (b) Coulombic efficiency (CE) for different DOD controls over 150 cycles. *dQ/dV* plots cycled under DOD100% during amorphising *c*-Si on (c) delithiation (charge) and (d) lithiation (discharge) at 0.1 C ($1^{st}$ cycle) and 0.2 C ($2^{nd}$ cycle), and at recovery points at 0.1 C on (e) delithiation and (f) lithiation. Note that at the recovery points, all the electrodes are cycled at 0.1 C regardless of the ever-cycled DOD pathways. The different Li–Si processes are labeled as Si#d X/Si#c X and Gr#d X/Gr#c X for Si and Gr for the #$X^{th}$ discharge/charge process; the *dQ/dV* profiles are stacked with a constant pitch to show the different processes more clearly. The notations for Li–Si processes are summarised in Supplementary Table S 2. (g) Specific capacity and (h) CE for DOD90% switched to DOD100% at different cycle numbers; the switching points for red, green, blue, and yellow solid lines are the $23^{rd}$, $44^{th}$, $65^{th}$, and $86^{th}$, respectively. (i) CE as a function of the specific capacity for different DOD controls. Coloured solid lines show the profiles for switching from DOD90% to DOD100% at different cycle numbers in (g,h). CE error bars for DOD70-90% are all within a range of ±0.1 % and omitted in (b,h,i).

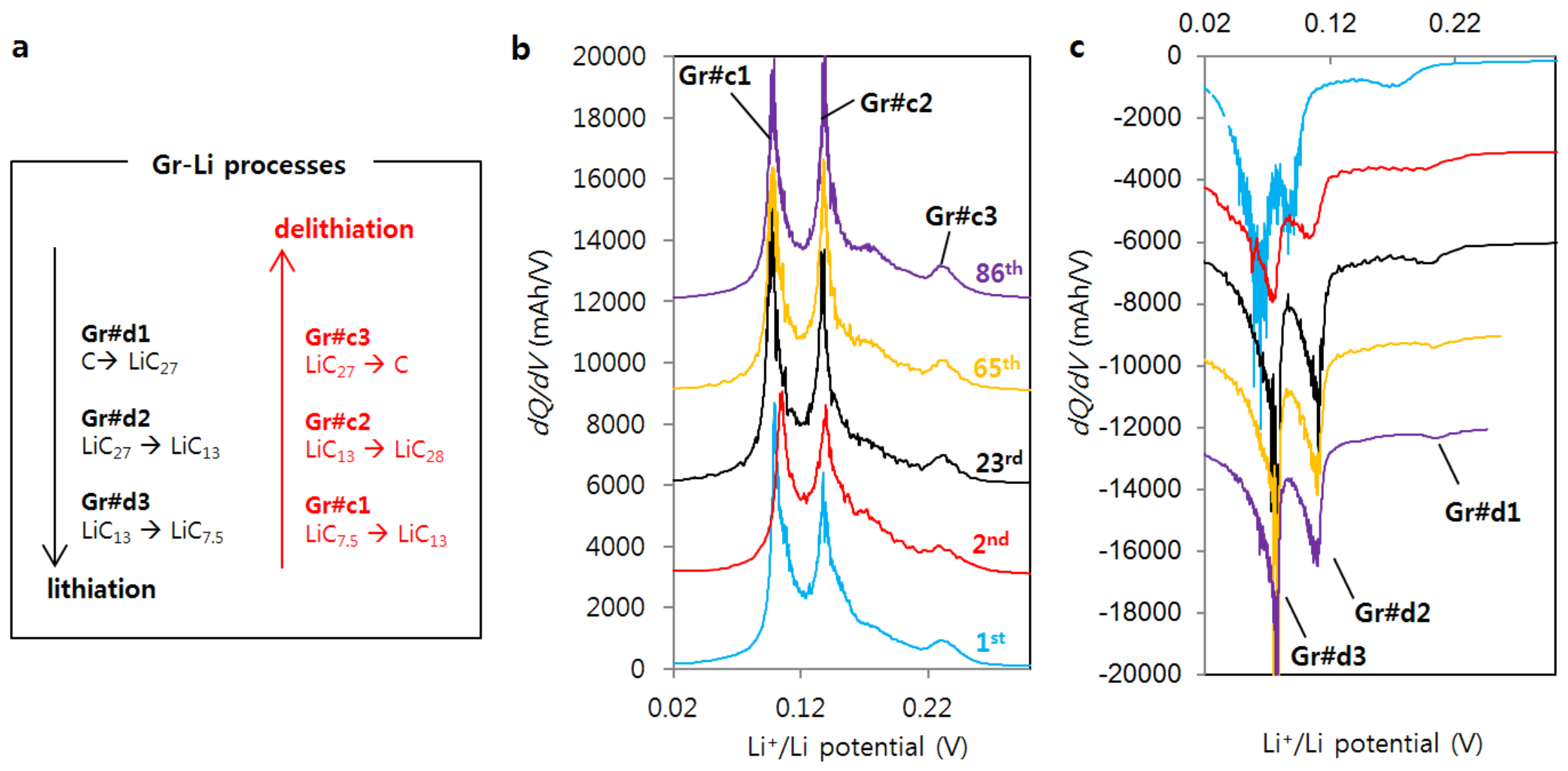

**Supplementary Fig.S 5**

Li–Gr processes for bare graphite (Gr) electrodes in 2032-type coin half cells at recovery points at 0.1 C with depth of discharge 100% (DOD100%) under CCCV-CC cycling conditions (current cutoff at 0.01 C in the CV domain). (a) Schematic showing Li–Gr reactions. *dQ/dV* profiles on (b) delithiation and (c) lithiation. Li–Gr processes during discharge/charge are labeled as Gr#d-X/Gr#c-X, for the #X$^{th}$ discharge/charge process; the *dQ/dV* profiles are stacked with a constant pitch to show the different processes more clearly. Blue, red, black, yellow, and purple profiles correspond to the 1$^{st}$, 2$^{nd}$, 23$^{rd}$, 65$^{th}$, and 86$^{th}$ cycles, respectively.

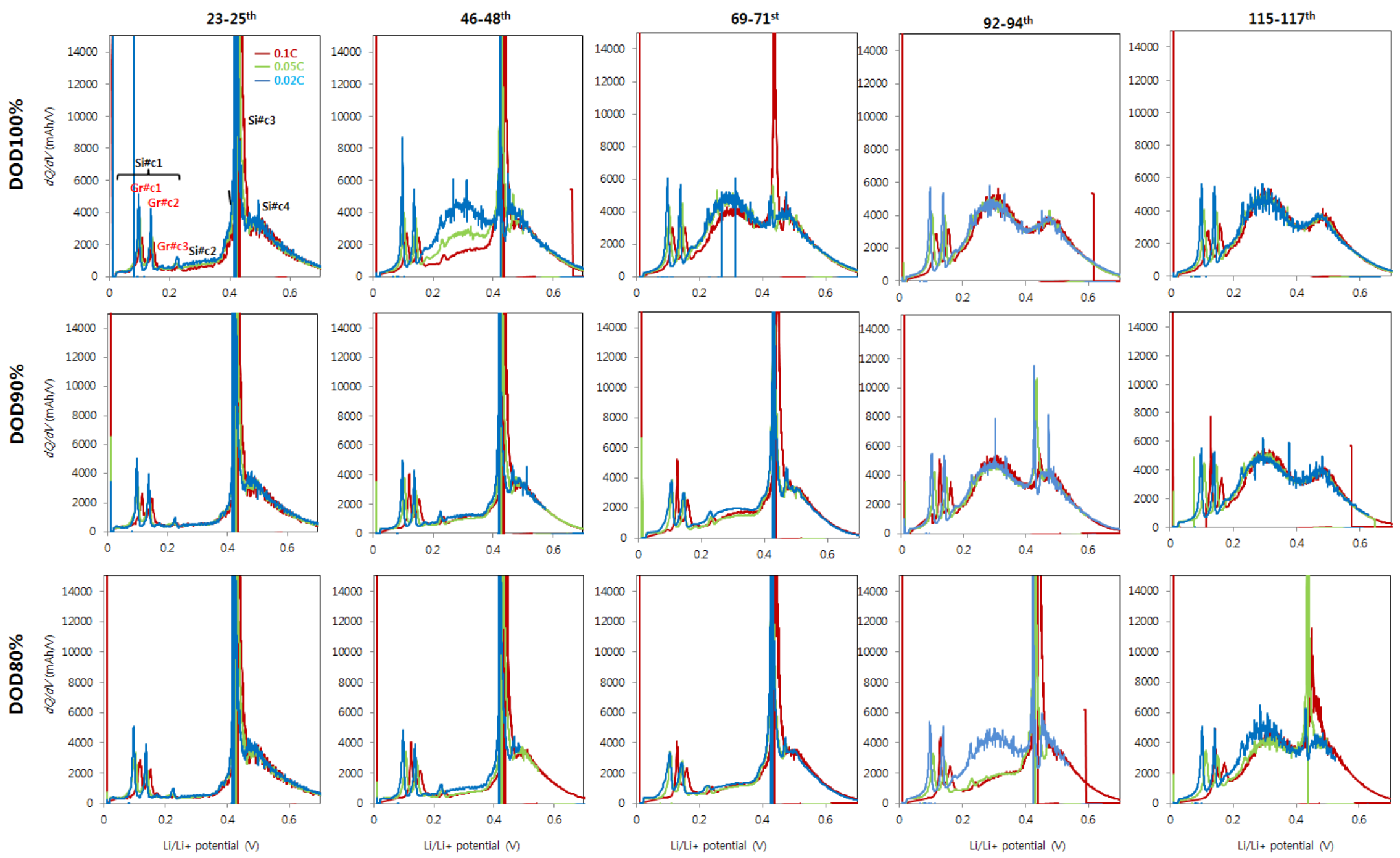

**Supplementary Fig.S 6**

Evolution of Si#c1–3 processes in *dQ/dV* profiles on delithiation of type-A electrodes (2250 mAh/g, initial reversible capacity) in 2032-type coin half cells at recovery points (0.1 C with depth of discharge 100% (DOD100%), regardless of ever-cycled DOD) over 117 cycles with different depth of discharge DOD controls under CCCV-CC cycling conditions (current cutoff at 0.01 C in the CV domain). The notations for Li–Si processes (Si#dX/Si#cX represents the $X^{th}$ Li–Si discharge/charge process) are summarised in Supplementary Table S 2. At the recovery points, three lower current rates (0.02, 0.05, and 0.1 C) are used consecutively to observe the rate dependence of the processes. Blue, green, and red profiles correspond to rates of 0.02, 0.05, and 0.1 C, respectively.

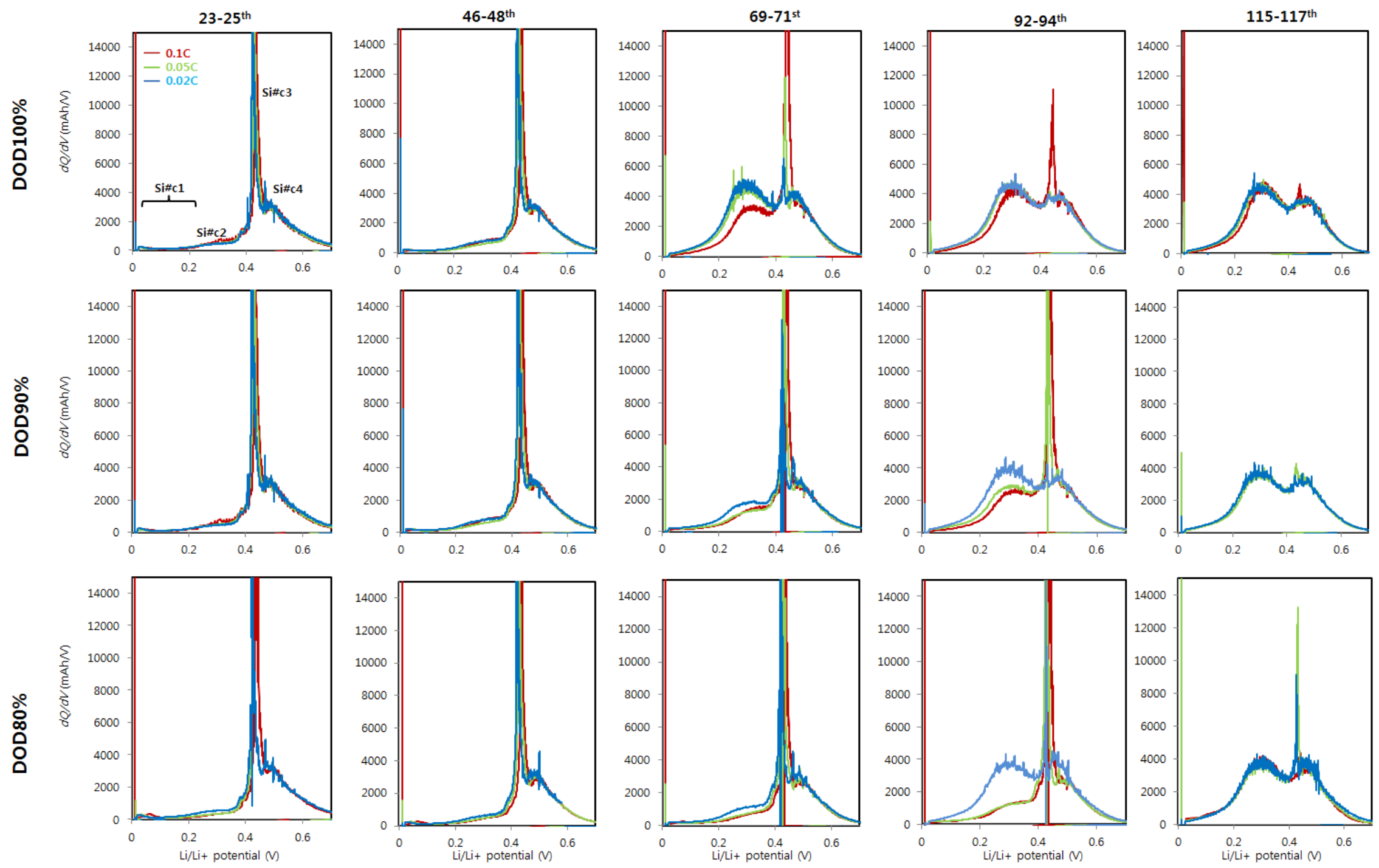

**Supplementary Fig.S 7**

Evolution of Si#c1–3 processes in *dQ/dV* profiles on delithiation of type-B electrodes (3350 mAh/g, initial reversible capacity) in 2032-type coin half cells at recovery points (0.1 C with depth of discharge 100% (DOD100%), regardless of ever-cycled DOD) over 117 cycles with different depth of discharge DOD controls under CCCV-CC cycling conditions (current cutoff at 0.01 C in the CV domain). The notations for Li–Si processes (Si#dX/Si#cX represents the $X^{th}$ Li–Si discharge/charge process) are summarised in Supplementary Table S 2. At the recovery points, three lower current rates (0.02, 0.05, and 0.1 C) are used consecutively to observe the rate dependence of the processes. Blue, green, and red profiles correspond to rates of 0.02, 0.05, and 0.1 C, respectively.

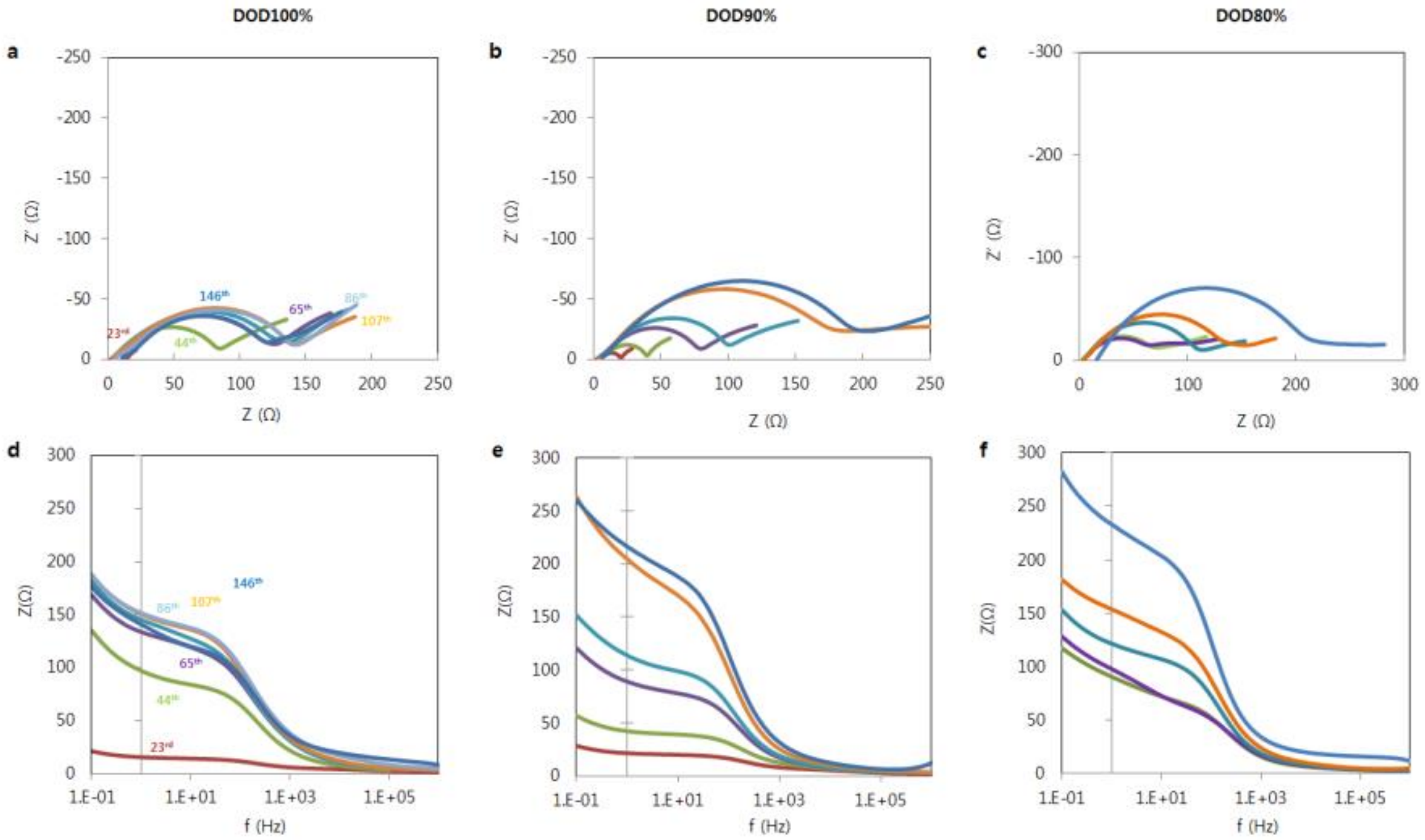

**Supplementary Fig.S 8**

Electrochemical impedance spectroscopy for lithiated type-A electrodes in 2032-type coin half cells cycled under depth of discharge 80–100% (DOD80–100%) over 146 cycles. The cell is cycled at 1 C until it reaches a cycle number prior to the recovery points (0.1 C with DOD100%, regardless of ever-cycled DOD), and then switched to 0.05 C at the recovery points under CCCV, holding the voltage at 10 mV on lithiation for at least for 24 h until the residual current becomes less than 0.001 C. The frequency is swept from 1 MHz to 0.1 Hz with a fluctuating voltage of ±5 mV. Red, green, purple, light blue, orange, and dark blue profiles correspond to the 23$^{rd}$, 44$^{th}$, 65$^{th}$, 86$^{th}$, 107$^{th}$, and 146$^{th}$ cycles, respectively.

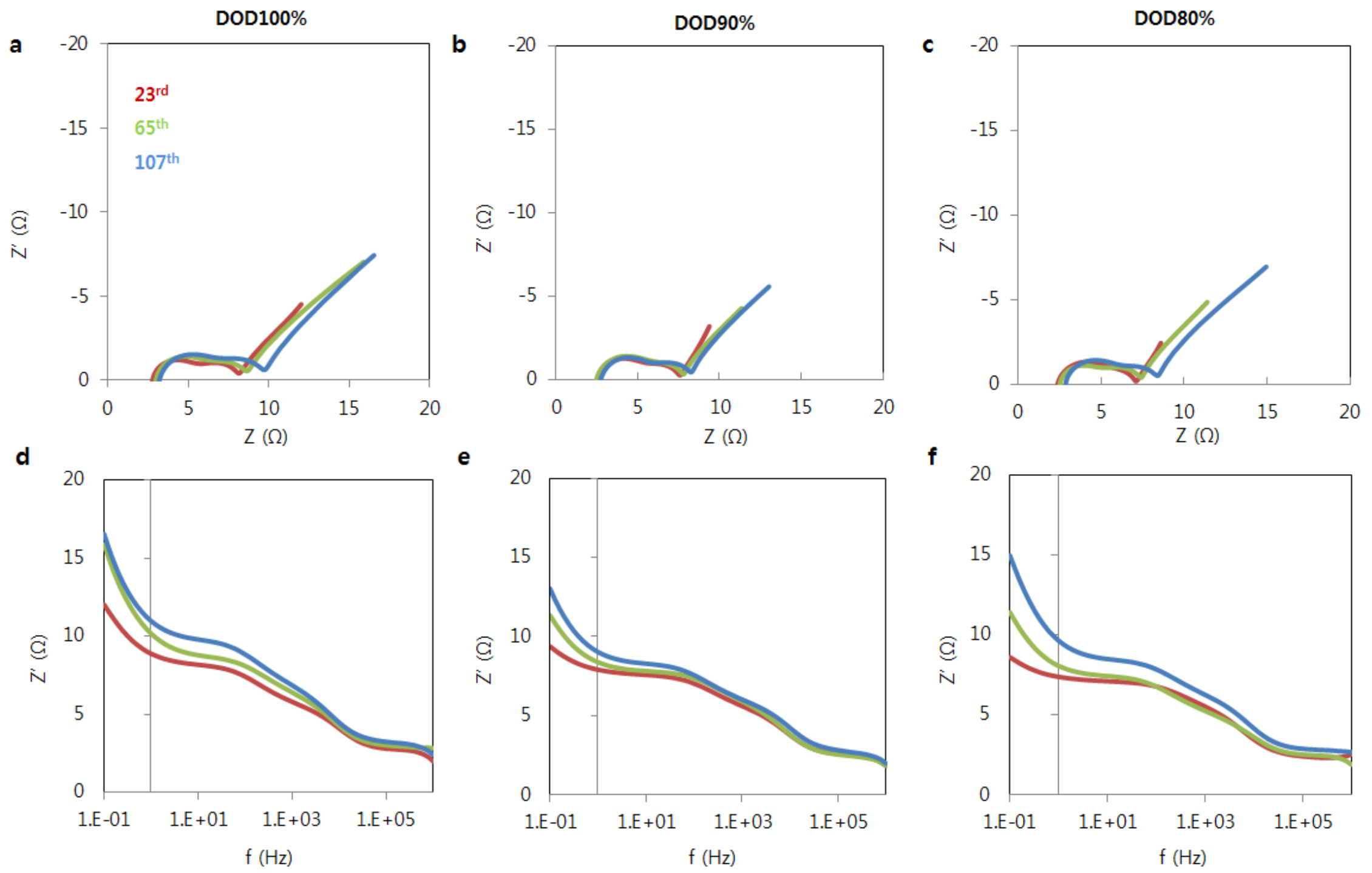

**Supplementary Fig.S 9**

Electrochemical impedance spectroscopy for lithiated type-A electrodes in 2032-type *symmetric* cells cycled under DOD80% for lithiated type-A electrodes in 2032-type coin cells. Two identical cells are cycled at 1 C under different depths of discharge (DOD) until it reaches a cycle number prior to the recovery points (0.1 C with DOD100%, regardless of ever-cycled DOD), and then switched to 0.05 C at the recovery points under CCCV, holding the voltage at 10 mV on lithiation for at least for 24 h until the residual current becomes less than 0.001 C. The anodes in two coin cells are then disassembled in an Ar-filled glovebox and reassembled into a newly assembled symmetric coin cell with newly filled the electrolyte and the separator. The frequency is swept from 1 MHz to 0.1 Hz with a fluctuating voltage of ±5 mV. Red, green, and blue profiles correspond to the 23$^{rd}$, 65$^{th}$, and 107$^{th}$ cycles, respectively.

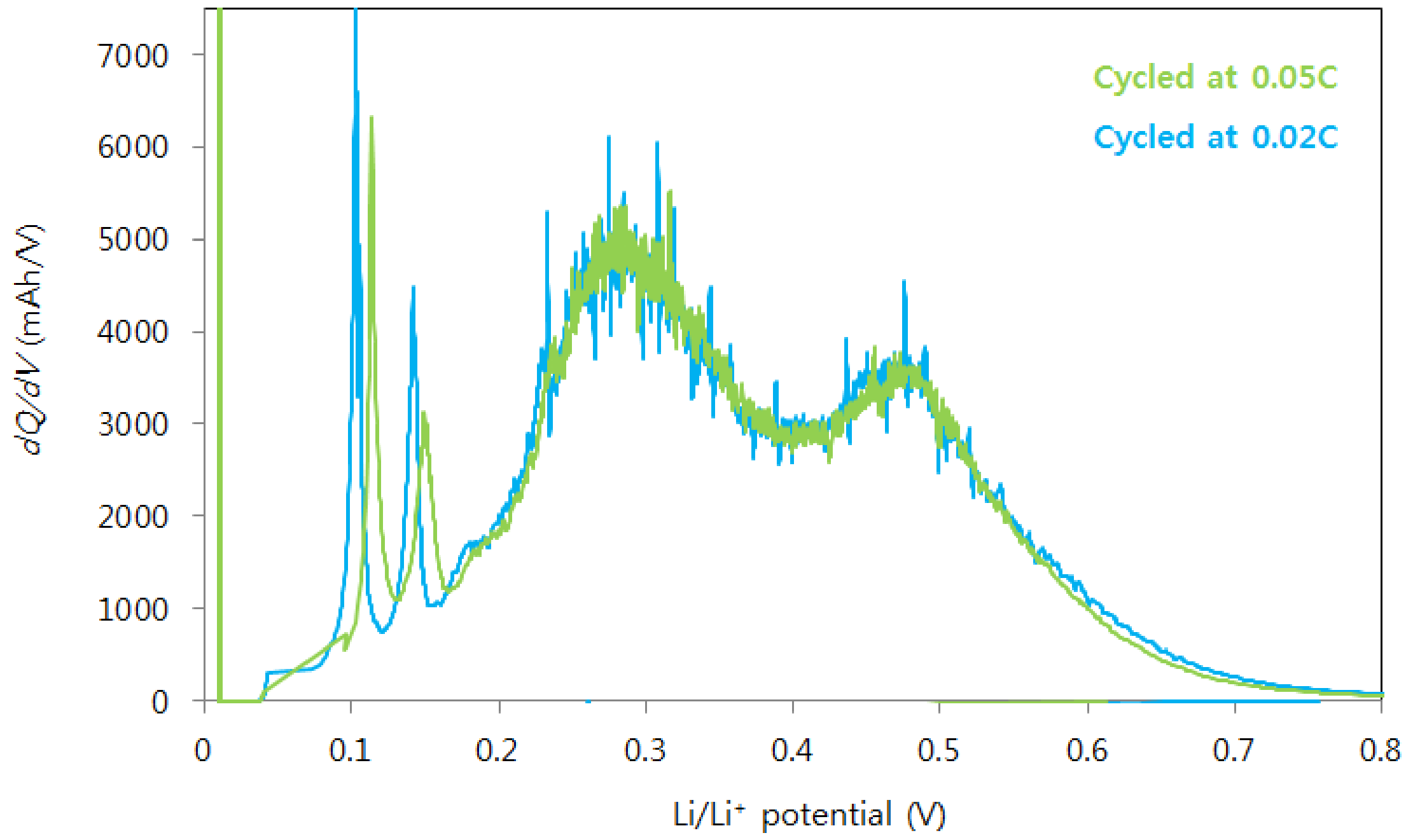

**Supplementary Fig.S 10**

*dQ/dV* profiles on delithiation of type-A electrodes at 0.05 C (green) and 0.02 C (blue) in a newly assembled 2032-type half coin cell after 107 cycles so as to investigate effects of the cycled Li metal on the electrochemistry. The type-A electrode is cycled in a 2032-type coin half-cell for 107 cycles under CCCV-CC with 0.01 C current cutoff in the CV domain. After cycling, the cell is disassembled in an Ar-filled glovebox and the half-cell is reassembled with fresh polished Li-metal with the newly injected electrolyte. Green, and blue profiles correspond to 0.05 and 0.02 C, respectively. Si#c3 is absent under such quasi-thermodynamic cycling conditions. The notations for Li–Si processes (Si#d-X/Si#c-X represents the $X^{th}$ Li–Si discharge/charge process) are summarised in Supplementary Table S 2.

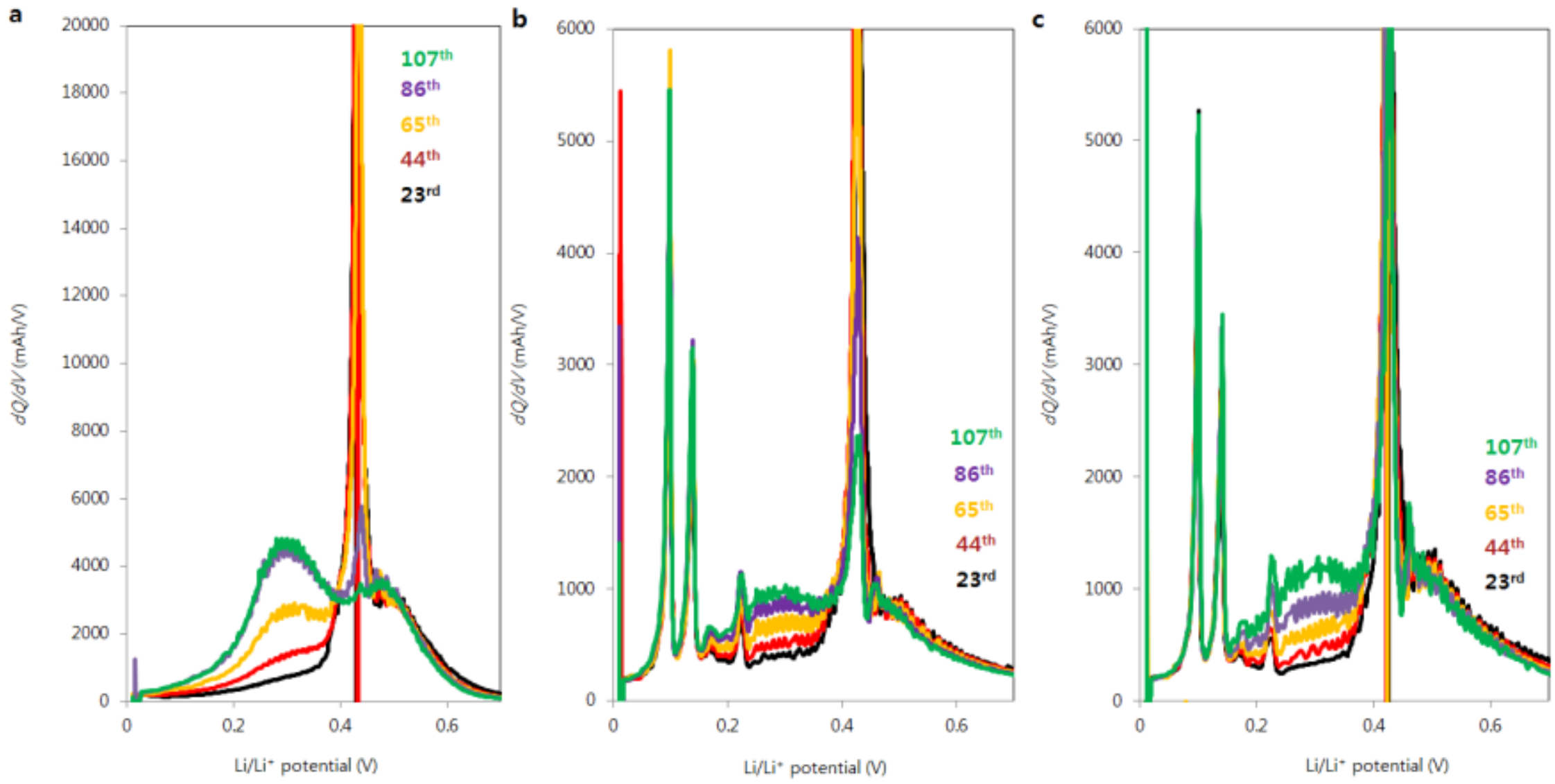

**Supplementary Fig.S 11**

*dQ/dV* profiles on delithiation of electrodes made of different Si sources at the recovery points. Spherical crystalline Si nanoparticle-based (<100 nm, Sigma Aldrich) composite with different Si concentrations (a) 2620 mAh/g and (b) 1640 mAh/g. (c) Smaller sized crystalline Si nanoparticle-based (30–50 nm, Hwananotech) composite (1210 mAh/g). Electrodes are cycled in 2032-type coin half cells for 107 cycles under 100% depth of discharge 100% (DOD100%). Black, red, yellow, purple, and green profiles correspond to the 23$^{rd}$, 44$^{th}$, 65$^{th}$, 86$^{th}$, and 107$^{th}$ cycles, respectively. The results indicate that the change of the reaction sequence occurs not only in the Si nanoparticles in the main experiments but also in a variety of Si sources.

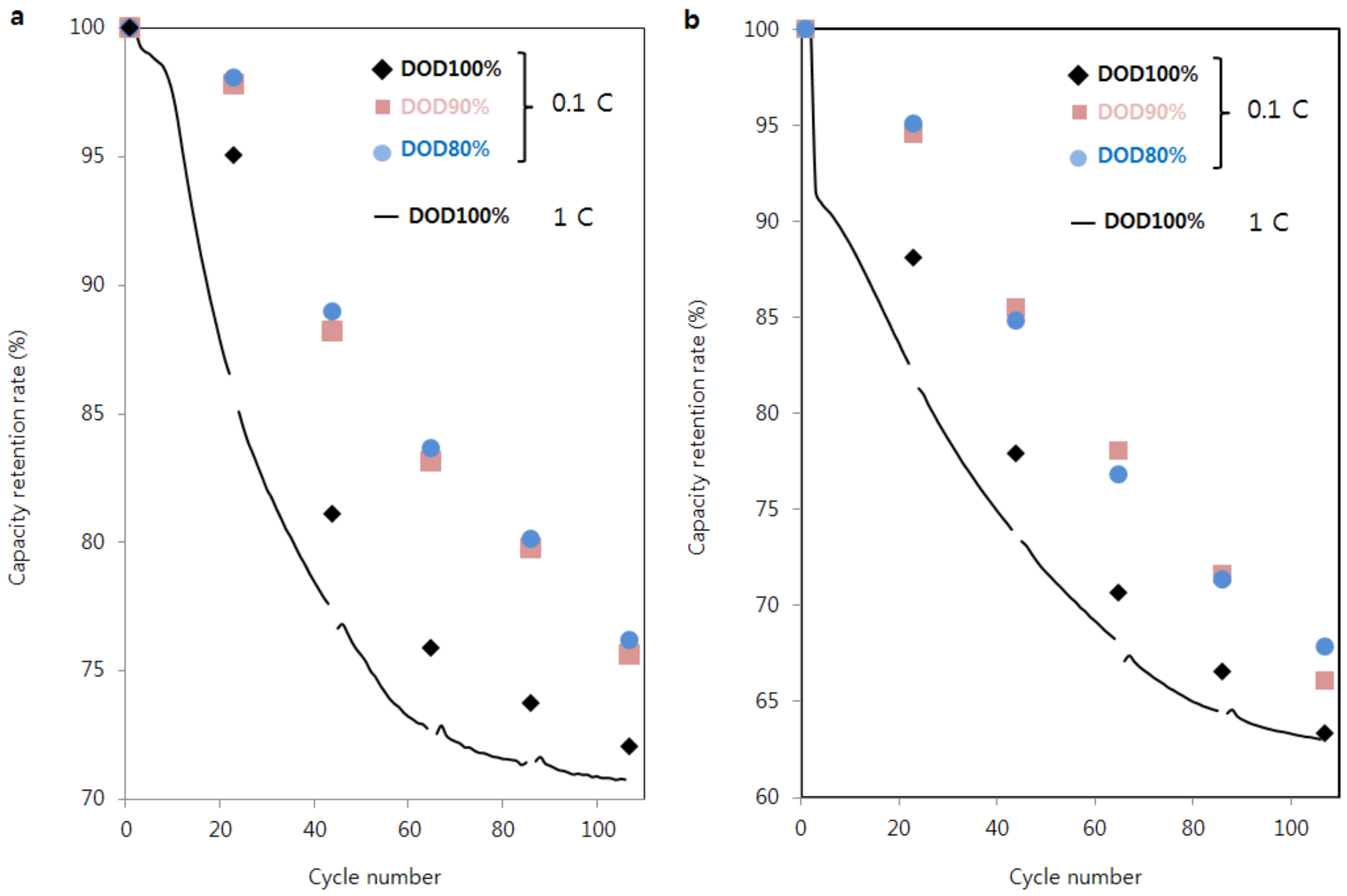

**Supplementary Fig.S 12**

Capacity retention rate (CRR) for (a) type-A and (b) type-B electrodes (initial reversible capacities of 2250 and 3350 mAh/g, respectively) in 2032-type coin half cells. The solid dots show CRR at recovery points (0.1 C with DOD100%, regardless of ever-cycled depth of discharge **(**DOD) controls) so as to highlight the survived Si difference depending on different DOD cycling protocols. Black, pink, and blue dots correspond to DOD100%, 90%, and 80%, respectively. The black solid lines show reference CRR under DOD100% at 1 C.

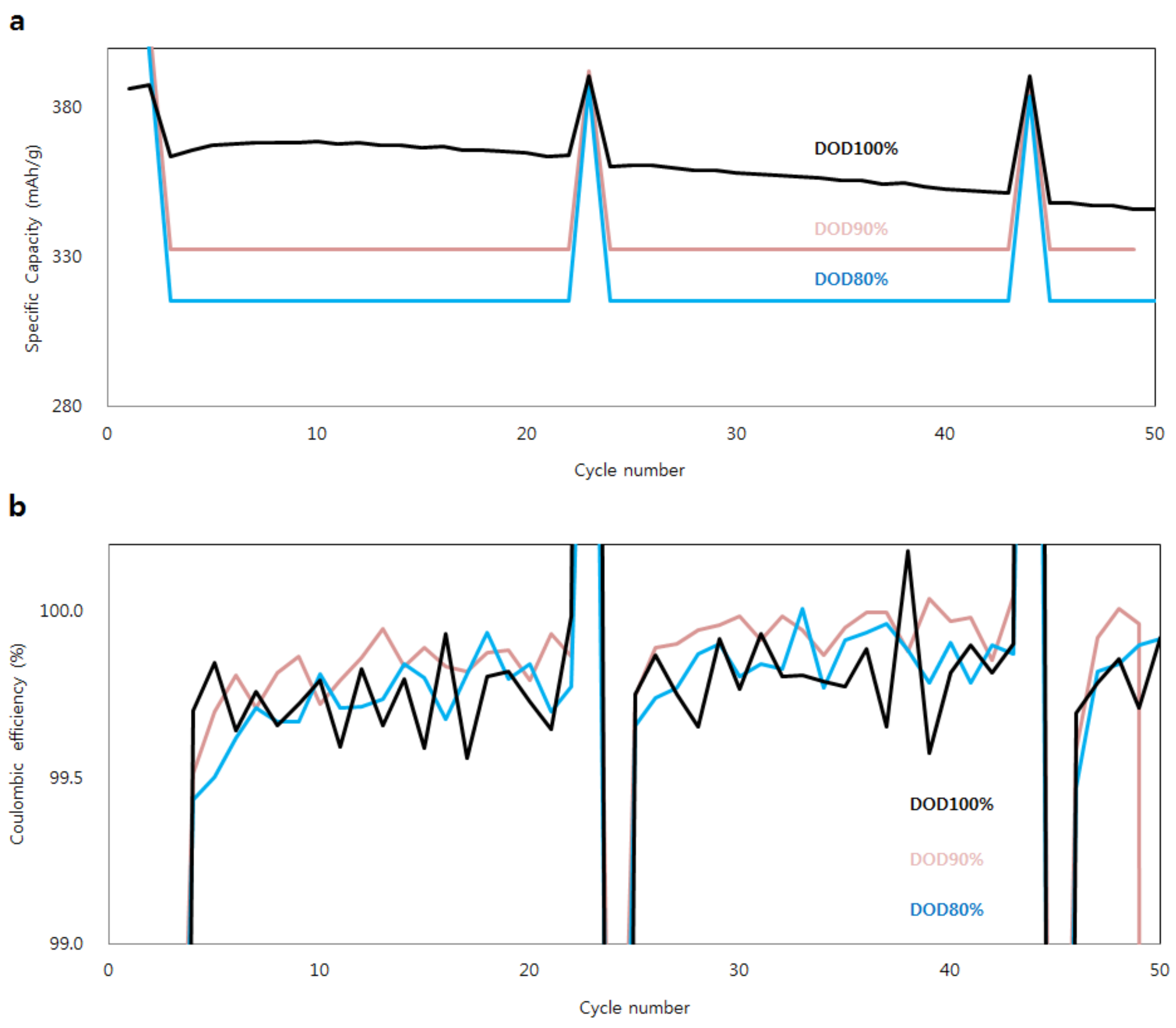

**Supplementary Fig.S 13**

Electrochemical performance of bare graphite (Gr) electrodes in 2032-type coin half cells w ith the same cell design as the ones in type-A and –B, cycled under different depth of dis charge **(**DOD) controls under CCCV-CC cycling conditions (current cutoff at 0.01 C in the CV domain). (a) Specific capacity and (b) Coulombic efficiency. Black, pink, and blue lines correspond to DOD100%, 90%, and 80%, respectively.

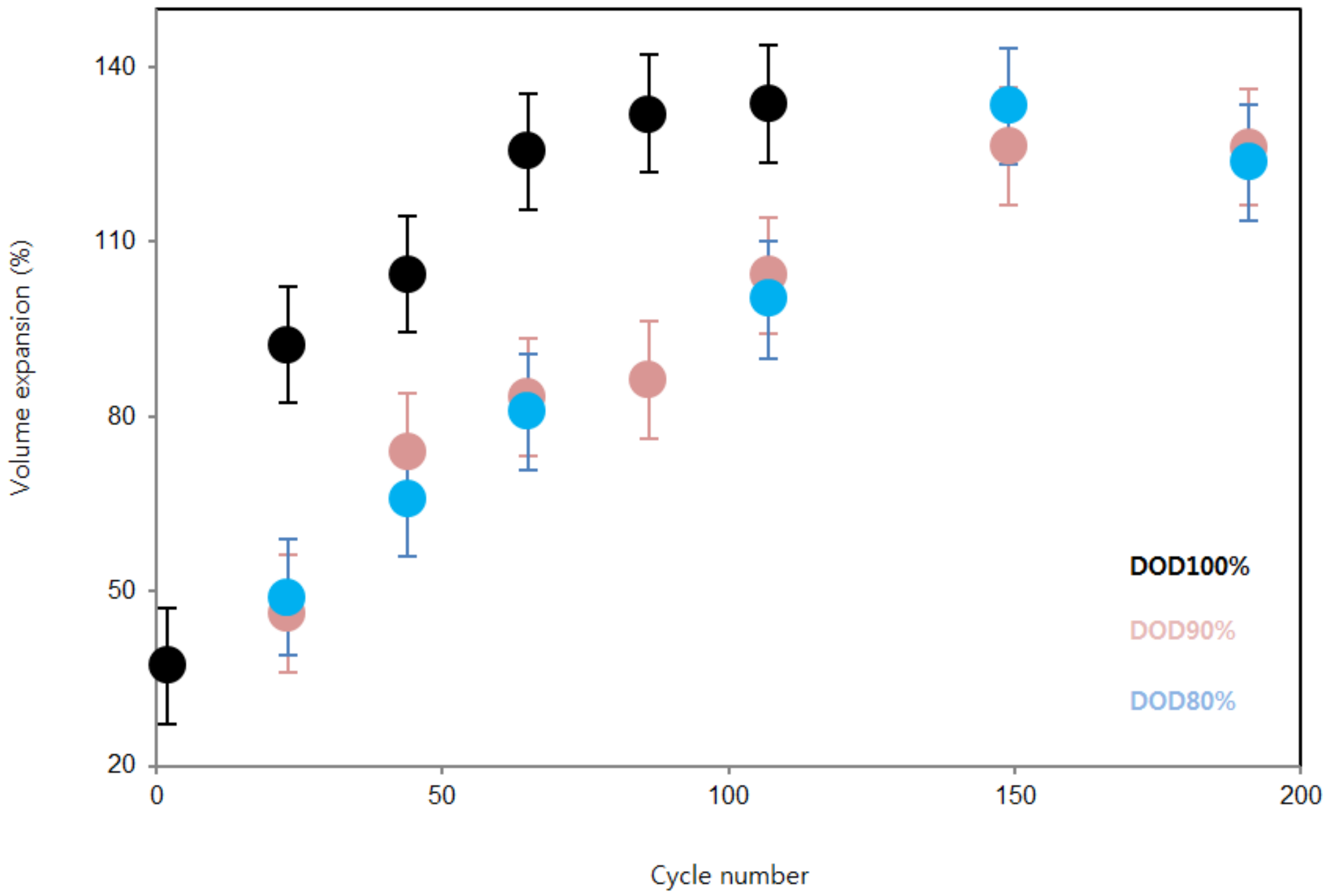

**Supplementary Fig.S 14**

Volume expansion rate for fully delithiated type-A electrodes (2250 mAh/g, initial reversible capacity) under different depth of discharge (DOD) controls plotted over number of cycle at recovery points in 2032 coin half cells. Black, pink, and blue dots correspond to DOD100%, 90%, and 80%, respectively.

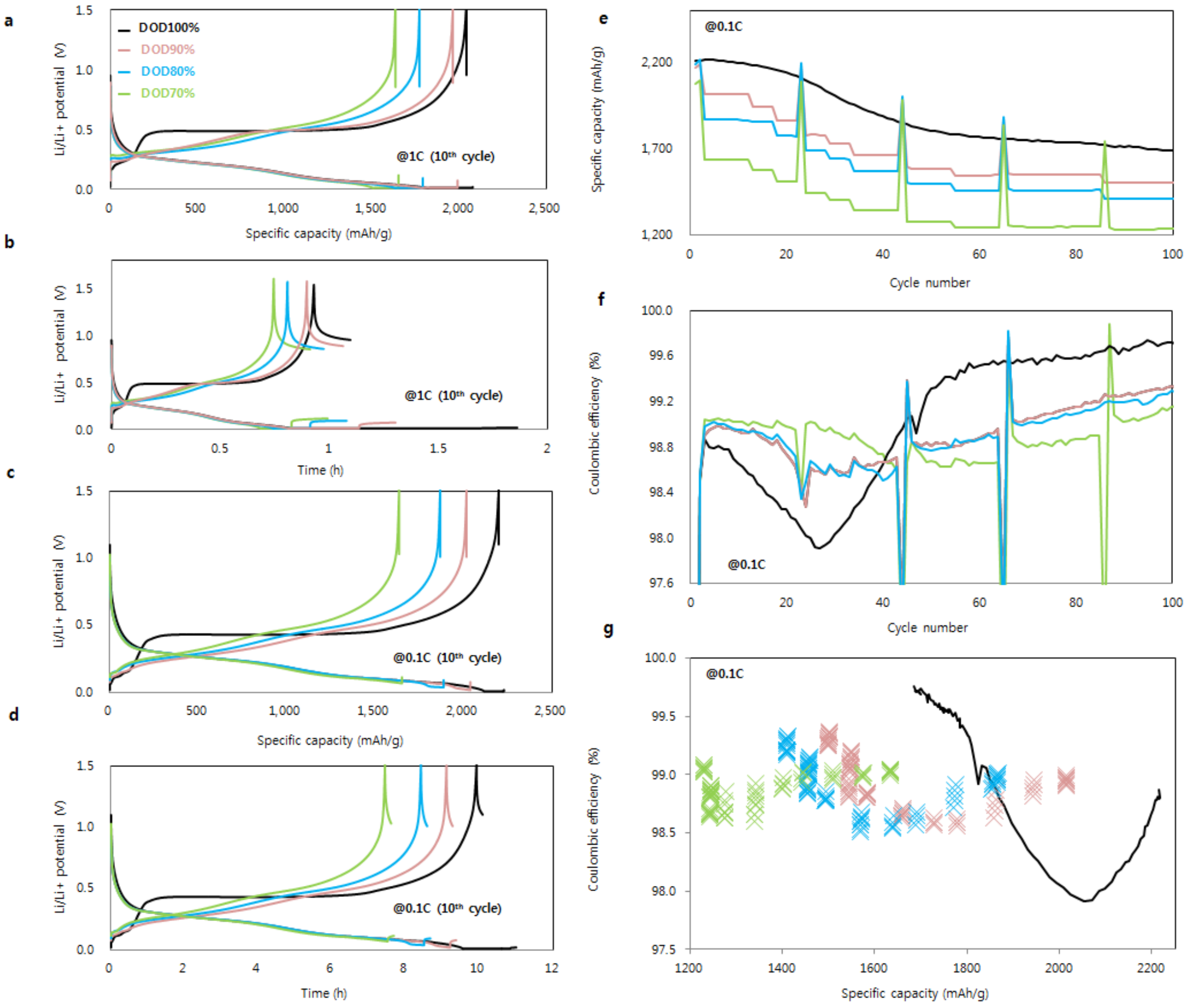

**Supplementary Fig.S 15**

$Li/Li^+$ potential in 2032-type coin half cells for type-A electrodes as a function of (a,c) specific capacity and (b,d) time in the 10th cycle at (a,b) 1 C and (c,d) 0.1 C. The total duration of lithiation is not proportional to depth of discharge **(**DOD) at 1 C, while it is proportional at 0.1 C. (e) Specific capacity and (f) Coulombic efficiency (CE) cycled at 0.1 C with different DOD controls. (g) CE as a function of specific capacity for type-A electrodes under different DOD controls. Black, pink, blue, and green profiles correspond to DOD100%, 90%, 80%, and 70%, respectively. CE error bars are all within a range of ±0.1 % and omitted in the figure.

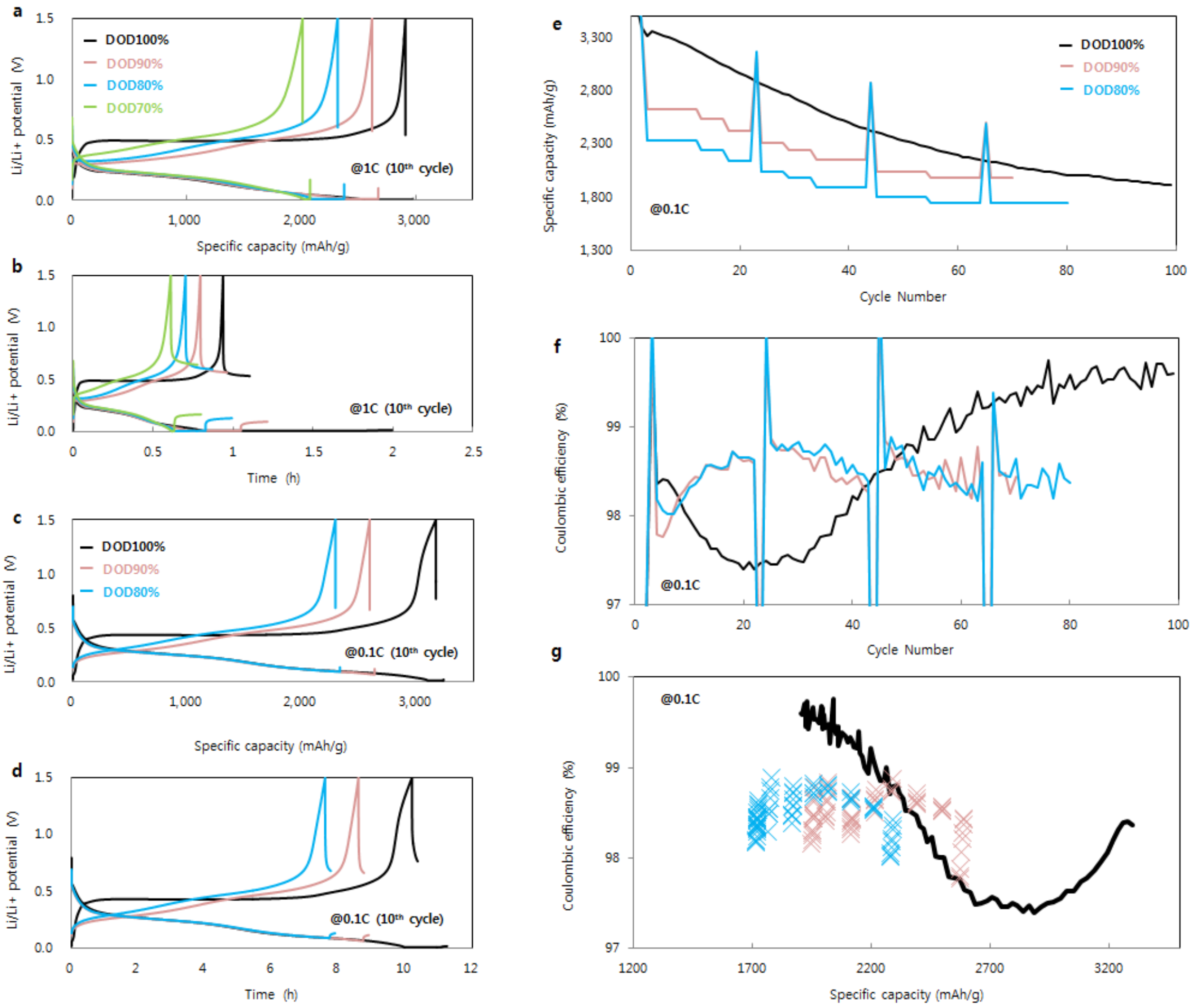

**Supplementary Fig.S 16**

Li/Li+ potential in 2032-type coin half cells for type-B electrodes as a function of (a,c) specific capacity and (b,d) time in the 10th cycle at (a,b) 1 C and (c,d) 0.1 C. The total duration of lithiation is not proportional to depth of discharge (DOD) at 1 C, while it is proportional at 0.1 C. (e) Specific capacity and (f) Coulombic efficiency (CE) cycled at 0.1 C with different DOD controls. (g) CE as a function of specific capacity for type-B electrodes under different DOD controls. Black, pink, blue, and green profiles correspond to DOD100%, 90%, 80%, and 70%, respectively. CE error bars are all within a range of ±0.1 % and omitted in the figure.

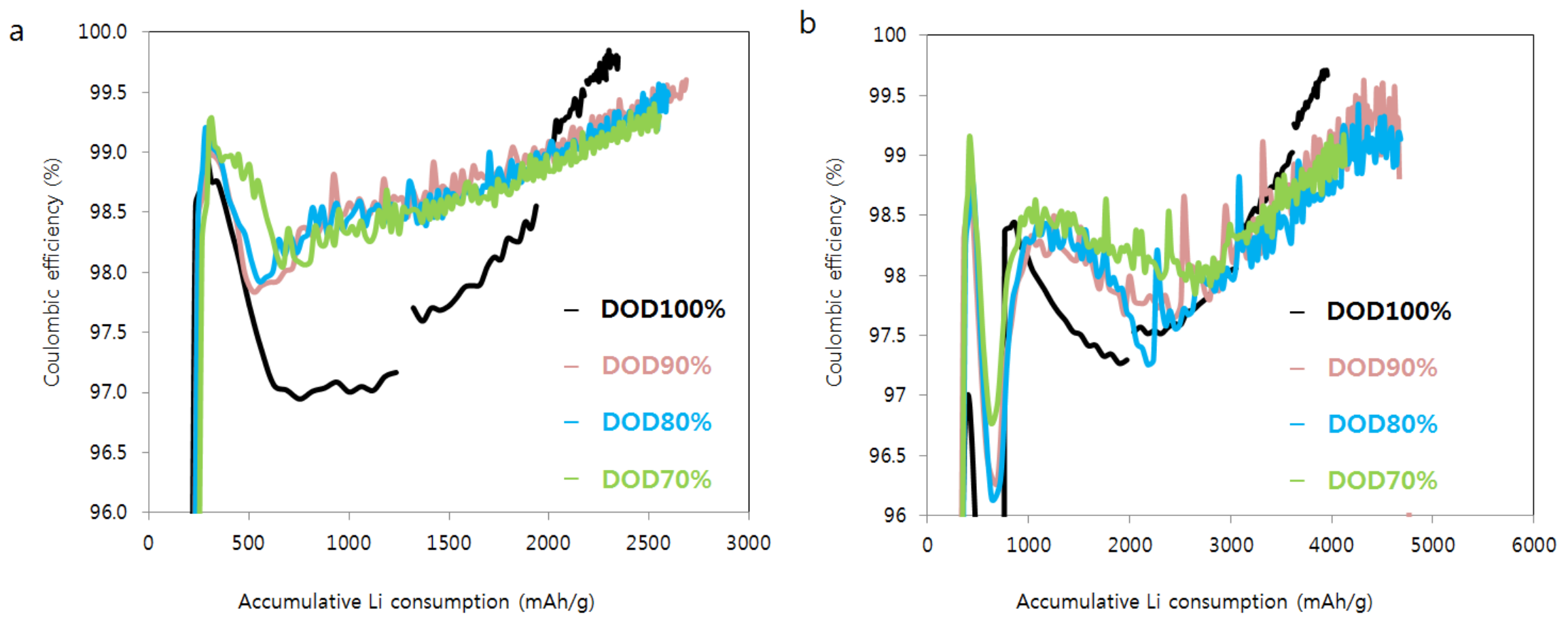

**Supplementary Fig.S 17**

Coulombic efficiency (CE) cycled over 190 cycles in 2032-type coin half-cell as a function of accumulated Li consumption for (a) type-A and (b) type-B electrodes. Accumulative Li consumption is calculated based on Figure 1(a,b) and Supplementary Fig.S-2(a,b), i.e. a reversible capacity multiplied by corresponding (100-CE)/100 is accumulatively added.

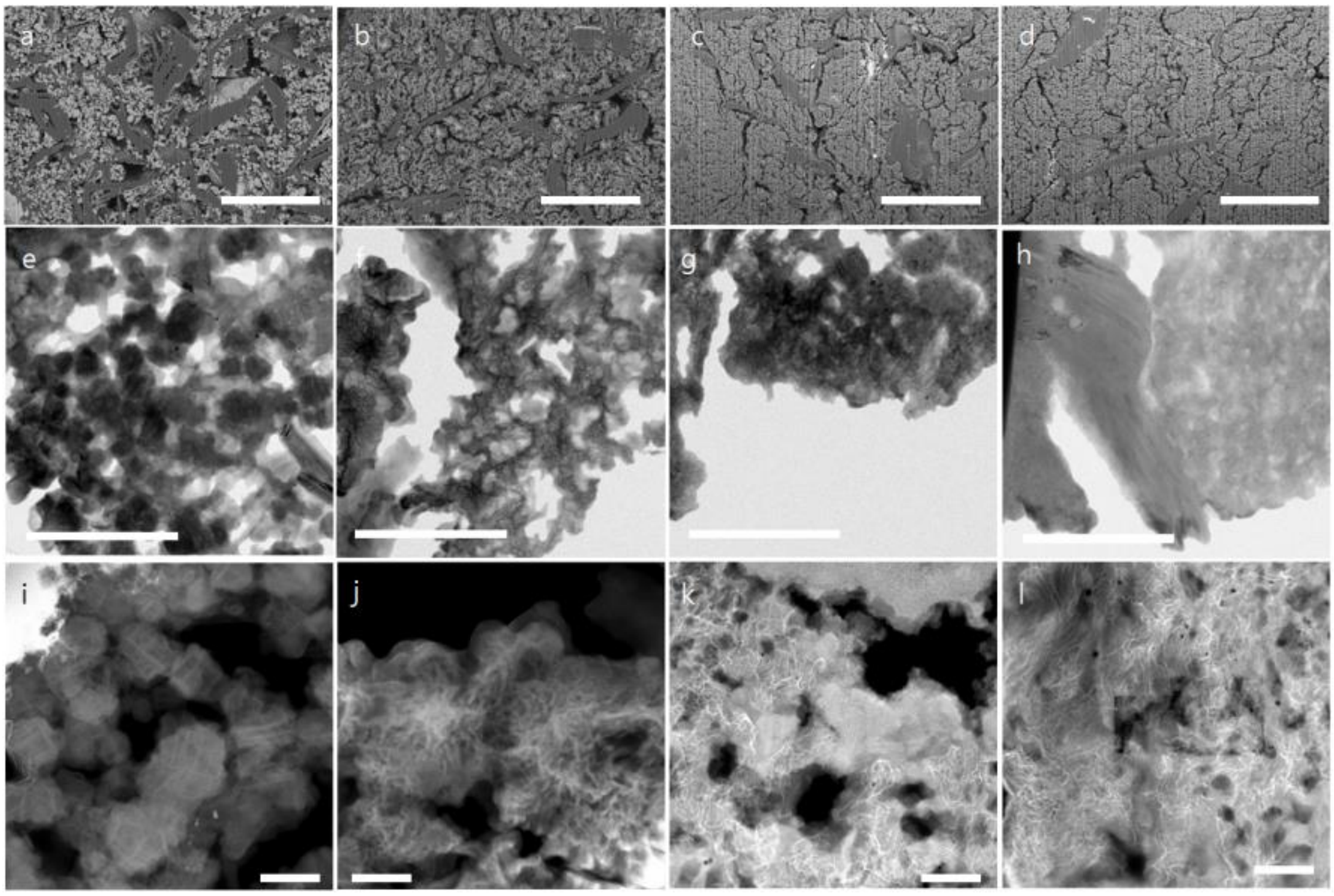

**Supplementary Fig.S 18**

Electron microscopy images of fully delithiated type-A electrodes cycled in 2032-type coin half cells under DOD100% over 107 cycles. (a–d) Back scattering electron (BSE) images of electrode cross sections sliced via focused ion beam (FIB), (e–h) bright-field (BF) transmission electron microscopy (TEM) images, and (i–l) high-angle annular dark field (HAADF) images of the sliced electrodes. (a,e,i) 3$^{rd}$ (after the amorphisation of *c*-Si), (b,f,j) 23$^{rd}$, (c,g,k) 65$^{th}$, and (d,h,l) 107$^{th}$ cycle (see *TEM* in Methods). Prior to observation, the coin cell battery is cycled *ex situ* under CCCV-CC at 1 C (0.01 C current cutoff) for the target number of cycles, terminating the schedule at 1.5 V under CC. The electrodes are washed with dimethyl carbonate (DMC) in Ar-filled glovebox, vacuum dried, and transferred to the TEM holder without exposure to ambient air. The scale bars in (a–d), (e–h), and (i–l) are 5 µm, 1 µm, and 200 nm, respectively.

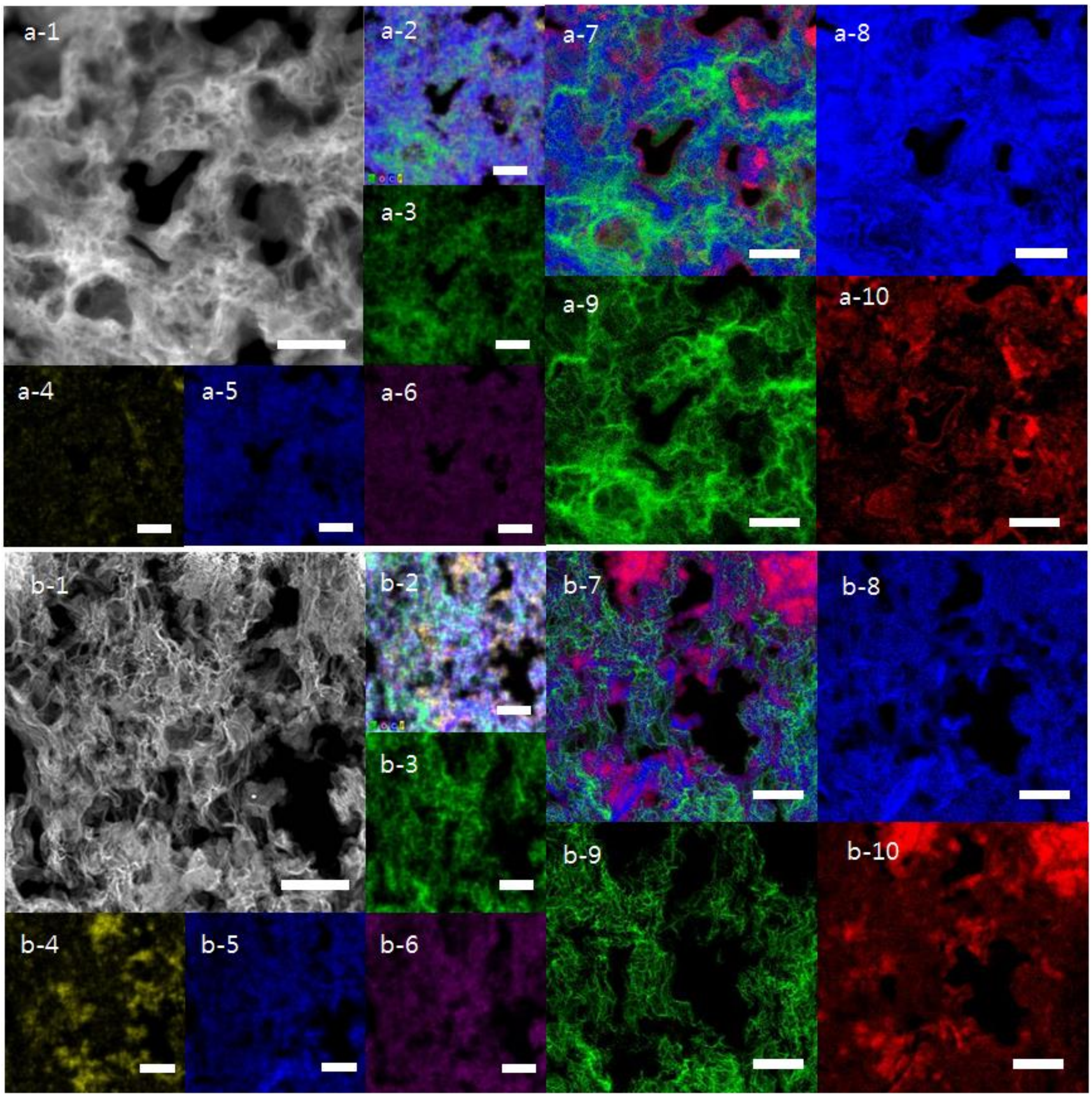

**Supplementary Fig.S 19**

High-angle annular dark field (HAADF) images and corresponding energy dispersive X-ray spectroscopy (EDS) and electron energy loss spectroscopy (EELS) spectrum image (SI) elemental mapping of delithiated type-A electrodes cycled in 2032-type coin half cells under DOD100% for (a) 23 and (b) 65 cycles. (a1, b1) HAADF images, (a2–6, b2–6) EDS elemental mapping; green, yellow, blue, and purple show Si, F, C, and O, respectively, and (a7–10, b7–10) EELS SI elemental mapping; blue, green, and red show C, Si, and Li, respectively. Prior to observation, the coin half-cell is cycled *ex situ* under CCCV-CC at 1 C (0.01 C current cutoff) for the target number of cycles, terminating the schedule at 1.5 V under CC. The electrodes are washed with dimethyl carbonate (DMC) in an Ar-filled glovebox, vacuum dried, and transferred to the TEM holder without exposure to ambient air (see Methods). The scale bars are all 200 nm.

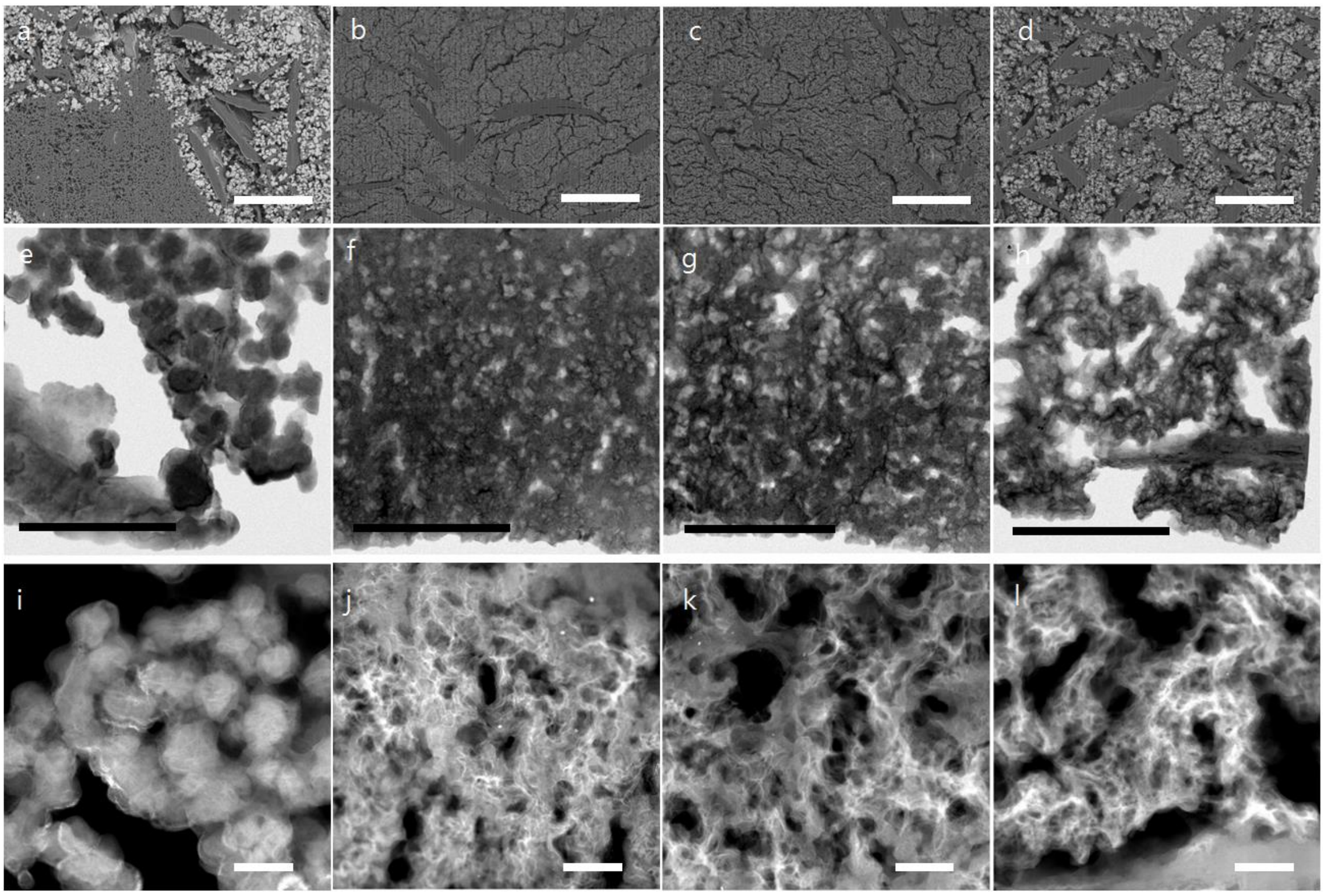

**Supplementary Fig.S 20**

Electron microscopy images of fully delithiated type-A electrodes cycled in 2032-type coin half cells under DOD90% over 107 cycles. (a–d) Back scattering electron (BSE) images of electrode cross sections sliced via focused ion beam (FIB), (e–h) bright-field (BF) transmission electron microscopy (TEM) images, and (i–l) high-angle annular dark field (HAADF) images of the sliced electrodes. (a,e,i) $3^{rd}$ (after the amorphisation of *c*-Si), (b,f,j) $23^{rd}$, (c,g,k) $65^{th}$, and (d,h,l) $107^{th}$ cycle (see *TEM* in Methods). Prior to observation, the coin cell battery is cycled *ex situ* under CCCV-CC at 1 C (0.01 C current cutoff) for the target number of cycles, terminating the schedule at 1.5 V under CC. The electrodes are washed with dimethyl carbonate (DMC) in Ar-filled glovebox, vacuum dried, and transferred to the TEM holder without exposure to ambient air. The scale bars in (a–d), (e–h), and (i–l) are 5 µm, 1 µm, and 200 nm, respectively.

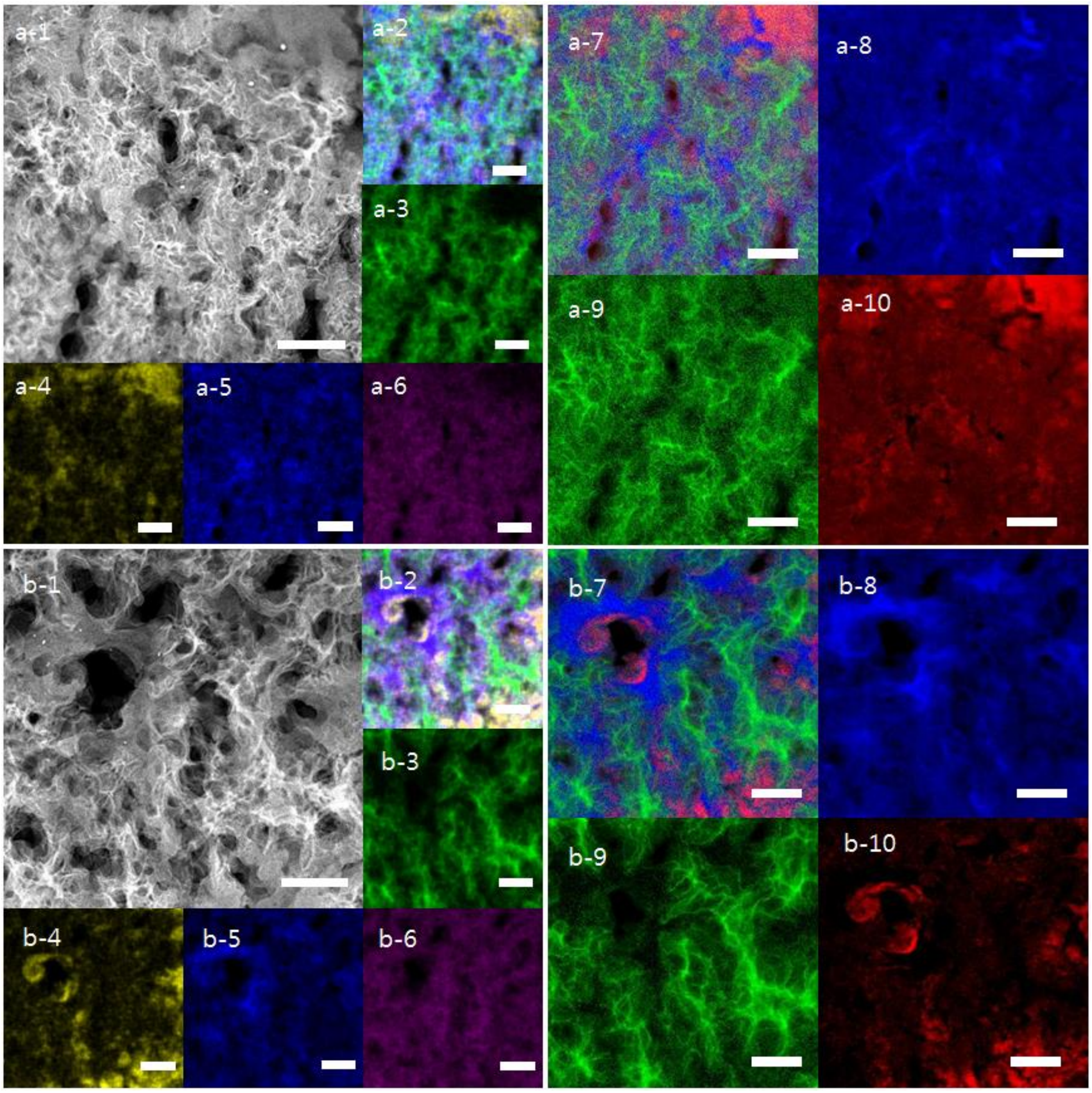

**Supplementary Fig.S 21**

High-angle annular dark field (HAADF) images and corresponding energy dispersive X-ray spectroscopy (EDS) and electron energy loss spectroscopy (EELS) spectrum image (SI) elemental mapping of delithiated type-A electrodes cycled in 2032-type coin half cells under DOD90% for (a) 23 and (b) 65 cycles. (a1, b1) HAADF images, (a2–6, b2–6) EDS elemental mapping; green, yellow, blue, and purple show Si, F, C, and O, respectively, and (a7–10, b7–10) EELS SI elemental mapping; blue, green, and red show C, Si, and Li, respectively. Prior to observation, the coin half-cell is cycled *ex situ* under CCCV-CC at 1 C (0.01 C current cutoff) for the target number of cycles, terminating the schedule at 1.5 V under CC. The electrodes are washed with dimethyl carbonate (DMC) in an Ar-filled glovebox, vacuum dried, and transferred to the TEM holder without exposure to ambient air (see Methods). The scale bars are all 200 nm.

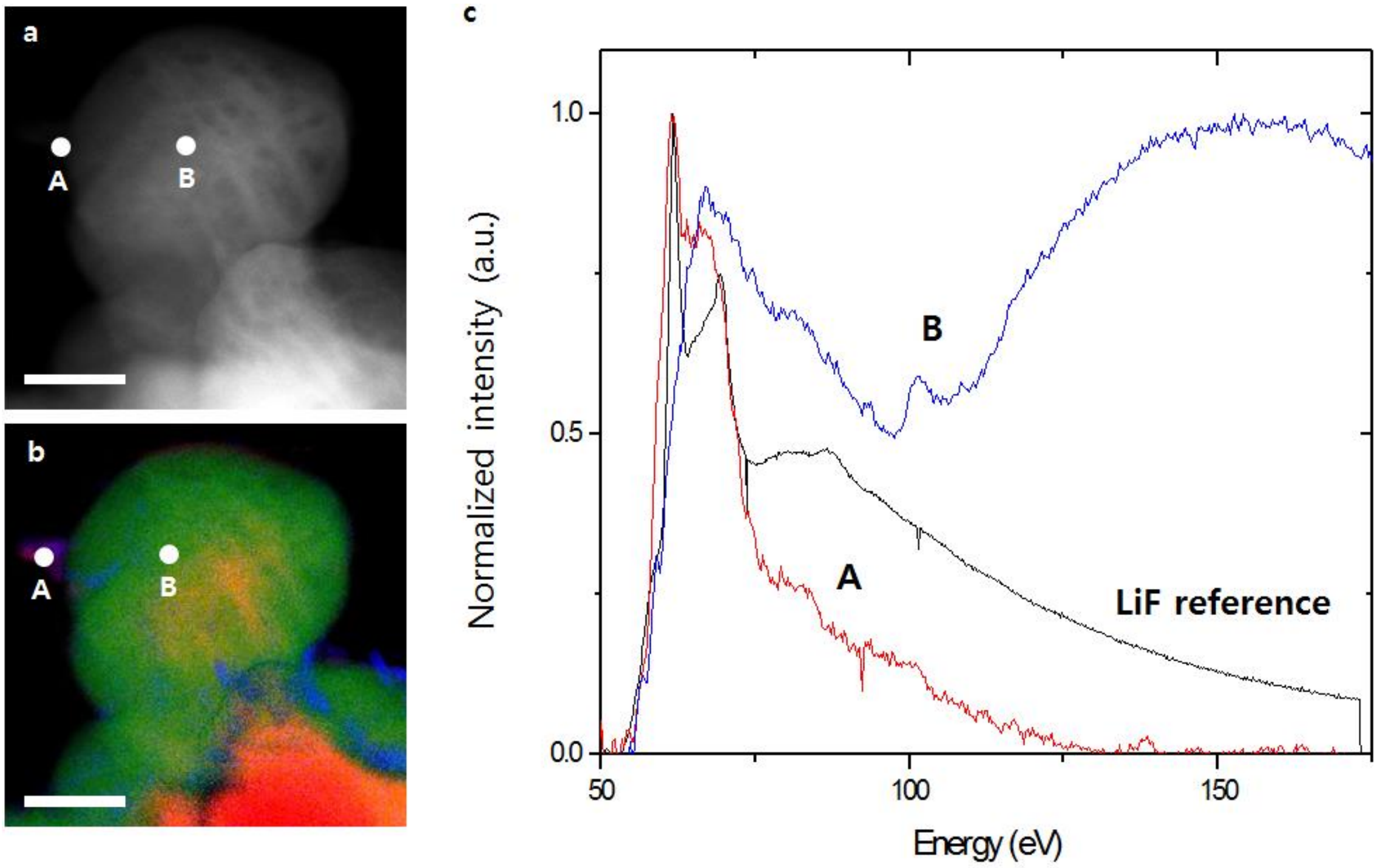

**Supplementary Fig.S 22**

Electron microscopy image and electron energy loss spectroscopy (EELS) spot analysis of a delithiated type-A electrode after the amorphisation of *c*-Si. (a) high-angle annular dark field (HAADF) image, (b) overlapped EELS spectrum image (SI) elemental mapping, in which green, blue, and red represent Si, C, and Li, respectively. The scale bars are 50 nm. (c) EELS spectra for the two spots shown in (a,b). These profiles show that LiF is concentrated near the surface of the spherical particle.

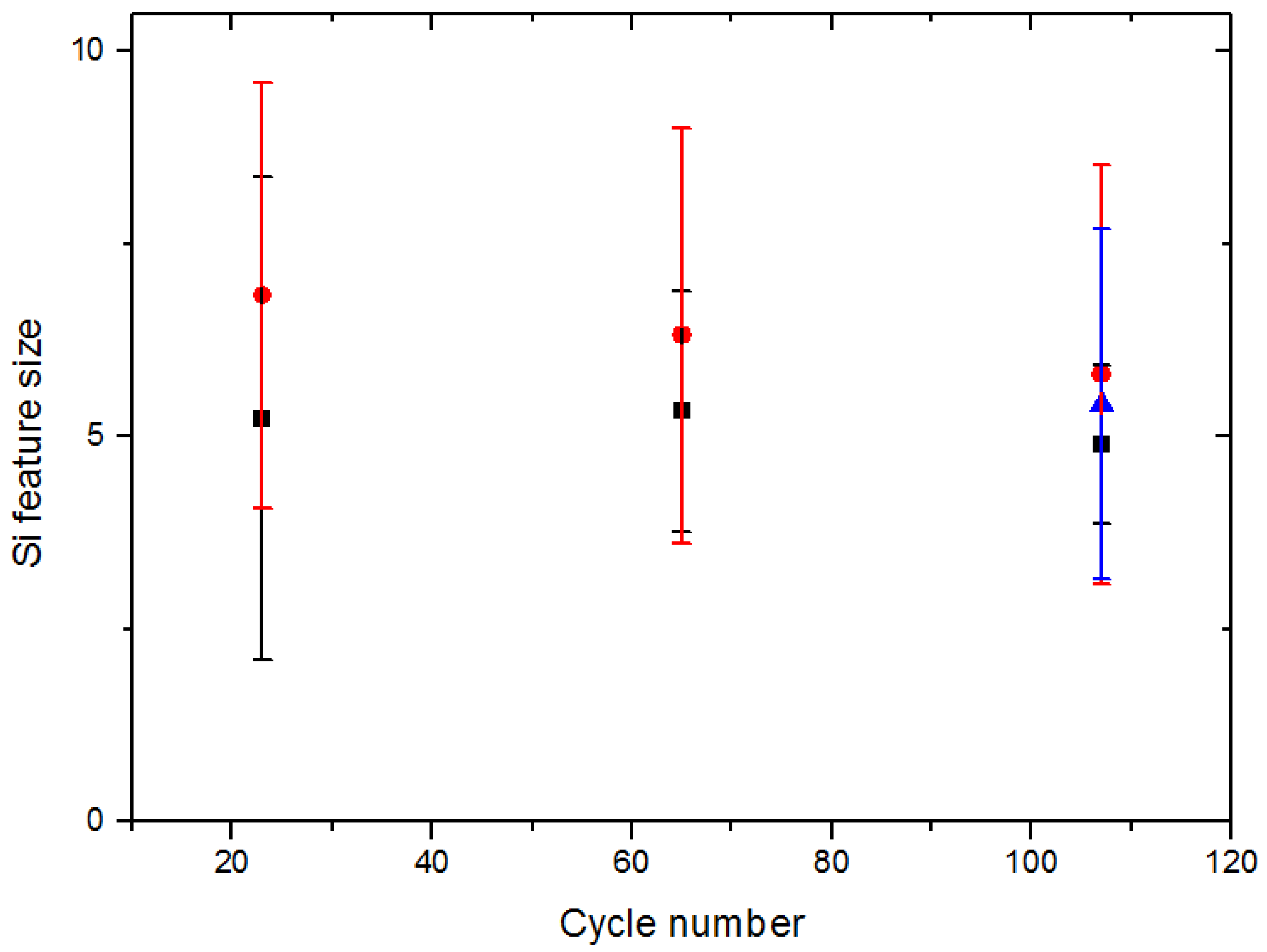

**Supplementary Fig.S 23**

Size of representative Si particles at various cycle numbers for type-A electrodes under different depth of discharge (DOD) controls. The size is determined from 100 randomly picked spots in the TEM images of fully delithiated type-A electrodes. The error bars show the standard deviations. Black, red, and blue data points correspond to depth of discharge 100% (DOD100%), DOD90%, and DOD80%, respectively.

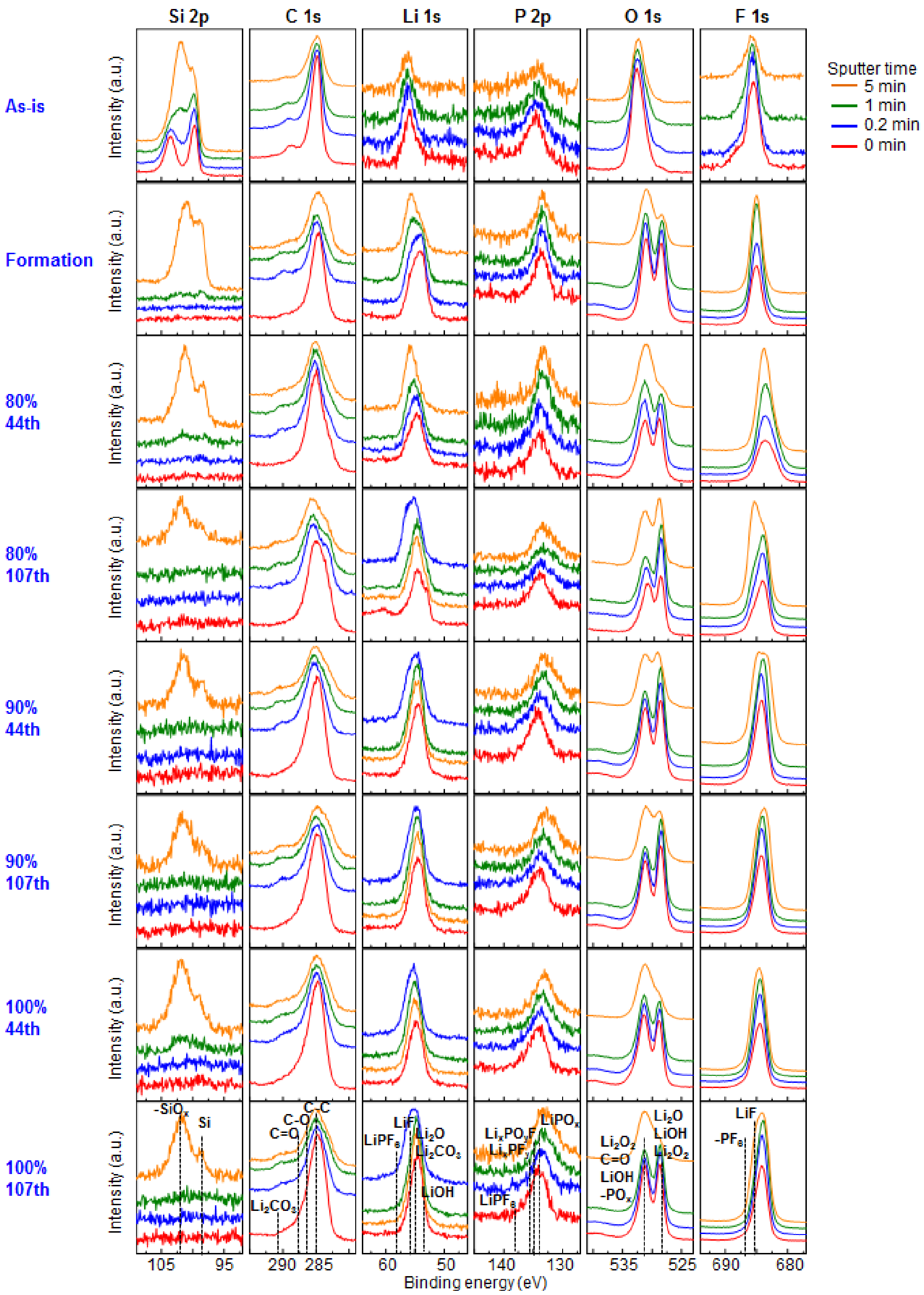

**Supplementary Fig.S 24**

Si-2p, C-1s, Li-1s, P-2p, O-1s, and F-1s XPS spectra with different sputtering times for type-A electrodes before cycling (1st row), after the amorphisation of *c*-Si (2nd row), and delithiated after the 44th and 107th cycles under DOD80–100% (3rd–8th rows). Prior to observation, the coin cells are cycled under CCCV-CC at 1 C (0.01 C current cutoff) for the target number of cycles, terminating the schedule by holding the potential at 1.5 V until the current decays to less than 0.001 C. The electrodes are washed using dimethyl carbonate (DMC) in Ar-filled glovebox, vacuum dried, and transferred to in-house XPS holder without exposure to ambient air.

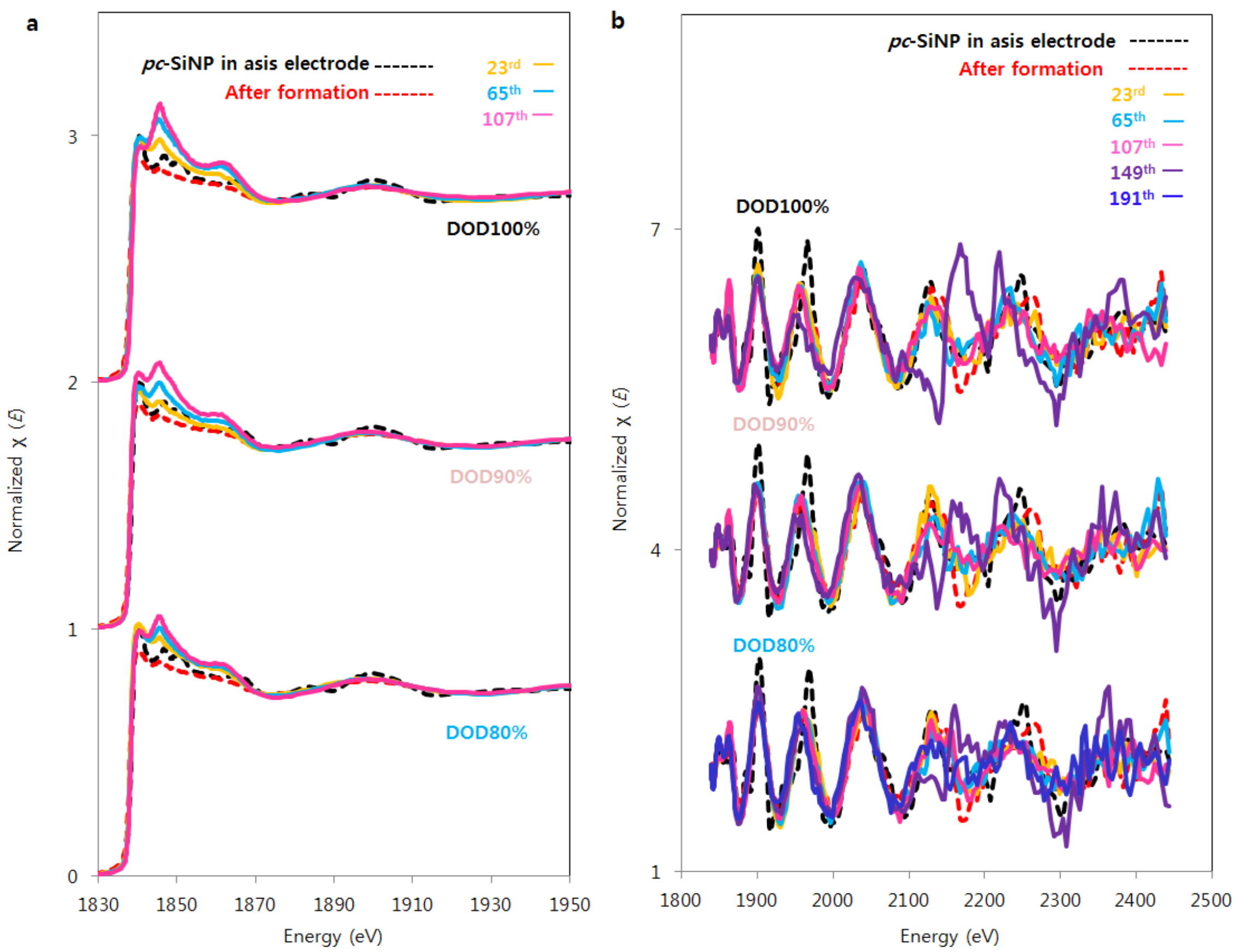

**Supplementary Fig.S 25**

X-ray absorption near edge structure (XANES) and extended X-ray absorption fine structure (EXAFS) profiles for the Si K-edge for fully delithiated type-A electrodes cycled in 2032-type coin half cells cycled under depth of discharge 80–100% (DOD80–100%) over 190 cycles. (a) Stacked XANES profiles and (b) stacked EXAFS profiles. Black and red dotted lines show electrodes before cycling and after the amorphisation of *c*-Si, respectively. Prior to observation, the coin cells made of type-A electrodes are cycled under CCCV-CC at 1 C (0.01 C current cutoff) for the target number of cycles, terminating the schedule by holding the potential at 1.5 V until the current decays to less than 0.001 C. The electrodes are washed using dimethyl carbonate (DMC) in Ar-filled glovebox, vacuum dried and transferred to the XAFS holder without exposure to ambient air (see Methods).

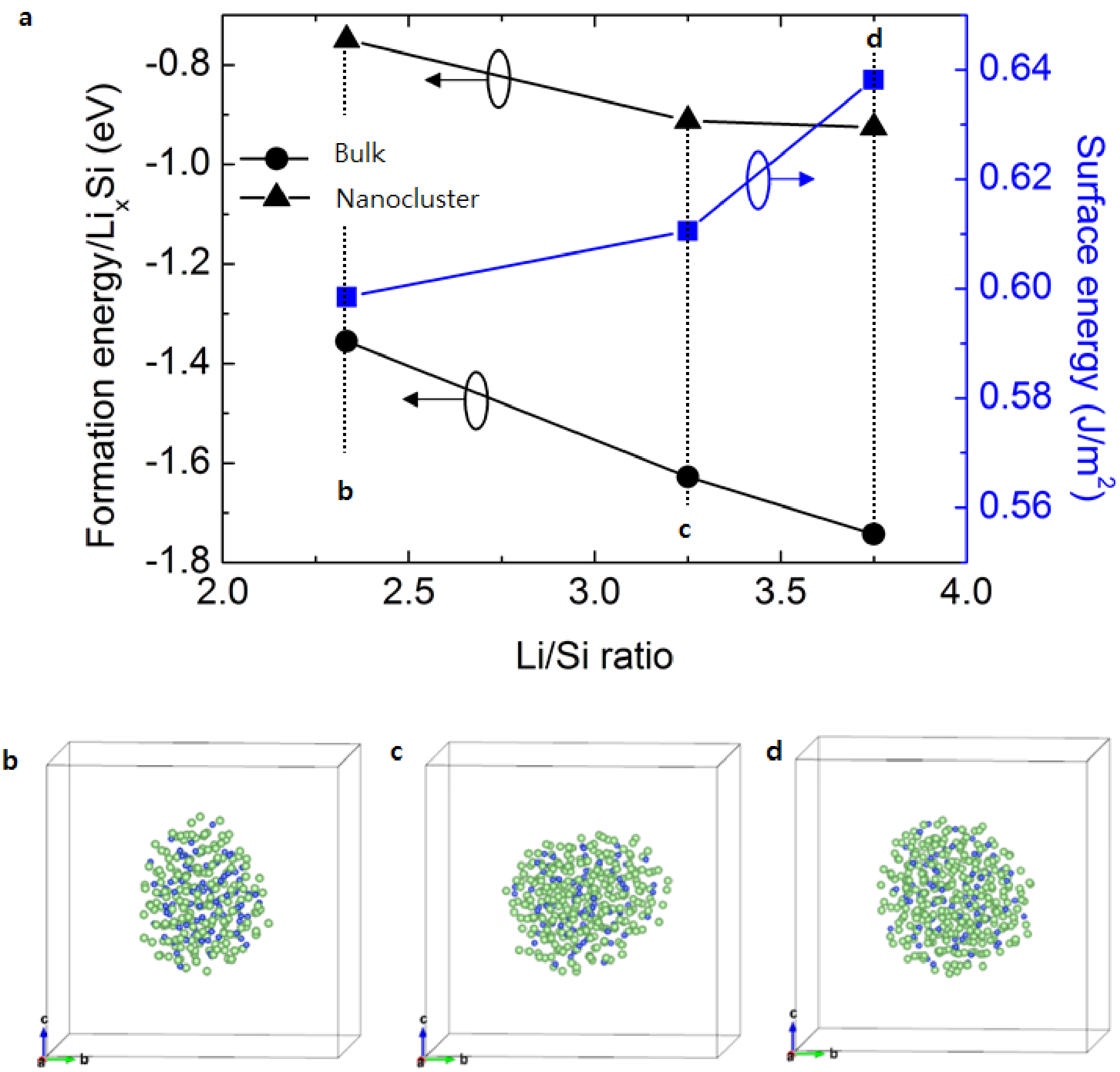

**Supplementary Fig.S 26**

(a) Formation energy and surface energy calculated with density functional theory (DFT, see *Computational simulations* in Methods) as a function of fractional Li concentration (x = Li/Si) for bulk *a*-$Li_xSi$ and ~2 nm Si nanoclusters for x = 2.33, 3.25, and 3.75. (b–d) Schematics of nano-$Li_xSi$ clusters at x = 2.33, 3.25, and 3.75. The final structures are created by performing a classical molecular dynamics simulation, implemented in the Large-scale Atomic/Molecular Massively Parallel Simulator (Lammps) package with the reactive force field (ReaxFF).